\documentclass[a4paper,11pt]{article}
\usepackage{jheppub} 
 \usepackage{amssymb,amsthm}
 \usepackage{amsmath,amssymb,amsfonts,bm,amscd}
 \usepackage{graphicx}
 
 \usepackage{caption, subcaption}
 
 \usepackage{tikz}
 
\def\CI{{\cal I}}

\def\diag{\mathop{\rm diag}\nolimits}

\def\SO{\mathop{\rm SO}}

\def\beq#1\eeq{\begin{align}#1\end{align}}

\usepackage{enumitem}

\def\fg{\mathfrak{g}}
\def\fh{\mathfrak{h}}

\def\fe{\mathfrak{e}}

\def\fsl{\mathfrak{sl}}
\def\fsu{\mathfrak{su}}
\def\fso{\mathfrak{so}}
\def\fsp{\mathfrak{sp}}

\def\half{\frac{1}{2}}

\title{$W$ algebras, Cosets and VOAs for 4d $\mathcal{N}=2$ SCFTs from M5 branes}

\author[a,b]{Dan Xie}
\author[a]{ Wenbin Yan}

\affiliation[a]{Yau Mathematics Science center, Tsinghua University, Beijing, 10084, China}
\affiliation[b]{Department of Mathematics, Tsinghua University, Beijing, 10084, China}

\abstract{We identify vertex operator algebras (VOAs) of a class of  Argyres-Douglas (AD) matters with two types of non-abelian flavor symmetries. They are the $W$ algebra defined using nilpotent orbit with partition $[q^m,1^s]$. Gauging above AD matters, we can find VOAs for more general $\mathcal{N}=2$ SCFTs engineered from 6d $(2,0)$ theories. For example, the VOA for general $(A_{N-1}, A_{k-1})$ theory is found as the coset of a collection of above $W$ algebras. Various new interesting  properties of 2d VOAs  such as  level-rank duality, conformal embedding, collapsing levels, coset constructions for known VOAs can be derived from 4d theory.}

\begin{document} 
\maketitle
\flushbottom

\section{Introduction}

\begin{figure}
	
	\tikzset{every picture/.style={line width=0.75pt}} 
	\centering
	\begin{tikzpicture}[x=0.75pt,y=0.75pt,yscale=-1,xscale=1]
	
	\draw   (125.52,135.74) .. controls (125.52,95.57) and (158.08,63) .. (198.26,63) .. controls (238.43,63) and (271,95.57) .. (271,135.74) .. controls (271,175.92) and (238.43,208.48) .. (198.26,208.48) .. controls (158.08,208.48) and (125.52,175.92) .. (125.52,135.74) -- cycle ;
	
	\node at (195,40) {{\LARGE$\mathfrak{g}$}};
	\draw   (192,89.48) -- (203.52,89.48) -- (203.52,101) -- (192,101) -- cycle ;
	\node at (178,172) {\Large $f$};
	\draw   (192.53,165.18) -- (203.58,175.9)(203.49,164.94) -- (192.62,176.13) ;
	\draw    (285,140) -- (345.5,140.47) ;
	\node at (178,97) {\Large $\Phi$};
	\draw [shift={(347.5,140.48)}, rotate = 180.44] [color={rgb, 255:red, 0; green, 0; blue, 0 }  ][line width=0.75]    (10.93,-3.29) .. controls (6.95,-1.4) and (3.31,-0.3) .. (0,0) .. controls (3.31,0.3) and (6.95,1.4) .. (10.93,3.29)   ;
	\draw (420,141) node  [align=right] {{\LARGE $\mathrm{VOA}(\mathfrak{g},\Phi,f)$}};
	\end{tikzpicture}
	\caption{A mapping of a 6d $(2,0)$ configuration to a 2d VOA, here $\fg$ is a simple Lie algebra, $\Phi$ is an irregular singularity, and $f$ represents a regular singularity.}
	\label{4d-to-2d}
\end{figure}
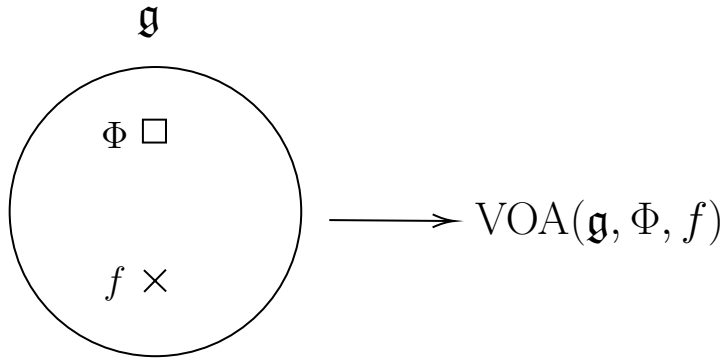

A remarkable correspondence between a four dimensional $\mathcal{N}=2$ 
superconformal field theory (SCFT) and a two dimensional vertex operator algebra (VOA) is found in \cite{Beem:2013sza}. which provides a promising organizing principle for the whole landscape of $\mathcal{N}=2$ theories (see \cite{Beem:2014rza, Lemos:2014lua, Lemos:2015orc, Cecotti:2015lab, 
Arakawa:2016hkg, 
Bonetti:2016nma, Song:2016yfd, Fredrickson:2017yka, Cordova:2017mhb, Song:2017oew, Buican:2017fiq, Beem:2017ooy, Pan:2017zie, Fluder:2017oxm, Choi:2017nur, Arakawa:2017fdq, Niarchos:2018mvl, Feigin:2018bkf, Creutzig:2018lbc, Bonetti:2018fqz} for some further developments). 
Once a 4d/2d pair is found, one can use the 
2d theory to learn 4d theory and vice versa. For example, one can compute the Schur index of 4d SCFT by calculating the vacuum character of 2d VOA which is often much easier to work out, meanwhile 4d result also motivates the study of certain 2d VOAs which received little attention before \cite{arakawa2018quasi}.

If our 4d SCFTs can be enginnered from string/M theory,  
it is possible to gain more insights about 4d/2d pair. In the past few years, People have found a large class of 4d $\mathcal{N}=2$ SCFTs 
by putting 6d $(2,0)$ theory on a Riemann surface with various defects \cite{Gaiotto:2009we,Gaiotto:2009hg,Xie:2012hs,Wang:2015mra,Wang:2018gvb}, so we should have a map between such 6d configuration and a 
2d VOA, which is schematically depicted in figure \ref{4d-to-2d}. The 2d VOA for the theory defined using only regular punctures was studied in \cite{Beem:2014rza,arakawa2018chiral,Lemos:2014lua}. The most important step is to understand the VOA
for the theory defined by three full punctures (maximal flavor symmetry), since the general cases can 
be found from following two correspondence between operations in 4d theory and operations in 2d VOAs:
\begin{itemize}
\item On the 4d side one can reduce the full puncture to a generic puncture labeled by a 
nilpotent orbit $f$. The 2d counterpart of such operation corresponds to the quantum Drinfeld-Sokolov (qDS) reduction of the original VOA \cite{Beem:2014rza}, 
as sketched in figure \ref{qDS}.

\item If a theory is formed by conformally gauging various matters together, 
the VOA is formed by performing cosets on those VOAs \cite{Beem:2013sza,Beem:2014rza} as in figure \ref{gauging}. 
\end{itemize}

Things become more interesting and complicated if we consider Argyres-Douglas (AD) theories which are engineered using one irregular singularity $\Phi$
and one regular singularity $f$ on a sphere as in figure \ref{4d-to-2d}. The correspondence between 2d VOAs and certain AD theories were discussed in \cite{Buican:2015hsa, Cecotti:2015lab, Buican:2015tda, Song:2015wta, Buican:2016arp,Cordova:2015nma,Xie:2016evu,Cordova:2016uwk,Creutzig:2017qyf,Cordova:2017mhb,Cordova:2017ohl,Song:2017oew, Buican:2017uka, Wang:2018gvb,Creutzig:2018lbc}. More generally it was conjectured in \cite{Song:2017oew,Xie:2016evu,Wang:2018gvb} that if there is no mass parameter associated with the irregular singularity,  the corresponding VOA is just the vacuum module of a $W$ algebra denoted by $W^{k'}(\fg,f)$ which is obtained through the quantum Hamiltonian reduction from the vacuum $\hat{\fg}$-module of level $k^{'}$ \cite{Frenkel:1992ju}, where $k'$ 
is determined by the data $\Phi$.  Again, the choice of the generic regular puncture $f$ determines the qDS reduction type.

\begin{figure}
	\centering
	\begin{tikzpicture}
	
	\node at (-2.5,0) {4d};
	\draw (0,0) circle [radius=2];
	\draw (0-0.2,1-0.2) rectangle (0.2,1.2);
	\draw (0,-1) circle [radius=0.2];
	\draw [fill] (0,-1) circle [radius=0.05];
	\node [below] at (0,-1.2) {full};
	
	\draw [thick,->] (-0.8+3,0)--(1.8+3,0);
	\node [above] at (0.5+3,0) {closing puncture};
	
	\draw (0+7,0) circle [radius=2];
	\draw (0-0.2+7,1-0.2) rectangle (0.2+7,1.2);
	\draw [fill] (0+7,-1) circle [radius=0.05];
	\node [below] at (0+7,-1.2) {generic $f$};
	
	\node at (-2.5,-3) {2d};
	\node at (0,-3) {$\mathrm{VOA}_1$};
	\draw [thick,->] (-0.8+3,-3)--(1.8+3,-3);
	\node [above] at (0.5+3,-3) {DS reduction};
	\node at (7,-3) {$\mathrm{VOA}_2$};
	
	\end{tikzpicture}
	
	\caption{Closing puncture in Class S construction corresponds to qDS reduction of 2d VOA.}
	\label{qDS}
\end{figure}

\begin{figure}
	\centering
	\begin{tikzpicture}
	
	\node at (-2.5,0) {4d};
	\draw [thick] (0,0) circle [radius=0.5];
	\node at (0,0) {$G$};
	\draw [thick] (-0.5,0) -- (-1.5,0);
	\draw [thick] (0.5,0) -- (1.5,0);
	\node at (-1.8,0) {$T_1$};
	\node at (1.8,0) {$T_2$};
	
	\node at (-2.5,-2) {2d};
	\node [above] at (0,-2) {$V_1\oplus V_2$};
	\draw (-0.7,-2) -- (0.7,-2);
	\node [below] at (0,-2) {$G_{-2h^\vee}$};
	\node at (0,-3) {$V_{k_1}(G)\subset V_1$, $V_{k_2}(G)\subset V_2$};
	\node at (0,-3.5) {$k_1+k_2=-2h^{\vee}$};
	
	\end{tikzpicture}
	
	\caption{Four dimensional conformal gauging is interpreted as cosets of two dimensional VOAs. $T_1$ and $T_2$ are four dimensional matter
	with non-abelian flavor symmetry $G$, and one gauge them to get a new conformal field theory with exact marginal deformation. Here $V_1$ and $V_2$ are 
	VOAs for matter $T_1$ and $T_2$, and each of them has an affine vertex subalgebra $V_{k_i}(G)$.}
	\label{gauging}
\end{figure}
 
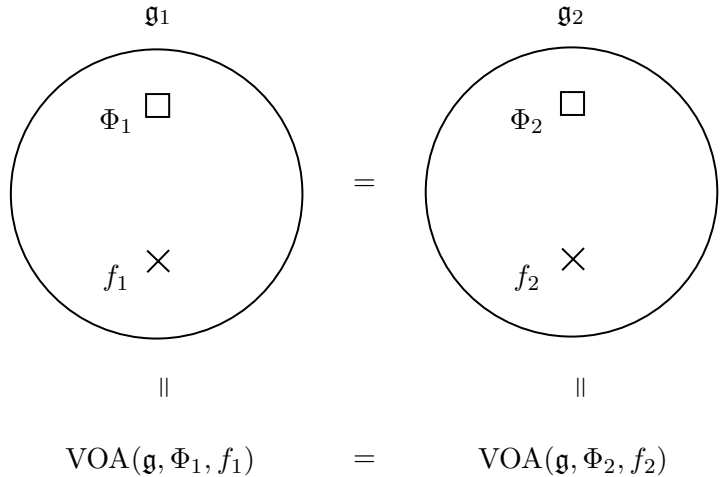
\begin{figure}
	\centering	
	\tikzset{every picture/.style={line width=0.75pt}} 
	\begin{tikzpicture}[x=0.75pt,y=0.75pt,yscale=-1,xscale=1]
	
	\draw   (118.52,103.74) .. controls (118.52,63.57) and (151.08,31) .. (191.26,31) .. controls (231.43,31) and (264,63.57) .. (264,103.74) .. controls (264,143.92) and (231.43,176.48) .. (191.26,176.48) .. controls (151.08,176.48) and (118.52,143.92) .. (118.52,103.74) -- cycle ;
	\draw   (186,53.48) -- (197.52,53.48) -- (197.52,65) -- (186,65) -- cycle ;
	\draw   (186.53,132.18) -- (197.58,142.9)(197.49,131.94) -- (186.62,143.13) ;
	\draw   (325.52,102.74) .. controls (325.52,62.57) and (358.08,30) .. (398.26,30) .. controls (438.43,30) and (471,62.57) .. (471,102.74) .. controls (471,142.92) and (438.43,175.48) .. (398.26,175.48) .. controls (358.08,175.48) and (325.52,142.92) .. (325.52,102.74) -- cycle ;
	\draw   (393,52.48) -- (404.52,52.48) -- (404.52,64) -- (393,64) -- cycle ;
	\draw   (393.53,131.18) -- (404.58,141.9)(404.49,130.94) -- (393.62,142.13) ;
	
	\draw (295,99) node   {$=$};
	\draw (295,238) node   {$=$};
	\draw (171,67) node   {$\Phi _{1}$};
	\draw (376,67) node   {$\Phi _{2}$};
	\draw (171,146) node   {$f_{1}$};
	\draw (376,146) node   {$f_{2}$};
	\draw (194,201) node [rotate=-90]  {$=$};
	\draw (402,201) node [rotate=-90]  {$=$};
	\draw (192,14) node   {$\mathfrak{g}_{1}$};
	\draw (398,14) node   {$\mathfrak{g}_{2}$};
	\draw (193,238) node   {$\mathrm{VOA}(\mathfrak{g},\Phi_1,f_1)$};
	\draw (399,238) node   {$\mathrm{VOA}(\mathfrak{g},\Phi_2,f_2)$};	
	\end{tikzpicture}
	\caption{Equivalence of 6d $(2,0)$ configuration implies the equivalence of 2d VOAs.}
	\label{equival}
\end{figure}

There remains the question on determining the VOA for remaining cases,
and the main purpose of this paper is to partially solve this problem by using following two facts:
\begin{itemize}
	\item Irregular singularities with mass deformations often have exact marginal deformations and their weakly coupled gauge theory descriptions 
	are found in \cite{Xie:2016uqq,Xie:2017vaf,Xie:2017aqx}. They are described by gauging AD matters with 
	at least two types of non-abelian flavor symmetries\footnote{One can get more non-abelian flavor symmetries from closing regular puncture, but gauge theory description found in \cite{Xie:2017vaf} has to use the non-abelian flavor symmetry arising from irregular singularity.}. Therefore once we find the VOA for 
	such AD matter, the VOA of the full theory can be found by the coset construction.
	\item A crucial observation for this paper is that all AD matters with two non-abelian flavor symmetries studied in \cite{Xie:2017vaf,Xie:2017aqx} can be engineered by a different realization whose VOAs are known as certain $W$ algebra studied in \cite{Song:2017oew,Xie:2016evu,Wang:2018gvb}. The $W$ algebra takes the form $W^{k'}(\fg,[q^m,1^s])$ with $k'$ depends on parameter $(q,m,s)$, see the summary in section \ref{summary}. 
\end{itemize}
There are several new interesting features about VOAs of AD matters studied in this paper:
\begin{enumerate}[label=\alph*)]
\item The VOA has an affine VOA $V_{k_1}(\fg_1)\oplus V_{k_2}(\fg_2)$ as its subalgebra, where $V_k(\fg)$ is the affine Kac-Moody (AKM) vertex algebra\footnote{By affine Kac-Moody and $W$ algebra, we mean the irreducible vertex operator algebra constructed from the vacua module of AKM and W algebra.} of Lie algebra $\fg$ with level $k$. This affine VOA has the same central charge as the $W$ algebra and therefore we found a large number of new possible conformal embeddings of VOAs. 
\item The simple fact that a theory can be engineered in various ways can often tell us interesting properties about VOAs, see figure \ref{equival}. For instance, we can derive new level-rank type duality.
\item S-duality of 4d theory implies the equivalence between different cosets 
constructions of a single VOA. 
\end{enumerate}
Once VOAs for AD matters are known, 
we are able to write down VOAs for more general theories engineered from M5 branes. For example, the VOA for 
$(A_{N-1}, A_{k-1})$ theory with arbitrary $N$ and $k$  is found by using its weakly coupled gauge theory descriptions in section \ref{sec:weaklycoupledAD}.

This paper is organized as the following. Section \ref{sec:results} reviews known results about the mapping between AD theories engineered from M5 branes and VOAs. Section \ref{sec:ADwithTwoFlavors} studies VOAs corresponding to AD matters with two distinct non-abelian flavor symmetries. Section \ref{sec:Higgs} focuses on the associated variety of the VOA for a given AD matter, which determines the Higgs branch chiral ring of the theory. Section \ref{sec:VOAwithDef} describes weakly coupled descriptions of AD theories and the coset 
construction of the corresponding VOA. Section \ref{sec:ConformalEmbedding} discusses  conformal embeddings
and VOAs for most general AD theories. Finally, a summary is 
given in section \ref{sec:conclusion}.

\section{Known results}
\label{sec:results}

A four dimensional $\mathcal{N}=2$ SCFT has a bosonic symmetry group $SO(2,4)\times SU(2)_R\times U(1)_R \times G_F$, where $SO(2,4)$ is the four dimensional conformal group, $SU(2)_R\times U(1)_R$ is the R symmetry group which exists for every $\mathcal{N}=2$ SCFT, and $G_F$ is the flavor symmetry group which might be absent for some theories. The representation theory of 4d $\mathcal{N}=2$ superconformal algebra is studied in \cite{Dolan:2002zh}, in which short representations  (where some of states in the representation are annihilated by a fraction of supercharges) were completely classified. Important half-BPS operators include primary operators of multiplets ${\cal E}_{r}$ and $\hat{B}_R$.

The moduli space of vacua of a 4d $\mathcal{N}=2$ SCFT is extremely rich. It consists of a 
Coulomb branch, whose low energy effective theory involves abelian gauge theory in general. The Coulomb branch is parameterized by expectation values of primary operators of ${\cal E}_{r}$ multiplets, and the low energy effective theory is described by a Seiberg-Witten geometry\cite{Seiberg:1994rs,Seiberg:1994aj}. The set of rational numbers $[r_1,\ldots, r_s]$ of $U(1)_r$ charges of ${\cal E}_{r}$ (unitarity implies that $r_i>1$) is an important set associated to a 4d $\mathcal{N}=2$ SCFT. The $U(1)_r$ symmetry acts non-trivially on the Coulomb branch while $SU(2)_R\times G_F$ symmetry acts trivially. 

Some theories also have a Higgs branch where the gauge group is completely broken in general should the SCFT has a gauge theory description. The Higgs branch, being a conical hyperkhaler manifold, is parameterized by expectation values of primary operators of $\hat{B}_R$ multiplets. One of important questions about the Higgs branch is to determine the affine chiral ring of the cone. The $SU(2)_R\times G_F$ symmetry acts non-trivially on Higgs branch, while $U(1)_R$ symmetry acts trivially.

Besides the Coulomb branch spectrum (just a set of rational numbers $r_i>1$) and 
the Higgs branch chiral ring, one would also like to determine three interesting 
quantities of a 4d $\mathcal{N}=2$ SCFT: central charges $a_{4d}$ and $c_{4d}$ which is defined using the energy-momentum tensor, and the flavor central charge $k_G$. 

There is an interesting set of short multiplets called Schur sector\cite{Gadde:2011uv} which consists of Higgs branch operators $\hat{B}_R$, energy momentum tensors, and etc. Moreover, one can define a Schur index which counts those operators. It was proposed in \cite{Beem:2013sza} that 
one can get a 2d VOA from the Schur sector of a 4d $\mathcal{N}=2$ SCFT, and the basic 4d/2d correspondence used in current paper is\cite{Beem:2013sza}:
\begin{itemize}
	\item There is an AKM subalgebra ($V_{k_{2d}}(\fg)$) in 2d VOA, where $\fg$ is the Lie algebra of four dimensional flavor symmetry $G_F$.
	\item The 2d central charge $c_{2d}$ and the level of AKM algebra $k_{2d}$ are related to the 4d central charge $c_{4d}$ and the flavor central charge $k_F$ as
	\begin{equation}
	c_{2d}=-12 c_{4d},~~k_{2d}=-k_F\footnote{Our normalization of $k_F$ is half of that of \cite{Beem:2013sza,Beem:2014rza}.}.
	\label{eq:centralchargerelation}
	\end{equation}
	\item The (normalized) vacuum character of 2d VOA is the 4d Schur index $\CI(q)$.
\end{itemize}
Many 4d/2d pairs are found in \cite{Beem:2014rza,arakawa2018chiral,Lemos:2014lua,Song:2017oew,Xie:2016evu,Wang:2018gvb, Buican:2015hsa, Buican:2015tda, Song:2015wta, Buican:2016arp, Cordova:2015nma, Buican:2017uka, Cordova:2016uwk, Cordova:2017mhb,Cordova:2017ohl,Creutzig:2017qyf,Creutzig:2018lbc}. Various interesting properties of 4d theory using 2d VOA are studied in \cite{Lemos:2015awa,Lemos:2015orc,Lemos:2016xke, Buican:2015hsa, Buican:2015tda, Buican:2017rya,Buican:2019huq,Beem:2018duj,Beem:2017ooy,Fredrickson:2017yka,Kozcaz:2018usv,Song:2015wta,Song:2016yfd,Song:2017oew,Fluder:2017oxm,Agarwal:2018zqi,Nishinaka:2016hbw,Nishinaka:2018zwq,Kiyoshige:2018wol}.  See also \cite{Feigin:2018bkf,Dedushenko:2017tdw, Nishinaka:2018zwq} for  VOAs corresponding to 6d $(2,0)$ theory on a four manifold. Moreover VOAs corresponding to three dimensional $\mathcal{N}=4$ theory are discussed in \cite{Costello:2018fnz,Costello:2018swh}.

VOAs arise from study of the chiral part of two dimensional conformal field theories.
A general definition has been given by mathematicians\cite{kac1998vertex}, but it seems difficult 
to construct them abstractly. On the other hand,   
a VOA is often defined as an irreducible (which is also called simple) vacuum module of a particular algebra, which seems much more tractable. The simplest case 
is the Virasoro algebra whose elements are the modes of energy momentum tensor $T(z)=\sum L_n z^{-n-2}$: 
\begin{equation}
[L_n,L_m]=(n-m)L_{n+m}+{c\over 12}n(n^2-1)\delta_{n+m,0}.
\end{equation}
Another important algebra is the AKM algebra associated with a simple Lie algebra $\fg$,
\begin{equation}
[J_n^a,J_m^b]=if^{ab}_cJ_{n+m}^c+k n \delta_{ab}\delta_{n+m,0},
\end{equation}
with $f^{ab}_c$ the structure constant of $\fg$.
One can construct an energy-momentum tensor using the Sugawara construction, and the simple vacuum module is indeed an VOA\cite{frenkel1992vertex}.
The representation theory of above two algebras has been studied thoroughly in the literature (see \cite{francesco2012conformal} and many others).

Finally, one can have intricated algebras built from a set of higher spin fields $(W_{d_1},\ldots, W_{d_i})$, which are called $W$ algebras\cite{bouwknegt1993w}. The full algebra content and its representation theory of a $W$ algebra is very complicated, however, if a $W$ algebra can be derived from the qDS reduction of an AKM algebra\cite{Frenkel:1992ju}, one can actually derive lots of important information of this $W$ algebra from the representation theory of the AKM algebra. 

In general it is difficult to work out the full Schur sector of a four dimensional $\mathcal{N}=2$ SCFT. However, using a surprising fact that almost all the known 2d VOA for the above 4d/2d mapping involves the $W$ algebra derived from qDS reduction\cite{arakawa2017introduction}, 
one can hope that a lot can be learned about 4d theories by using existing knowledge of 2d VOAs. 

\subsection{AD theories correspond to $W^{k'}( \mathfrak{g},f)$ algebras}
\label{sec:ADandWk}
One can engineer a large class of four dimensional $\mathcal{N}=2$ SCFTs by starting with a 6d $(2,0)$ theory of type $\mathfrak{j}=ADE$ on a sphere with an irregular singularity and a regular singularity\cite{Gaiotto:2009we,Gaiotto:2009hg,Xie:2012hs,Wang:2015mra,Wang:2018gvb}\footnote{See the appendix \ref{sec:app:ADs} for relations between this construction and other constructions}. 
The Coulomb branch is captured by a Hitchin system with singular boundary conditions near the singularity. The Higgs field of the Hitchin system near the irregular singularity takes the following form,
\begin{equation}
\Phi={T\over z^{2+{k\over b}}}+\ldots.
\end{equation}
Here $T$ is determined by a positive principle grading of Lie algebra $\mathfrak{j}$\cite{reeder2012gradings}, and is a regular semi-simple element of $\mathfrak{j}$. $k>-b$ and  is an integer. Subsequent terms are chosen 
such that they are compatible with the leading order term (essentially the grading determines the choice of these terms). We call them $J^{(b)}[k]$ type irregular puncture. Theories constructed using only above irregular singularity can also be engineered using a three dimensional singularity in type IIB string theory as summarized in table \ref{table:sing}\cite{Xie:2015rpa}. 

\begin{table}[!htb]
\begin{center}
	\begin{tabular}{ |c|c|c|c| }
		\hline
		$ \mathfrak{j}$& $b$  & Singularity   \\ \hline
		$A_{N-1}$&$N$ &$x_1^2+x_2^2+x_3^N+z^k=0$\\ \hline
		$~$  & $N-1$ & $x_1^2+x_2^2+x_3^N+x_3 z^k=0$\\ \hline
		
		$D_N$  &$2N-2$ & $x_1^2+x_2^{N-1}+x_2x_3^2+z^k=0$ \\     \hline
		$~$  & $N$ &$x_1^2+x_2^{N-1}+x_2x_3^2+z^k x_3=0$ \\     \hline
		
		$E_6$&12  & $x_1^2+x_2^3+x_3^4+z^k=0$   \\     \hline
		$~$ &9 & $x_1^2+x_2^3+x_3^4+z^k x_3=0$   \\     \hline
		$~$  &8 & $x_1^2+x_2^3+x_3^4+z^k x_2=0$    \\     \hline
		
		$E_7$& 18  & $x_1^2+x_2^3+x_2x_3^3+z^k=0$   \\     \hline
		$~$&14   & $x_1^2+x_2^3+x_2x_3^3+z^kx_3=0$    \\     \hline

		$E_8$ &30   & $x_1^2+x_2^3+x_3^5+z^k=0$  \\     \hline
		$~$  &24  & $x_1^2+x_2^3+x_3^5+z^k x_3=0$  \\     \hline
		$~$  & 20  & $x_1^2+x_2^3+x_3^5+z^k x_2=0$  \\     \hline
	\end{tabular}
\end{center}
\caption{Three-fold isolated quasihomogenous singularities of cDV type corresponding to the $J^{(b)}[k]$  irregular punctures of the regular-semisimple type in \cite{Wang:2015mra}. These 3d singularity is very useful in extracting the Coulomb branch spectrum, see \cite{Xie:2015rpa}. }
\label{table:sing}
\end{table}

One can add another regular singularity which is labeled by a nilpotent orbit $f$ of $\mathfrak{j}$ (We use Nahm labels such that the trivial orbit corresponding to regular puncture with maximal flavor symmetry). A detailed discussion about these defects can be found in  \cite{Chacaltana:2012zy}. If there is no mass parameter
encoded in the irregular singularity $\Phi$ (which means that the $z^{-1}$ term in $\Phi$ is not allowed, and we also assume that singular parts of $\Phi$ are diagonalized near the irregular singularity), the 
VOA which corresponds to this 4d SCFT is the vacuum module of following $W$ algebra\footnote{We always use the irreducible vacuum module as the VOA corresponding to 4d theory.}\cite{Xie:2016evu,Song:2017oew},
\begin{equation}
W^{k'}(\mathfrak{j},f),~~~~k'=-h^{\vee}+{b\over k+b}.
\end{equation}
Here $h^\vee$ is the dual Coxeter number of $\mathfrak{j}$, and the $W$ algebra is defined as the qDS reduction from 
the affine VOA\cite{kac2003quantum}.

To get non-simply laced flavor groups, we need to consider the outer-automorphism twist of ADE Lie algebra and its Langlands dual. A systematic study of these AD theories 
was performed in \cite{Wang:2018gvb}. Denoting the twisted Lie algebra of $\mathfrak{j}$ as $\fg^{\vee}$ and its Langlands dual as $\fg$, outer-automorphisms and twisted algebras of $\mathfrak{j}$ are summarized in table \ref{table:outm}. The irregular singularity of regular semi-simple type is also classified as in table \ref{table:SW} with the following form,
\begin{equation}
\Phi={T^t\over z^{2+{k\over b}}}+\ldots
\end{equation} 
Here $T^t$ is an element of Lie algebra $\fg^{\vee}$ or other parts of the decomposition of $\mathfrak{j}$ under outer automorphism. $k>-b$, and
the novel thing  is that $k$ could take half-integer value or 
in  thirds ($\fg=G_2$). One can also represent those irregular singularities by 
3-fold singularities as in table \ref{table:SW}.

\begin{table}[htb]
	\begin{center}
		\begin{tabular}{ |c|c| c|c|c|c| }
			\hline
			$  j $ ~&$A_{2N}$ &$A_{2N-1}$ & $D_{N+1}$  &$E_6$&$D_4$ \\ \hline
			Outer-automorphism $o$  &$Z_2$ &$Z_2$& $Z_2$  & $Z_2$&$Z_3$\\     \hline
			Invariant subalgebra  $\fg^\vee$ &$B_N$&$C_N$& $B_{N}$  & $F_4$&$G_2$\\     \hline
			Flavor symmetry $\fg$ &$C_N^{(1)}$&$B_N$& $C_{N}^{(2)}$  & $F_4$&$G_2$\\     \hline
		\end{tabular}
	\end{center}
	\caption{Outer-automorphisms of simple Lie algebras $ j$, its invariant subalgebra $ g^\vee$ and flavor symmetry $ g$ from the Langlands dual of $g^\vee$.}
	\label{table:outm}
\end{table}

\begin{table}[!htb]
	\begin{center}
		\begin{tabular}{|c|c|c|c|}
			\hline
			$j$ with twist & $b_t$ & SW geometry at SCFT point & $\Delta[z]$ \\ \hline
			$A_{2N}/Z_2$ & $4N+2$ &$x_1^2+x_2^2+x^{2N+1}+z^{k+{1\over2}}=0$ & ${4N+2\over 4N+2k+3}$  \\ \hline
			~& $2N$ & $x_1^2+x_2^2+x^{2N+1}+xz^k=0$ & ${2N\over k+2N}$ \\ \hline
			$A_{2N-1}/Z_2$& $4N-2$ & $x_1^2+x_2^2+x^{2N}+xz^{k+{1\over2}}=0$ & ${4N-2\over 4N+2k-1}$ \\ \hline
			~ & $2N$ &$x_1^2+x_2^2+x^{2N}+z^{k}=0$ & ${2N\over 2N+k}$  \\ \hline
			$D_{N+1}/Z_2$& $2N+2$ & $x_1^2+x_2^{N}+x_2x_3^2+x_3z^{k+{1\over2}}=0$ & ${2N+2\over 2k +2N+3}$ \\ \hline
			~ & $2N$ &$x_1^2+x_2^{N}+x_2x_3^2+z^{k}=0$ & ${2N\over k+2N}$  \\ \hline
			$D_4/Z_3$ & $12$ &$x_1^2+x_2^{3}+x_2x_3^2+x_3z^{k\pm {1\over3}}=0$ & ${12\over 12+3k\pm1}$  \\ \hline
			~& $6$ &$x_1^2+x_2^{3}+x_2x_3^2+z^{k}=0$ & ${6\over 6+k}$  \\ \hline
			$E_6/Z_2$& $18$ &$x_1^2+x_2^{3}+x_3^4+x_3z^{k+{1\over2}}=0$ & ${18\over 18+2k+1}$  \\ \hline
			~& $12$ &$x_1^2+x_2^{3}+x_3^4+z^{k}=0$ & ${12\over 12+k}$  \\ \hline
			~ & $8$ &$x_1^2+x_2^{3}+x_3^4+x_2z^{k}=0$ & ${8\over 12+k}$  \\ \hline
		\end{tabular}
		\caption{Seiberg-Witten geometry of twisted theories at the SCFT point.}
		\label{table:SW}
	\end{center}
\end{table}
We could again add 
a twisted regular puncture labeled also by a nilpotent orbit $f$ of $\fg$. 
If there is no mass parameter  in the irregular singularity,
the corresponding VOA is given by following $W$ algebra\cite{Wang:2018gvb},
\begin{equation}
\label{eq:ADtoWalgebra}
\boxed{W^{k'}(\fg,f),~~k'=-h^{\vee}+{1\over n} {b\over k+b}},
\end{equation}
where $h^{\vee}$ is the dual Coxeter number of $\fg$, $n$ is the number listed in table \ref{table:lie}, and $k$ is restricted to 
the value such that no mass parameter is in the irregular singularity.

The Seiberg-Witten geometry of these theories are identified as the spectral curve 
of the Hitchin system\cite{hitchin1987stable}.
\begin{equation}
\det(x-\Phi)=0,
\end{equation}
and one can read off the Coulomb branch spectrum from an associated Newton polygon\cite{Xie:2012hs,Wang:2015mra,Wang:2018gvb}, which is also reviewed in the appendix \ref{sec:app:Newton}.

\begin{table}[h]
	\begin{center}
		\begin{tabular}{|c|c|c|c|c|}
			\hline
			~&dimension & $h$ & $h^{\vee}$&$n$ \\ \hline
			$A_{N-1}$&$N^2-1$& $N$ & $N$&1  \\ \hline
			$B_N$ & $(2N+1)N$ & $2N$& $2N-1$&2 \\ \hline
			$C_N^{(1)}$ &  $(2N+1)N$&  $2N$ &$N+1$&4 \\ \hline
			$C_N^{(2)}$ &  $(2N+1)N$&  $2N$ &$N+1$&2 \\ \hline
			$D_N$ & $N(2N-1)$ & $2N-2$ & $2N-2$&1 \\ \hline
			$E_6$& 78 & 12 & 12 &1\\ \hline
			$E_7$& 133 & 18 & 18 &1\\ \hline
			$E_8$& 248 & 30 & 30 &1\\ \hline
			$F_4$& 52 & 12 & 9 &2\\ \hline
			$G_2$& 14 & 6 & 4&3 \\ \hline
			
		\end{tabular}
	\end{center}
	\caption{Lie algebra data. $h$ is the Coexter number and $h^{\vee}$ is the dual Coexter number.}
	\label{table:lie}
\end{table}

\subsection{AD theories correspond to $B_{p+1}(g)$ and $W_{p+1}(g)$ algebras}
Consider the following irregular singularity of $\mathfrak{j}=ADE$ $(2,0)$ theory,
\begin{equation}
\Phi={T\over z^{2+p}}+\ldots 
\end{equation}
Notice that there are $l$ (the rank of $\mathfrak{j}$) mass parameters in this singularity. We add a trivial regular singularity ($f$ is regular nilpotent orbit), then these theories can be engineered by following three-fold singulariities,
\begin{equation}
f_{ADE}(x,y,z)+w^{ph^{\vee}}=0,
\end{equation}
where $h^{\vee}$ is the dual Coxeter number of $\mathfrak{j}$, and $f_{ADE}(x,y,z)$ is the famous two dimensional ADE singularity.
The 4d central charge is computed using following formula\cite{Xie:2014yya},
\begin{equation}
c_{4d}={\mu \alpha_{max}\over 12}+{r\over 6},
\end{equation} 
where $\alpha_{max}$ is the maximal scaling dimension of Coulomb branch spectrum, $r$ is the rank of Coulomb branch, and $\mu=2r+f_0$ with $f_0$ the number of mass parameters. Using the result  found in \cite{Xie:2016evu}, we have,
\begin{equation}
\mu=l (ph^{\vee}-1),~~~\alpha_{max}={p h^{\vee} \over p+1},~~~f_0=l,
\end{equation}
then the central charge takes following form,
\begin{equation}
c_{4d}(\mathfrak{j},p)={1\over 12}\left(-2l+ {(h^{\vee}+1) l h^{\vee} p^2\over p+1}\right).  
\end{equation}

When $\mathfrak{j}=A_{N-1}$, it was propsed in \cite{Creutzig:2018lbc} that the corresponding VOA 
is the $B_{p+1}(A_{N-1})$ algebra constructed in \cite{feigin2010logarithmic}. For general $\mathfrak{j}=ADE$, the central charge of  $B_{p+1}(\mathfrak{j})$ algebra is,
\begin{equation}
c_{2d}(B_{p+1}(\mathfrak{j}))=2l+ h^{\vee} \dim(\mathfrak{j})\left(2-(p+1)-{1\over p+1}\right).
\end{equation}
One finds that $c_{2d}(B_{p+1}(\mathfrak{j}))=-12c_{4d}(\mathfrak{j},p)$ for general $\mathfrak{j}$\footnote{We used the fact that $dim(\mathfrak{j})=l(h^{\vee}+1)$, here $h^{\vee}$ is the dual Coxeter number and $l$ is the rank of $\mathfrak{j}$.}, therefore we conjecture that the VOA of 
above 4d SCFT is given by the $B_{p+1}(\mathfrak{j})$ algebra.

We could also consider twisted AD theories with the following Higgs field,
\begin{equation}
\Phi={T^t\over z^{2+p}}+\ldots
\end{equation}
Using the method proposed in \cite{Wang:2018gvb}, the 4d central charge is,
\begin{equation}
c_{4d}={1\over12}\left(-2l+{(h^{\vee}l(h^{\vee}+1))(n(p+1)-1)^2\over n(p+1)}\right).
\label{nonsim}
\end{equation}
$n=1$ for simply-laced cases, and values of $n$ for non simply-laced cases are summarized in table \ref{table:lie}\footnote{Notice that $n$ can take two values for $C_N$ type Lie algebra.}. This implies that it should be possible to generalize the construction in \cite{feigin2010logarithmic} to non-simply laced Lie algebra whose central charge is given by $-12c_{4d}$ with $c_{4d}$ in equation \ref{nonsim}.

One can also add another full puncture to get a theory with $G$ flavor symmetry whose flavor central charge is,
\begin{equation}
k_G=h^{\vee}-{1\over n(p+1)}.
\end{equation}
The central charge for these theories are,
\begin{equation}
c_{4d}={(h^{\vee}-{1\over n(p+1)})l(h^{\vee}+1)n(p+1)\over 12}-{l\over 12}.
\end{equation}
This suggests that there should be a VOA $W_{p+1}(\fg)$ with an affine vertex operator subalgebra $V_{-k_G}(\fg)$. Some suggestions on the construction of this class of VOA are given in \cite{Creutzig:2018lbc}, and  we will give a coset construction in later sections. The qDS reduction of $W_p(\fg)$ produces the $B_p(\fg)$ VOA.

\section{VOA for AD matter with two non-abelian flavor symmetries}
\label{sec:ADwithTwoFlavors}

Theories considered in the last section usually carry only one type of non-abelian flavor symmetries, which is determined by the regular puncture.  
When the order of the irregular singularity considered in the last section is integral, one can consider degenerating cases and could get another type of non-abelian flavor symmetries\cite{Xie:2012hs}.

It was realized in \cite{Xie:2017aqx,Xie:2017vaf} that besides the theory considered in the last section, we can get new 4d $\mathcal{N}=2$ SCFT by considering more general irregular singularities (taking $\fg=A_{N-1}$ for example),
\begin{equation}
\Phi={T\over z^{2+{k\over n}}}
\end{equation}
with $T$ being the following diagonal matrix,
\begin{equation}
T=\diag(I_{n\times n},\underbrace{0,\ldots,0}_{N-n}).
\end{equation}
Here $I_{n\times n}$ being a diagonal matrix with eigenvalues $(1,w,\ldots, w^{n-1})$ with $w$ the $n$th root of unity. 
To get a SCFT, coefficients of subsequent terms in the Higgs field have to take the same form as the leading one. Such irregular singularity carries a flavor symmetry $U(N-n)$. One can also 
add an extra  regular puncture so that there are two distinct types of non-abelian flavor symmetries.  

The purpose of this section is to find the VOA for the above class of theories when $(k,n)=1$. The key observation is that the same theory can be realized by a different 
$(2,0)$ configuration whose VOA is already found in section \ref{sec:results}. We first study the $A_{N-1}$ case in detail in section \ref{sec:Acase}. The same construction is then generalized to other classical Lie algebras as well.

\subsection{$A_{N-1}=\mathfrak{sl}_N$: $SU(N)\times U(n_1)$ flavor symmetry}
\label{sec:Acase}
The regular puncture of $A_{N-1}$ theory is classified by a size $N$ Young tableaux (or the partition of $N$), which also gives a nilpotent orbit of the $A_{N-1}$ Lie algebra. Given a Young tableaux $[h_1^{r_1},h_2^{r_2},\ldots,h_s^{r_s}]$, the flavor symmetry is,
\begin{equation}
G_F=(\prod U(r_i))/U(1).
\end{equation} 
We are interested in regular punctures with partitions like $[m^q,1^{s}]$, whose 4d flavor symmetry is $SU(q)\times SU(s) \times U(1)$ with flavor central charges,
\begin{equation}
k_{SU(s)}=m+s-[z],~~k_{SU(q)}=s+qm-m[z],
\label{flavorA}
\end{equation}
where $[z]$ is the scaling dimension of $z$ coordinate in the spectral curve of $A$ type Hitchin system. 

Now consider the following configuration,
\begin{equation}
\label{eq:ADconfigA1}
\fg=\mathfrak{sl}_{n_1+n},~~~~~\Phi={T\over z^{2+{k\over n}}},~~~~f=[\underbrace{1,\ldots,1}_{n+n_1}]=[1^{n+n_1}],
\end{equation}
with $k$ and $n$ coprime. Here $T$ takes the following form $T=[I_{n_1\times n_1},0,\ldots,0]$ with $I_{n\times n}$ being a diagonal matrix with eigenvalues $(1,w,\ldots, w^{n-1})$ with $w$ being the $n$th root of unity.
This theory has a $U(n_1)\times SU(n_1+n)$ flavor symmetry with 4d flavor central charges\cite{Xie:2017aqx}\footnote{We ignore the subscript $2d$ and $4d$ when apparenat.},
\begin{equation}
k_{SU(n_1)}=n_1+{n\over n+k},~~~~k_{SU(n_1+n)}=n_1+n-{n\over n+k}.
\label{flavorAA}
\end{equation}
The Newton polygon for this theory is shown on the left of figure \ref{Atype}, from which
one can read the Coulomb branch spectrum using the procedure in the appendix \ref{sec:app:Newton}.

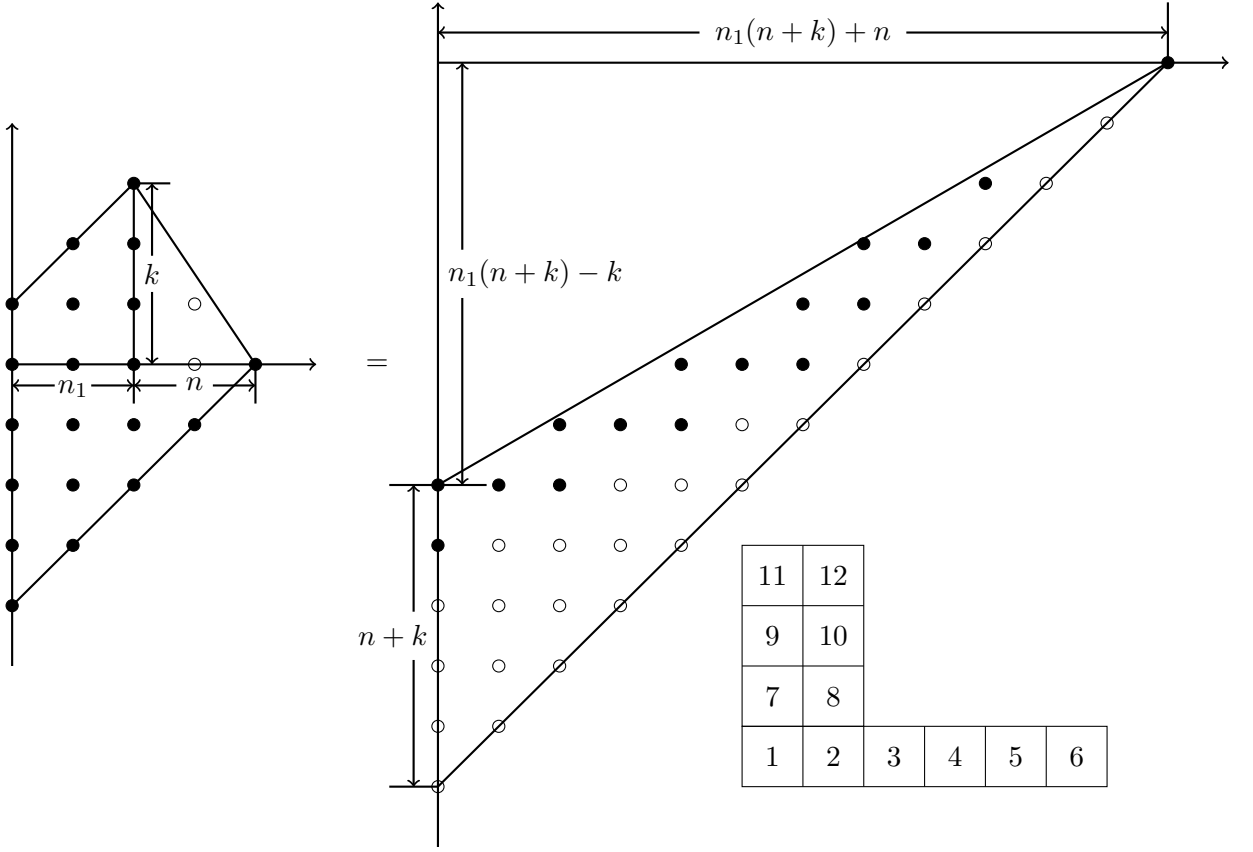
\begin{figure}
	\centering
	\begin{tikzpicture}[scale=0.8]
	
	\draw [thick,->] (0,0)--(5,0);
	\draw [thick,->] (0,-5)--(0,4);
	\draw [thick] (0,1)--(2,3);
	\draw [thick] (2,3)--(4,0);
	\draw [thick] (4,0)--(0,-4);
	
	\draw [fill] (2,3) circle [radius=0.1];
	\draw [fill] (1,2) circle [radius=0.1];
	\draw [fill] (2,2) circle [radius=0.1];
	
	\draw [fill] (0,1) circle [radius=0.1];
	\draw [fill] (1,1) circle [radius=0.1];
	\draw [fill] (2,1) circle [radius=0.1];
	\draw (3,1) circle [radius=0.1];
	
	\draw [fill] (0,0) circle [radius=0.1];
	\draw [fill] (1,0) circle [radius=0.1];
	\draw [fill] (2,0) circle [radius=0.1];
	\draw (3,0) circle [radius=0.1];
	\draw [fill] (4,0) circle [radius=0.1];
	
	\draw [fill] (0,-1) circle [radius=0.1];
	\draw [fill] (1,-1) circle [radius=0.1];
	\draw [fill] (2,-1) circle [radius=0.1];
	\draw [fill] (3,-1) circle [radius=0.1];
	
	\draw [fill] (0,-2) circle [radius=0.1];
	\draw [fill] (1,-2) circle [radius=0.1];
	\draw [fill] (2,-2) circle [radius=0.1];
	
	\draw [fill] (0,-3) circle [radius=0.1];
	\draw [fill] (1,-3) circle [radius=0.1];
	
	\draw [fill] (0,-4) circle [radius=0.1];
	
	\node at (1,-0.4) {$n_1$};
	\draw [thick,->] (0.7,-0.35)--(0,-0.35);
	\draw [thick,->] (1.3,-0.35)--(2,-0.35);
	\draw [thick] (2,0)--(2,-0.65);
	\draw [thick] (2,0)--(2,3);
	
	\node at (3,-0.32){$n$};
	\draw [thick,->] (2.7,-0.35)--(2,-0.35);
	\draw [thick,->] (3.3,-0.35)--(4,-0.35);
	\draw [thick] (4,0)--(4,-0.65);
	
	\node at (2.3,1.5){$k$};
	\draw [thick,->] (2.3,1.8)--(2.3,3);
	\draw [thick,->] (2.3,1.2)--(2.3,0);
	\draw [thick] (2,3)--(2.6,3);
	
	\node at (6,0){$=$};
	
	\draw [thick,->] (0+7,-13+5)--(0+7,1+5);
	\draw [thick,->] (0+7,0+5)--(13+7,0+5);
	\draw [thick] (0+7,-7+5)--(12+7,0+5);
	\draw [thick] (12+7,0+5)--(0+7,-12+5);
	
	\draw [fill] (12+7,0+5) circle [radius=0.1];
	
	\draw (11+7,-1+5) circle [radius=0.1];
	
	\draw (10+7,-2+5) circle [radius=0.1];
	\draw [fill] (9+7,-2+5) circle [radius=0.1];
	
	\draw [fill] (8+7,-3+5) circle [radius=0.1];
	\draw (9+7,-3+5) circle [radius=0.1];
	\draw [fill] (7+7,-3+5) circle [radius=0.1];
	
	\draw (8+7,-4+5) circle [radius=0.1];
	\draw [fill] (7+7,-4+5) circle [radius=0.1];
	\draw [fill] (6+7,-4+5) circle [radius=0.1];
	
	\draw (7+7,-5+5) circle [radius=0.1];
	\draw [fill] (6+7,-5+5) circle [radius=0.1];
	\draw [fill] (5+7,-5+5) circle [radius=0.1];
	\draw [fill] (4+7,-5+5) circle [radius=0.1];
	
	\draw (6+7,-6+5) circle [radius=0.1];
	\draw (5+7,-6+5) circle [radius=0.1];
	\draw [fill] (4+7,-6+5) circle [radius=0.1];
	\draw [fill] (3+7,-6+5) circle [radius=0.1];
	\draw [fill] (2+7,-6+5) circle [radius=0.1];
	
	\draw (5+7,-7+5) circle [radius=0.1];
	\draw (4+7,-7+5) circle [radius=0.1];
	\draw  (3+7,-7+5) circle [radius=0.1];
	\draw [fill] (2+7,-7+5) circle [radius=0.1];
	\draw [fill] (1+7,-7+5) circle [radius=0.1];
	\draw [fill] (0+7,-7+5) circle [radius=0.1];
	
	\draw (4+7,-8+5) circle [radius=0.1];
	\draw (3+7,-8+5) circle [radius=0.1];
	\draw (2+7,-8+5) circle [radius=0.1];
	\draw (1+7,-8+5) circle [radius=0.1];
	\draw [fill] (0+7,-8+5) circle [radius=0.1];
	
	\draw  (3+7,-9+5) circle [radius=0.1];
	\draw (2+7,-9+5) circle [radius=0.1];
	\draw  (1+7,-9+5) circle [radius=0.1];
	\draw (0+7,-9+5) circle [radius=0.1];
	
	\draw  (2+7,-10+5) circle [radius=0.1];
	\draw (1+7,-10+5) circle [radius=0.1];
	\draw (0+7,-10+5) circle [radius=0.1];
	
	\draw  (1+7,-11+5) circle [radius=0.1];
	\draw (0+7,-11+5) circle [radius=0.1];
	
	\draw  (0+7,-12+5) circle [radius=0.1];
	
	\node at (6+7,0.5+5) {$n_1(n+k)+n$}; 
	\draw [thick,->] (4.3+7, 0.5+5)--(0+7,0.5+5);
	\draw [thick,->] (6+1.7+7, 0.5+5)--(12+7,0.5+5);
	\draw [thick] (12+7,0+5)--(12+7,1+5);
	
	\node [right] at (0+7,-3.5+5) {$n_1(n+k)-k$};
	\draw [thick,->] (0.4+7,-3.5+0.4+5)--(0.4+7,0+5);
	\draw [thick,->] (0.4+7,-3.5-0.4+5)--(0.4+7,-7+5);
	\draw [thick] (0+7,-7+5)--(0.8+7,-7+5);
	
	\node [left] at (0+7,-7-2.5+5) {$n+k$};
	\draw [thick,->] (-0.4+7,-7-2.5+0.4+5)--(-0.4+7,-7+5);
	\draw [thick,->] (-0.4+7,-7-2.5-0.4+5)--(-0.4+7,-12+5);
	\draw [thick] (0+7,-7+5)--(-0.8+7,-7+5);
	\draw [thick] (0+7,-12+5)--(-0.8+7,-12+5);
	
	\draw (0+5+7,0-12+5) rectangle (6+5+7,1-12+5);
	\draw (1+5+7,0-12+5) -- (1+5+7,0-11+5);
	\draw (2+5+7,0-12+5) -- (2+5+7,0-11+5);
	\draw (3+5+7,0-12+5) -- (3+5+7,0-11+5);
	\draw (4+5+7,0-12+5) -- (4+5+7,0-11+5);
	\draw (5+5+7,0-12+5) -- (5+5+7,0-11+5);
	\draw (0+5+7,1-12+5) rectangle (2+5+7,4-12+5);
	\draw (1+5+7,1-12+5) -- (1+5+7,4-12+5);
	\draw (0+5+7,3-12+5) -- (2+5+7,3-12+5);
	\draw (0+5+7,2-12+5) -- (2+5+7,2-12+5);
	
	\node at (0+0.5+5+7,0+0.5-12+5) {$1$};
	\node at (1+0.5+5+7,0+0.5-12+5) {$2$};
	\node at (2+0.5+5+7,0+0.5-12+5) {$3$};
	\node at (3+0.5+5+7,0+0.5-12+5) {$4$};
	\node at (4+0.5+5+7,0+0.5-12+5) {$5$};
	\node at (5+0.5+5+7,0+0.5-12+5) {$6$};
	\node at (0+0.5+5+7,1+0.5-12+5) {$7$};
	\node at (1+0.5+5+7,1+0.5-12+5) {$8$};
	\node at (0+0.5+5+7,2+0.5-12+5) {$9$};
	\node at (1+0.5+5+7,2+0.5-12+5) {$10$};
	\node at (0+0.5+5+7,3+0.5-12+5) {$11$};
	\node at (1+0.5+5+7,3+0.5-12+5) {$12$};
	
	\end{tikzpicture}
	\caption{Equivalence of two different $(2,0)$ configurations of $A$ type theory.}
	\label{Atype}
\end{figure}

To find its VOA, we would like to find a different realization of this theory whose VOA is known through results in the last section, which has the following data,
\begin{equation}
\label{eq:ADconfigA2}
\fg=\mathfrak{sl}_{n_1(n+k)+n},~~~\Phi={T\over z^{2+{k-n_1(n+k)\over n_1(n+k)+n}}},~~~f=[(n+k-1)^{n_1},1^{n_1+n}].
\end{equation}
One can show that these two realizations have the same Coulomb branch and flavor symmetries.
Now we compute 4d flavor central charges of this new realization of the AD theory \ref{eq:ADconfigA2}. Notice $[z]={n_1(n+k)+n\over n+k}$ and using the formula \ref{flavorA}, we have,
\begin{equation}
\begin{split}
&k_{SU(n_1+n)}=n+2n_1-{n_1(n+k)+n\over n+k}=n+n_1-{n\over n+k},\\
&k_{SU(n_1)}=n_1(n+k)+n-(n+k-1){n_1(n+k)+n\over n+k}=n_1+{n\over n+k},
\end{split}
\end{equation}
which is exactly the same as flavor central charges (equation \ref{flavorAA}) of the previous description \ref{eq:ADconfigA1}. Therefore, we have compelling reasons to believe that \ref{eq:ADconfigA1} and \ref{eq:ADconfigA2} give the same AD theory and they correspond to the same 2d VOA. 

The later realization \ref{eq:ADconfigA2} is just the AD theory discussed in \ref{sec:ADandWk}. According to equation \ref{eq:ADtoWalgebra}, the corresponding VOA is the following $W$ algebra,
\begin{equation}
\boxed{VOA_A=W^{-n_1(n+k)-n+{n_1(n+k)+n\over n+k}}(\mathfrak{sl}_{n_1(n+k)+n},[(n+k-1)^{n_1},1^{n_1+n}])}
\label{VOAA}
\end{equation}
Using the central charge formula \ref{eq:centralA} in the appendix \ref{sec:app:central}, the central charge of $VOA_A$ is
\begin{equation}
\label{eq:ccVOAA}
c(VOA_A)=-\left(n^2-1\right) (k+n-1)-n_1(n+n_1) (3 k+3 n-2).
\end{equation}
Also by the 4d/2d correspondence, there is an affine $\mathfrak{u}(1)\oplus V_{-(n_1+{n\over n+k})}(\fsu_{n_1}) \oplus V_{-(n+n_1)+{n\over n+k}}(\fsu_{n+n_1})$ subalgebra within $VOA_A$.  

\textbf{Example}: Let us look at the example in figure \ref{Atype}. For the left Newton polygon, we have $n_1=2,n=2, k=3$. Using the formula \ref{CoulombA}, we have following Coulomb branch spectrum $[\frac{8}{5},\frac{6}{5},\frac{13}{5},\frac{11}{5},\frac{9}{5},\frac{7}{5},\frac{18}{5},\frac{16}{5},\frac{14}{5},\frac{12}{5}]$. For the right Newton polygon, we have the data $h(l)=[0,1,1,1,1,1,2,2,3,3,4,4]$, and we have $n_1'=0, n'=12, k'=-7$.  Equation \ref{CoulombAA} gives the spectrum 
$[\frac{8}{5},\frac{13}{5},\frac{18}{5},\frac{6}{5},\frac{11}{5},\frac{16}{5},\frac{9}{5},\frac{14}{5},\frac{7}{5},\frac{12}{5}]$. One can see that two 
Coulomb branch spectra match, although they are encoded in quite different ways. 

\subsubsection{New level-rank dualitys} \label{level}
Now consider the theory with $f$ being a regular nilpotent orbit (hence no flavor symmetry), and the irregular singularity,
\begin{equation}
\fg=\fsl_{n_1+n},~~~~~\Phi={T\over z^{2+{k\over n}}},~~f=[n_1+n].
\end{equation}
Again, we take $n$ and $k$ to be coprime. This theory has a $U(n_1)$ flavor symmetry. Using the above result (equation \ref{VOAA}) for $f$ being a full puncture and the fact that closing of a puncture is equivalent to qDS, its VOA is 
\begin{equation}
W^{-n_1(n+k)-n+{n_1(n+k)+n\over n+k}}(\mathfrak{sl}_{n_1(n+k)+n},[(n+k-1)^{n_1},n+n_1]).
\end{equation}
Here we take $k>n_1$. The same theory can be realized by the following configuration,
\begin{equation}
\fg=\mathfrak{sl}_{k},~~~\Phi={T\over z^{2+{n\over k}}},~~f=[k-n_1,1^{n_1}].
\end{equation}
 
Using the correspondence \ref{eq:ADtoWalgebra} in section \ref{sec:ADandWk}, the corresponding VOA is then
\begin{equation}
W^{-k+{k\over n+k}}(\mathfrak{sl}_k,[k-n_1,1^{n_1}]).
\end{equation} 
Therefore, we find the following equivalence of two 2d VOAs from different realizations of the same 4d AD theories,
\begin{equation}
\begin{split}
&W^{-n_1(n+k)-n+{n_1(n+k)+n\over n+k}}(\mathfrak{sl}_{n_1(n+k)+n},[(n+k-1)^{n_1},n+n_1])
 \\   
 &=W^{-k+{k\over n+k}}(\mathfrak{sl}_k,[k-n_1,1^{n_1}]).
\end{split}
\end{equation}
We call this the level-rank duality as the rank and the level  are sort of exchanged for these two $W$ algebras. 

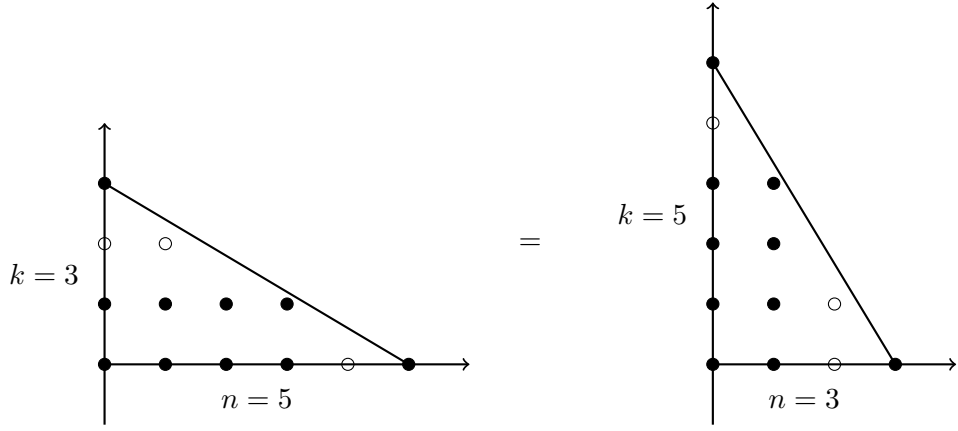
\begin{figure}
	\centering
	\begin{tikzpicture}[scale=0.8]
	\draw [thick,->] (0,-1)--(0,4);
	\draw [thick,->] (0,0)--(6,0);
	\draw [thick](0,3)--(5,0);
	
	\draw [fill] (0,0) circle [radius=0.1];
	\draw [fill] (1,0) circle [radius=0.1];
	\draw [fill] (2,0) circle [radius=0.1];
	\draw [fill] (3,0) circle [radius=0.1];
	\draw [fill] (5,0) circle [radius=0.1];
	\draw  (4,0) circle [radius=0.1];
	
	\draw [fill] (0,1) circle [radius=0.1];
	\draw [fill] (1,1) circle [radius=0.1];
	\draw [fill] (2,1) circle [radius=0.1];
	\draw [fill] (3,1) circle [radius=0.1];
	
	\draw  (0,2) circle [radius=0.1];
	\draw  (1,2) circle [radius=0.1];
	
	\draw [fill] (0,3) circle [radius=0.1];
	
	\node [left] at (-0.25,1.5) {$k=3$};
	\node [below] at (2.5,-0.25) {$n=5$};
	
	\node at (0+7,2) {$=$};
	
	\draw [thick,->] (0+10,-1)--(0+10,6);
	\draw [thick,->] (0+10,0)--(4+10,0);
	\draw [thick](0+10,5)--(3+10,0);
	
	\draw [fill] (0+10,0) circle [radius=0.1];
	\draw [fill] (10,1) circle [radius=0.1];
	\draw [fill] (10,2) circle [radius=0.1];
	\draw [fill] (10,3) circle [radius=0.1];
	\draw [fill] (10,5) circle [radius=0.1];
	\draw  (10,4) circle [radius=0.1];
	
	\draw [fill] (9+2,0) circle [radius=0.1];
	\draw [fill] (9+2,1) circle [radius=0.1];
	\draw [fill] (9+2,2) circle [radius=0.1];
	\draw [fill] (9+2,3) circle [radius=0.1];
	
	\draw  (12,0) circle [radius=0.1];
	\draw  (12,1) circle [radius=0.1];
	
	\draw [fill] (11+2,0) circle [radius=0.1];
	
	\node [left] at (10-0.25,2.5) {$k=5$};
	\node [below] at (2+9.5,-0.25) {$n=3$};
	
	\end{tikzpicture}
	\caption{Level rank duality example. The Coulomb branch spectrum of two configurations are the same. Notice that  the left hand side
	uses 6d $A_{4}$ $(2,0)$ theory while the right hand side uses 6d $A_{2}$ $(2,0)$ theory. }
	\label{Lrank}
\end{figure}

\textbf{Example}: Taking $n_1=0$, we have the following equivalence of VOAs:
\begin{equation}
W^{-n+{n\over n+k}}(sl_n,[n])=W^{-k+{k\over n+k}}(sl_k,[k]).
\end{equation}
This is the familiar level-rank duality discovered in \cite{altschuler1990level}. See figure \ref{Lrank} for an illustration
from the four dimensional theory point of view.

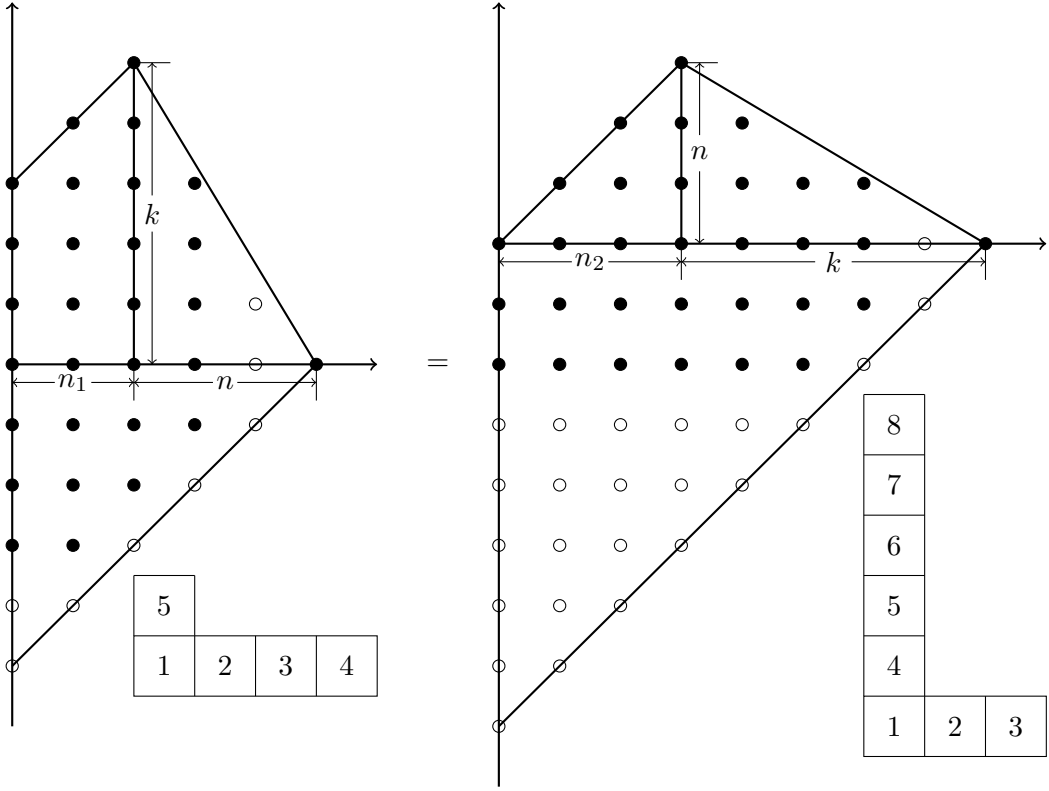
\begin{figure}
	\centering
	\begin{tikzpicture}[scale=0.8]
	
	\draw [thick,->] (0,-6)--(0,6);
	\draw [thick,->] (0,0)--(6,0);
	\draw [thick] (0,3)--(2,5);
	\draw [thick] (2,5)--(5,0);
	\draw [thick] (5,0)--(0,-5);
	\draw [thick] (2,5)--(2,0);
	
	\draw [fill] (0,3) circle [radius=0.1];
	\draw [fill] (0,2) circle [radius=0.1];
	\draw [fill] (0,1) circle [radius=0.1];
	\draw [fill] (0,0) circle [radius=0.1];
	\draw [fill] (0,-1) circle [radius=0.1];
	\draw [fill] (0,-2) circle [radius=0.1];
	\draw [fill] (0,-3) circle [radius=0.1];
	\draw (0,-4) circle [radius=0.1];
	\draw (0,-5) circle [radius=0.1];
	
	\draw [fill] (1,4) circle [radius=0.1];
	\draw [fill] (1,3) circle [radius=0.1];
	\draw [fill] (1,2) circle [radius=0.1];
	\draw [fill] (1,1) circle [radius=0.1];
	\draw [fill] (1,0) circle [radius=0.1];
	\draw [fill] (1,-1) circle [radius=0.1];
	\draw [fill] (1,-2) circle [radius=0.1];
	\draw [fill] (1,-3) circle [radius=0.1];
	\draw (1,-4) circle [radius=0.1];
	
	\draw [fill] (2,5) circle [radius=0.1];
	\draw [fill] (2,4) circle [radius=0.1];
	\draw [fill] (2,3) circle [radius=0.1];
	\draw [fill] (2,2) circle [radius=0.1];
	\draw [fill] (2,1) circle [radius=0.1];
	\draw [fill] (2,0) circle [radius=0.1];
	\draw [fill] (2,-1) circle [radius=0.1];
	\draw [fill] (2,-2) circle [radius=0.1];
	\draw (2,-3) circle [radius=0.1];
	
	\draw [fill] (3,3) circle [radius=0.1];
	\draw [fill] (3,2) circle [radius=0.1];
	\draw [fill] (3,1) circle [radius=0.1];
	\draw [fill] (3,0) circle [radius=0.1];
	\draw [fill] (3,-1) circle [radius=0.1];
	\draw (3,-2) circle [radius=0.1];
	
	\draw (4,1) circle [radius=0.1];
	\draw (4,0) circle [radius=0.1];
	\draw (4,-1) circle [radius=0.1];
	
	\draw [fill] (5,0) circle [radius=0.1];
	
	\draw (0+2,-5-0.5) rectangle (4+2,-4-0.5);
	\draw (0+1+2,-5-0.5) -- (0+1+2,-4-0.5);
	\draw (0+2+2,-5-0.5) -- (0+2+2,-4-0.5);
	\draw (0+3+2,-5-0.5) -- (0+3+2,-4-0.5);
	\draw (0+2,-4-0.5) -- (0+2,-3-0.5);
	\draw (0+2,-3-0.5) -- (1+2,-3-0.5);
	\draw (1+2,-3-0.5) -- (1+2,-4-0.5) ;
	\node at (0+0.5+2,-5) {$1$};
	\node at (1+0.5+2,-5) {$2$};
	\node at (2+0.5+2,-5) {$3$};
	\node at (3+0.5+2,-5) {$4$};
	\node at (0+0.5+2,-4) {$5$};
	
	\node  at (1,-0.3) {$n_1$};
	\draw [->] (0.7,-0.3) -- (0,-0.3);
	\draw [->] (1.3,-0.3) -- (2,-0.3);
	\draw (2,0) -- (2,-0.6);
	
	\node at (2.3,2.5) {$k$};
	\draw [->] (2.3,2.8) -- (2.3,5);
	\draw [->] (2.3,2.2) -- (2.3,0);
	\draw (2,5) -- (2.6,5);
	
	\node at (3.5, -0.3) {$n$};
	\draw [->] (3.3,-0.3) -- (2,-0.3);
	\draw [->] (3.7,-0.3) -- (5,-0.3);
	\draw (5,0) -- (5,-0.6);
	
	\node at (0+7,0) {$=$};
	
	\draw [thick,->] (0+8,-9+2)--(0+8,4+2);
	\draw [thick,->] (0+8,0+2)--(9+8,0+2);
	\draw [thick] (0+8,0+2)--(3+8,3+2);
	\draw [thick] (3+8,3+2)--(8+8,0+2);
	\draw [thick] (8+8,0+2)--(0+8,-8+2);
	\draw [thick] (3+8,3+2)--(3+8,0+2);
	
	\draw [fill] (0+8,2) circle [radius=0.1];
	\draw [fill] (0+8,1) circle [radius=0.1];
	\draw [fill] (0+8,0) circle [radius=0.1];
	\draw (0+8,-1) circle [radius=0.1];
	\draw (0+8,-2) circle [radius=0.1];
	\draw (0+8,-3) circle [radius=0.1];
	\draw (0+8,-4) circle [radius=0.1];
	\draw (0+8,-5) circle [radius=0.1];
	\draw (0+8,-6) circle [radius=0.1];
	
	\draw [fill] (1+8,3) circle [radius=0.1];
	\draw [fill] (1+8,2) circle [radius=0.1];
	\draw [fill] (1+8,1) circle [radius=0.1];
	\draw [fill] (1+8,0) circle [radius=0.1];
	\draw (1+8,-1) circle [radius=0.1];
	\draw (1+8,-2) circle [radius=0.1];
	\draw (1+8,-3) circle [radius=0.1];
	\draw (1+8,-4) circle [radius=0.1];
	\draw (1+8,-5) circle [radius=0.1];
	
	\draw [fill] (2+8,4) circle [radius=0.1];
	\draw [fill] (2+8,3) circle [radius=0.1];
	\draw [fill] (2+8,2) circle [radius=0.1];
	\draw [fill] (2+8,1) circle [radius=0.1];
	\draw [fill] (2+8,0) circle [radius=0.1];
	\draw (2+8,-1) circle [radius=0.1];
	\draw (2+8,-2) circle [radius=0.1];
	\draw (2+8,-3) circle [radius=0.1];
	\draw (2+8,-4) circle [radius=0.1];
	
	\draw [fill] (3+8,5) circle [radius=0.1];
	\draw [fill] (3+8,4) circle [radius=0.1];
	\draw [fill] (3+8,3) circle [radius=0.1];
	\draw [fill] (3+8,2) circle [radius=0.1];
	\draw [fill] (3+8,1) circle [radius=0.1];
	\draw [fill] (3+8,0) circle [radius=0.1];
	\draw (3+8,-1) circle [radius=0.1];
	\draw (3+8,-2) circle [radius=0.1];
	\draw (3+8,-3) circle [radius=0.1];
	
	\draw [fill] (4+8,4) circle [radius=0.1];
	\draw [fill] (4+8,3) circle [radius=0.1];
	\draw [fill] (4+8,2) circle [radius=0.1];
	\draw [fill] (4+8,1) circle [radius=0.1];
	\draw [fill] (4+8,0) circle [radius=0.1];
	\draw (4+8,-1) circle [radius=0.1];
	\draw (4+8,-2) circle [radius=0.1];
	
	\draw [fill] (5+8,3) circle [radius=0.1];
	\draw [fill] (5+8,2) circle [radius=0.1];
	\draw [fill] (5+8,1) circle [radius=0.1];
	\draw [fill] (5+8,0) circle [radius=0.1];
	\draw (5+8,-1) circle [radius=0.1];
	
	\draw [fill] (6+8,3) circle [radius=0.1];
	\draw [fill] (6+8,2) circle [radius=0.1];
	\draw [fill] (6+8,1) circle [radius=0.1];
	\draw (6+8,0) circle [radius=0.1];
	
	\draw (7+8,2) circle [radius=0.1];
	\draw (7+8,1) circle [radius=0.1];
	
	\draw [fill] (8+8,2) circle [radius=0.1];
	
	\draw (0+5+8+1,-6-1+0.5) rectangle (3+5+8+1,-5-1+0.5);
	\draw (1+5+8+1,-6-1+0.5) -- (1+5+8+1,-5-1+0.5);
	\draw (2+5+8+1,-6-1+0.5) -- (2+5+8+1,-5-1+0.5);
	\draw (0+5+8+1,-5-1+0.5) -- (0+5+8+1,0-1+0.5);
	\draw (1+5+8+1,-5-1+0.5) -- (1+5+8+1,0-1+0.5);
	\draw (0+5+8+1,-4-1+0.5) -- (1+5+8+1,-4-1+0.5);
	\draw (0+5+8+1,-3-1+0.5) -- (1+5+8+1,-3-1+0.5);
	\draw (0+5+8+1,-2-1+0.5) -- (1+5+8+1,-2-1+0.5);
	\draw (0+5+8+1,-1-1+0.5) -- (1+5+8+1,-1-1+0.5);
	\draw (0+5+8+1,0-1+0.5) -- (1+5+8+1,0-1+0.5);
	
	\node at (0+5+8+1.5,-6) {$1$};
	\node at (1+5+8+1.5,-6) {$2$};
	\node at (2+5+8+1.5,-6) {$3$};
	\node at (0+5+8+2-0.5,-5) {$4$};
	\node at (0+5+8+2-0.5,-4) {$5$};
	\node at (0+5+8+2-0.5,-3) {$6$};
	\node at (0+5+8+2-0.5,-2) {$7$};
	\node at (0+5+8+2-0.5,-1) {$8$};

	\node  at (1.5+8,-0.3+2) {$n_2$};
	\draw [->] (1.2+8,-0.3+2) -- (0+8,-0.3+2);
	\draw [->] (1.8+8,-0.3+2) -- (3+8,-0.3+2);
	\draw (3+8,0+2) -- (3+8,-0.6+2);
	
	\node at (3.3+8,1.5+2) {$n$};
	\draw [->] (3.3+8,1.8+2) -- (3.3+8,3+2);
	\draw [->] (3.3+8,1.2+2) -- (3.3+8,0+2);
	\draw (3+8,3+2) -- (3.6+8,3+2);
	
	\node at (5.5+8, -0.3+2) {$k$};
	\draw [->] (5.2+8,-0.3+2) -- (3+8,-0.3+2);
	\draw [->] (5.8+8,-0.3+2) -- (8+8,-0.3+2);
	\draw (8+8,0+2) -- (8+8,-0.6+2);
	
	\end{tikzpicture}
	\caption{An example of the generalized level rank duality from Newton polygon of four dimensional $\mathcal{N}=2$ theory. One can check that 
	these two configuraitons give the same Coulomb branch spectrum.}
    \label{LrankB}
\end{figure}

The above level-rank duality can be further generalized as follows. Consider two configurations,
\begin{equation}
\begin{split}
&A:~~~\fg=\fsl_{n_1+n},~~~~~\Phi={T_1\over z^{2+{k\over n}}},~~f=[n+n_1-n_2,1^{n_2}], \\
&B:~~~\fg=\fsl_{n_2+k},~~~~~\Phi={T_2\over z^{2+{n\over k}}},~~f=[k+n_2-n_1,1^{n_1}].
\end{split}
\end{equation}
Choose $n+n_1-n_2>1,~~k+n_2-n_1>1$ and $T_1=\diag(I_{n_1\times n_1},0_{n\times n})$, $T_2=\diag(I_{n_2\times n_2},0_{k\times k})$.
These two configurations give the same Coulomb branch spectrum, therefore we conjecture 
that they give the same 4d theory. The manifest flavor symmetry is $U(n_1)\times U(n_2)$ with following flavor central charges,
\begin{equation}
k_{SU(n_1)}=n_1+{n\over n+k},~~k_{SU(n_2)}=n_2+{k\over n+k}.
\end{equation}
The two seemingly different VOAs should 
be the same, so we have following (conjectured) equivalence,
\begin{equation}
\boxed{\begin{split}
&W^{-n_1(n+k)-n+{n_1(n+k)+n\over n+k}}\big(\fsl_{n_1(n+k)+n},[(n+k-1)^{n_1},n+n_1-n_2,1^{n_2}]\big) = \\
&W^{-n_2(n+k)-k+{n_2(n+k)+k\over n+k}}\big(\fsl_{n_2(n+k)+k},[(n+k-1)^{n_2},k+n_2-n_1,1^{n_1}]\big).
\end{split}}
\end{equation}
An example is illustrated in figure \ref{LrankB}.

\subsubsection{Conformal embedding in $W$ algebra}
Conformal embedding is defined as the following\cite{adamovic2017conformal}. Let V be a vertex algebra with a Virasoro (= conformal) vector $w_V$ and let W be a vertex subalgebra of V endowed with a Virasoro vector $w_W$. The embedding $W \subset V$ is called conformal if $w_W= w_V$. A necessary condition for conformal embedding is that $c_V=c_W$.

For our $W$ algebra $VOA_A$ defined in equation \ref{VOAA}, there is an affine vertex operator subalgebra $\mathfrak{u}(1)\oplus V_{-(n_1+{n\over n+k})}(\fsl_{n_1}) \oplus V_{-(n+n_1)+{n\over n+k}}(\fsl_{n+n_1})$. It is interesting to note 
that the central charge of $VOA_A$ is equal to the central 
charge of this affine vertex operator subalgebra. So the necessary condition for conformal embedding is achieved, and it 
is interesting to check whether the following embedding,
\begin{equation}
\begin{split}
&u(1)\oplus V_{-(n_1+{n\over n+k})}(\fsl_{n_1}) \oplus V_{-(n+n_1)+{n\over n+k}}(\fsl_{n+n_1}) \subset \\
& W^{-n_1(n+k)-n+{n_1(n+k)+n\over n+k}}(\fsl_{n_1(n+k)+n},(n+k-1)^{n_1},1^{n_1+n}])
\end{split}
\end{equation}
is indeed the conformal embedding (see \cite{Beem:2018duj} for physical discussions).

\textbf{Example}: Taking $n_1=1$ and $n+k-1=2$, we
have the following  embedding:
\begin{equation}
U(1)\oplus V_{-{2n\over 3}+1}(\fsu(n+1)) \subset W^{-{2(n+3)\over3}}(\fsl_{n+3},[2,1^{n+1}]),
\end{equation}
which is the conformal embedding studied in \cite{adamovic2017conformal}.

\subsubsection{Collapsing levels and decoupling of flavor symmetry} \label{collapsing}
We have found VOAs of some AD matters with two distinct non-abelian flavor symmetries by 
finding an alternative $(2,0)$ construction\footnote{This part is motivated by a question by T.Arakawa.}. 
Notice that the above construction does not give a new realization for theories
with just a $SU(n)$ flavor symmetry arising only from regular singularity. We would like to find a different realization for those theories too. Such realization is not very good for 4d theory as there appears to be more flavor symmetries which will decouple in the IR, but they do have interesting implications for VOAs.

The basic idea is the following. Start with a theory engineered by the following configuration,
\begin{equation}
\fg=\fsl_n,~~\Phi={T\over z^{2+{k\over n}}},~~~f=[1^n].
\end{equation}
We take $(k,n)=1$, and $T$ being the principle type (so no mass term is allowed in the irregular singularity) so that the VOA is the affine vertex operator algebra 
\begin{equation}
V^{k'}(\fsl_n),~~~k'=-n+{n\over k+n}.
\end{equation}
Now we would like to find another realization  whose assosiated VOA 
is known by the result of section \ref{sec:results} too, so it should take the following form,
\begin{equation}
\fg=\fsl_N,~~\Phi={T\over z^{2+{-N+n+k\over N}}},~~~f=[q^m,1^{n}].
\end{equation}
Naively, this configuration has flavor symmetry $SU(n)\times U(m)$. The flavor central charge for $SU(n)$ flavor group is $n+m-{N\over{n+k}}$. The necessary condition for the equivalence is that the flavor central charge for $SU(n)$ group should be the same,
\begin{equation}
n-{n\over n+k}=n+m-{N\over{n+k}},
\end{equation} 
therefore we find $N=m(n+k)+k$, so $q=(n+k)$. The flavor central charge
of $SU(m)$ group is 
\begin{equation}
m(n+k)+n-(n+k){m(n+k)+n\over n+k}=0.
\end{equation} 
Physically, we interpret that this result implies that the $U(m)$ flavor symmetry 
is decoupled in the IR 4d SCFT. These two configuration defines the same 4d SCFT
in the IR (One can check that they give the same Coulomb branch spectrum), and we have the following equivalence of VOAs,
\begin{equation}
V^{-n+{n\over k+n}}(\fsl_n)=W^{-m(n+k)-n+{m(n+k)+n\over n+k}}(\fsl_{m(n+k)+n},[(n+k)^m,1^{n}]).
\end{equation}
Mathematically, this means that the $W$ algebra collapses to its affine subalgebra\cite{adamovic2017conformal}.

We could generalize the above collapsing story as follows. Consider more general theory engineered by following data,
\begin{equation}
\fg=\fsl_{n_1+n},~~~~~\Phi={T\over z^{2+{k\over n}}},~~f=[1^{n_1+n}].
\end{equation}
Again $T=\diag(I_{n\times n},0^{n_1})$. 
This theory has flavor symmetry $U(n_1)\times SU(n+n_1)$. We found that its VOA is 
a $W$ algebra given by the nilpotent orbit $[(n+k-1)^{n_1},1^{n+n_1}]$.  
However, the theory can be engineered by the following configuration as well,
\begin{equation}
\fg=\fsl_{(m+n_1)(n+k)+n},~~~~~\Phi={T\over z^{2+{m\over N}}},~~f=[(n+k)^m,(n+k-1)^{n_1},1^{n+n_1}].
\end{equation}
We have $N=(m+n_1)(n+k)+n$ and $N+m=n+k$. The VOA for above configuration is 
\begin{equation}
\label{eq:collapseVOAA}
W^{k'}(\fsl_N,[(n+k)^m,(n+k-1)^{n_1},1^{n+n_1}]),~~~~~k'=-N+{N\over n+k}.
\end{equation}
This configuration has the naive flavor symmetry $U(m)\times U(n_1)\times SU(n+n_1)$. 
The flavor central charge for the $U(m)$ flavor group is zero, and therefore is decoupled in the IR. The above 
$W$ algebra \ref{eq:collapseVOAA} is therefore collapsed to $VOA_A$ defined in \ref{VOAA}.

\subsection{Classical Lie algebra}

We now discuss how to generaize the above construction to other Lie algebras. The idea is similar: we consider AD matters 
engineered from one 6d realization whose VOA is not known, then 
we find an equivalent realization whose VOA can be read from results in section \ref{sec:results}.

\subsubsection{$D_N=\fso_{2N}$: $SO(2N)\times Sp(n_1)$  flavor symmetry}

We start with AD theories engineered from 6d $D_N$ $(2,0)$ theory.
The regular puncture of $D_N$ theory is classified by nilpotent orbits of $\fso_{2N}$ Lie algebra and is labeled by a size $2N$ Young Tableaux  
whose \textbf{even} parts has \textbf{even} multiplicities. Given a Young Tableax 
$[r_1^{h_1},r_2^{h_2},\ldots]$, the flavor symmetry is 
\begin{equation}
G_F=\prod_{h_i~even}Sp(r_i)\times \prod_{h_i~odd}SO(r_i).
\end{equation}
Given the puncture $[m^q,1^s]$ with $(q,m,s)$ all even, the flavor symmetry is $Sp(q)\times SO(s)$ whose central charges are 
\begin{equation}
k_{SO(s)}=(s+q-2)-[z],~~k_{Sp(q)}={s+mq-m[z]\over 2}.
\label{Dtype}
\end{equation}
Here $[z]$ is the scaling dimension of $z$ coordinate in the spectral curve of $D_N$ type.

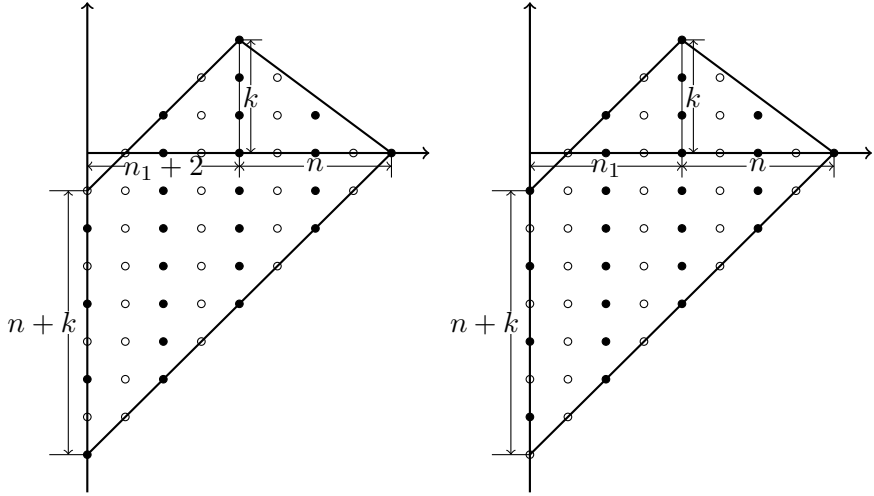
\begin{figure}
	\centering
	\begin{tikzpicture}[scale=0.5]
	
	\draw [thick,->] (0,0)--(9,0);
	\draw [thick,->] (0,-9)--(0,4);
	\draw [thick] (0,-1)--(4,3);
	\draw [thick] (4,3)--(8,0);
	\draw [thick] (8,0)--(0,-8);
	
	\draw [fill] (4,3) circle [radius=0.1];
	
	\draw (3,2) circle [radius=0.1];
	\draw [fill] (4,2) circle [radius=0.1];
	\draw (5,2) circle [radius=0.1];
	
	\draw [fill] (2,1) circle [radius=0.1];
	\draw (3,1) circle [radius=0.1];
	\draw [fill] (4,1) circle [radius=0.1];
	\draw (5,1) circle [radius=0.1];
	\draw [fill] (6,1) circle [radius=0.1];
	
	\draw (1,0) circle [radius=0.1];
	\draw [fill] (2,0) circle [radius=0.1];
	\draw (3,0) circle [radius=0.1];
	\draw [fill] (4,0) circle [radius=0.1];
	\draw (5,0) circle [radius=0.1];
	\draw [fill] (6,0) circle [radius=0.1];
	\draw (7,0) circle [radius=0.1];
	\draw [fill] (8,0) circle [radius=0.1];
	
	\draw (0,-1) circle [radius=0.1];
	\draw (1,-1) circle [radius=0.1];
	\draw [fill] (2,-1) circle [radius=0.1];
	\draw (3,-1) circle [radius=0.1];
	\draw [fill] (4,-1) circle [radius=0.1];
	\draw (5,-1) circle [radius=0.1];
	\draw [fill] (6,-1) circle [radius=0.1];
	\draw (7,-1) circle [radius=0.1];
	
	\draw [fill] (0,-2) circle [radius=0.1];
	\draw (1,-2) circle [radius=0.1];
	\draw [fill] (2,-2) circle [radius=0.1];
	\draw (3,-2) circle [radius=0.1];
	\draw [fill] (4,-2) circle [radius=0.1];
	\draw (5,-2) circle [radius=0.1];
	\draw [fill] (6,-2) circle [radius=0.1];
	
	\draw (0,-3) circle [radius=0.1];
	\draw (1,-3) circle [radius=0.1];
	\draw [fill] (2,-3) circle [radius=0.1];
	\draw (3,-3) circle [radius=0.1];
	\draw [fill] (4,-3) circle [radius=0.1];
	\draw (5,-3) circle [radius=0.1];
	
	\draw [fill] (0,-4) circle [radius=0.1];
	\draw (1,-4) circle [radius=0.1];
	\draw [fill] (2,-4) circle [radius=0.1];
	\draw (3,-4) circle [radius=0.1];
	\draw [fill] (4,-4) circle [radius=0.1];
	
	\draw (0,-5) circle [radius=0.1];
	\draw (1,-5) circle [radius=0.1];
	\draw [fill] (2,-5) circle [radius=0.1];
	\draw (3,-5) circle [radius=0.1];
	
	\draw [fill] (0,-6) circle [radius=0.1];
	\draw (1,-6) circle [radius=0.1];
	\draw [fill] (2,-6) circle [radius=0.1];
	
	\draw (0,-7) circle [radius=0.1];
	\draw (1,-7) circle [radius=0.1];
	
	\draw [fill] (0,-8) circle [radius=0.1];
	
	\node at (1+1,-0.4) {$n_1+2$};
	\draw [->] (1,-0.35)--(0,-0.35);
	\draw [->] (3,-0.35)--(2+2,-0.35);
	\draw  (4,0)--(4,-0.65);
	\draw  (4,0)--(4,3);
	
	\node at (6,-0.32){$n$};
	\draw [->] (2.7+3,-0.35)--(2+2,-0.35);
	\draw [->] (3.3+3,-0.35)--(4+4,-0.35);
	\draw (4+4,0)--(4+4,-0.65);
	
	\node at (2.3+2,1.5){$k$};
	\draw [->] (2.3+2,1.8)--(2.3+2,3);
	\draw [->] (2.3+2,1.2)--(2.3+2,0);
	\draw (2+2,3)--(2.6+2,3);
	
	\node [left] at (-0,-4.5) {$n+k$};
	\draw [->] (-0.5,-4.5+0.3)--(-0.5,-1);
	\draw (-1,-1)--(0,-1);
	\draw [->] (-0.5,-4.5-0.3)--(-0.5,-8);
	\draw  (-1,-8)--(0,-8);
	
	\end{tikzpicture}
	\begin{tikzpicture}[scale=0.5]
	
	\draw [thick,->] (0,0)--(9,0);
	\draw [thick,->] (0,-9)--(0,4);
	\draw [thick] (0,-1)--(4,3);
	\draw [thick] (4,3)--(8,0);
	\draw [thick] (8,0)--(0,-8);
	
	\draw [fill] (4,3) circle [radius=0.1];
	
	\draw (3,2) circle [radius=0.1];
	\draw [fill] (4,2) circle [radius=0.1];
	\draw (5,2) circle [radius=0.1];
	
	\draw [fill] (2,1) circle [radius=0.1];
	\draw (3,1) circle [radius=0.1];
	\draw [fill] (4,1) circle [radius=0.1];
	\draw (5,1) circle [radius=0.1];
	\draw [fill] (6,1) circle [radius=0.1];
	
	\draw (1,0) circle [radius=0.1];
	\draw [fill] (2,0) circle [radius=0.1];
	\draw (3,0) circle [radius=0.1];
	\draw [fill] (4,0) circle [radius=0.1];
	\draw (5,0) circle [radius=0.1];
	\draw [fill] (6,0) circle [radius=0.1];
	\draw (7,0) circle [radius=0.1];
	\draw [fill] (8,0) circle [radius=0.1];
	
	\draw [fill] (0,-1) circle [radius=0.1];
	\draw (1,-1) circle [radius=0.1];
	\draw [fill] (2,-1) circle [radius=0.1];
	\draw (3,-1) circle [radius=0.1];
	\draw [fill] (4,-1) circle [radius=0.1];
	\draw (5,-1) circle [radius=0.1];
	\draw [fill] (6,-1) circle [radius=0.1];
	\draw (7,-1) circle [radius=0.1];
	
	\draw (0,-2) circle [radius=0.1];
	\draw (1,-2) circle [radius=0.1];
	\draw [fill] (2,-2) circle [radius=0.1];
	\draw (3,-2) circle [radius=0.1];
	\draw [fill] (4,-2) circle [radius=0.1];
	\draw (5,-2) circle [radius=0.1];
	\draw [fill] (6,-2) circle [radius=0.1];
	
	\draw [fill] (0,-3) circle [radius=0.1];
	\draw (1,-3) circle [radius=0.1];
	\draw [fill] (2,-3) circle [radius=0.1];
	\draw (3,-3) circle [radius=0.1];
	\draw [fill] (4,-3) circle [radius=0.1];
	\draw (5,-3) circle [radius=0.1];
	
	\draw (0,-4) circle [radius=0.1];
	\draw (1,-4) circle [radius=0.1];
	\draw [fill] (2,-4) circle [radius=0.1];
	\draw (3,-4) circle [radius=0.1];
	\draw [fill] (4,-4) circle [radius=0.1];
	
	\draw [fill] (0,-5) circle [radius=0.1];
	\draw (1,-5) circle [radius=0.1];
	\draw [fill] (2,-5) circle [radius=0.1];
	\draw (3,-5) circle [radius=0.1];
	
	\draw (0,-6) circle [radius=0.1];
	\draw (1,-6) circle [radius=0.1];
	\draw [fill] (2,-6) circle [radius=0.1];
	
	\draw [fill] (0,-7) circle [radius=0.1];
	\draw (1,-7) circle [radius=0.1];
	
	\draw (0,-8) circle [radius=0.1];
	
	\node at (1+1,-0.4) {$n_1$};
	\draw [->] (1+0.7,-0.35)--(0,-0.35);
	\draw [->] (1.3+1,-0.35)--(2+2,-0.35);
	\draw [] (4,0)--(4,-0.65);
	\draw [] (4,0)--(4,3);
	
	\node at (6,-0.32){$n$};
	\draw [->] (2.7+3,-0.35)--(2+2,-0.35);
	\draw [->] (3.3+3,-0.35)--(4+4,-0.35);
	\draw [] (4+4,0)--(4+4,-0.65);
	
	\node at (2.3+2,1.5){$k$};
	\draw [->] (2.3+2,1.8)--(2.3+2,3);
	\draw [->] (2.3+2,1.2)--(2.3+2,0);
	\draw [] (2+2,3)--(2.6+2,3);
	
	\node [left] at (-0,-4.5) {$n+k$};
	\draw [->] (-0.5,-4.5+0.3)--(-0.5,-1);
	\draw [] (-1,-1)--(0,-1);
	\draw [->] (-0.5,-4.5-0.3)--(-0.5,-8);
	\draw [] (-1,-8)--(0,-8);
	
	\end{tikzpicture}
	\caption{Newton polygon for D-type and twisted D-type theories: there is a Coulomb branch parameter associated with each black lattice point of the Newton polygon. Unlike the A type case, the lattice points on odd $x$ (horizontal) axis is deleted.
	 The difference for untwisted and twisted case is that on $x=0$ axis: even points are kept for D type theory while odd points are kept for twisted D type theory.}
 \label{DandDtwist}
\end{figure}

Now consider following configuration,
\begin{equation}
\label{eq:ADconfigD1}
\fg=\fso(n+n_1+2),~~\Phi={T\over z^{2+{k\over n}}},~~f=[1^{n+n_1+2}],
\end{equation} 
with $(k,n)=1$, and $T$ takes a specific form, see \cite{Xie:2017aqx}. We choose $n$ to be even and consider the irregular singularity with which no flavor symmetry is associated. This also implies that  $n+k$ is odd. 
See figure \ref{DandDtwist} for its corresponding Newton polygon.
The flavor symmetry is $Sp(n_1)\times SO(n+n_1+2)$. The flavor central charges of them are 
\begin{equation}
\label{Dtypeflavor}
k_{SO(2N)}=n+n_1-{n\over n+k},~~k_{Sp(n_1)}={n_1+2\over2}+{n\over 2(n+k)}.
\end{equation}
The same theory can be engineered by following configuration,
\begin{equation}
\label{eq:ADconfigD2}
\fg=\fso_{n_1(n+k)+n+2},~~\Phi={T\over z^{2+{m\over  2N-2}}},~~f=[(n+k-1)^{n_1},1^{n+n_1+2}].
\end{equation}
We have $2N=n_1(n+k)+n+2$ and $2N-2+m=n+k$. It has the same Coulomb branch with realization \ref{eq:ADconfigD1}. Using $[z]={2N-2\over n+k}$ and equation \ref{Dtype}, flavor central charges of the $SO(n+n_1+2)\times Sp(n_1)$ flavor groups are
\begin{equation}
\begin{split}
& k_{SO(n+n_1+2)}={n+2n_1}-{n_1(n+k)+n\over n+k}=n+n_1-{n\over n+k},\\
& k_{Sp(n_1)}={1\over2}[n_1(n+k)+n+2-(n+k-1){n_1(n+k)+n\over n+k}]={1\over2}(n_1+2+{n\over n+k}),
\end{split}
\end{equation}
which is exactly the result found in other description \ref{eq:ADconfigD1} (cf. equation \ref{Dtypeflavor}).

The VOA for the realization \ref{eq:ADconfigD2} (See section \ref{sec:results}) is 
\begin{equation}
\boxed{\begin{split}
VOA_B=&W^{k'}(\fso_{n_1(n+k)+n+2},[(n+k-1)^{n_1},1^{n+n_1+2}]),\\
&k'=-[n_1(n+k)+n]+{n_1(n+k)+n \over n+k}.
\end{split}}
\end{equation}
Following equation \ref{eq:centralD}, the central charge of $VOA_B$ is
\begin{equation}
\label{eq:ccVOAB}
c(VOA_B)=-\half(n+1)(n+2)(n+k-1)-\half n_1(n+n_1+2)(3k+3n-2).
\end{equation}
It is then the corresponding VOA of AD theory \ref{eq:ADconfigD1}.

In previous discussions, we require $n$ to be even. We can also consider
the case where $n$ is odd. The AD matter with two non-abelian flavor symmetries are given by following configuration,
\begin{equation}
\label{eq:ADconfigD3}
\fg=\fso_{2n+n_1},~~~\Phi={T\over z^{2+{2k\over 2n}}},~~f=[1^{2n+n_1}].
\end{equation}
Here $n_1$ is even and $n$ is odd, and $T=diag(I_{2n\times 2n},0^{n_1})$, see \cite{Xie:2017aqx} for the specific form of diagonal matrix $I_{2n\times 2n}$. We also require $(n,k)=1$. The flavor symmetry is $SO(n_1)\times SO(2n+n_1) \times SO(2)$. Unlike the previous case, there is an extra $SO(2)$ flavor symmetry besides two simple flavor groups. The same theory 
can be described by following  configuration,
\begin{equation}
\label{eq:ADconfigD4}
\fg=\fso_{2N},~~~~\Phi={T\over z^{2+{2k'\over 2N}}},~~~~f=[(2n+2k-1)^{n_1},1^{2n+n1}].
\end{equation}
We have $N=n_1(n+k-1)+n+n_1$ and $N+k'=n+k$, hence
\begin{equation}
k'=k-(k+n)n_1
\end{equation} 
Here $N$ is odd, and there 
is a $SO(2)$ flavor symmetry in irregular singularity. Unfortunately, we 
do not know the VOA for this configuration yet.

\subsubsection{Twisted $D_N=\fso_{2N}$ theory: $Sp(2N-2)\times SO(n_1)$ flavor symmetry}
The regular puncture of twisted $D_N$ theory is classified by nilpotent orbits of $\fsp_{2N-2}$ Lie algebra and is labeled by a $2N-2$ size Young Tableaux whose \textbf{odd} parts has \textbf{even} multiplicities. Given a Young Tableax 
$[r_1^{h_1},r_2^{h_2},\ldots]$, the flavor symmetry is 
\begin{equation}
G_F=\prod_{h_i~even}SO(r_i)\times \prod_{h_i~odd}Sp(r_i).
\end{equation}
Given the partition $[m^q,1^s]$ with $(q,m,s)$ even, the flavor groups are $SO(q)\times Sp(s)$ with following central charge:  
\begin{equation}
k_{Sp(s)}={(s+q+2)\over 2}-{[z]\over2},~~k_{SO(q)}=s+qm-m[z].
\label{Ctype}
\end{equation}
here $[z]$ is the scaling dimension of the coordinate $z$ in spectral curve of twisted $D_N$ theory.

Next consider the twisted theory with the following data,
\begin{equation}
\label{eq:ADconfigTD1}
\fg=\fso(n+n_1)_{z_2},~~\Phi={T^t\over z^{2+{k\over n}}},~~f=[1^{n+n_1-2}].
\end{equation} 
We take $(n,k)=1$ and $n$ is even so that there is no flavor symmetry associated with the irregular part (this implies that $n+k$ is odd), and figure. \ref{DandDtwist} illustrates its Newton polygon. The flavor symmetry is $SO(n_1)\times Sp(n+n_1-2)$, and flavor 
central charges are 
\begin{equation}
k_{Sp(n+n_1-2)}={n+n_1\over2}-{n\over 2( n+k)},~~k_{SO(n_1)}=n_1-2+{n\over n+k}.
\label{twistedflavor}
\end{equation}
To find its VOA, we realize that there is another equivalent description:
\begin{equation}
\fg=(\fso_{n_1(n+k)+n})_{z_2},~~\Phi={T^t\over z^{2+{m\over  2N}}},~~f=[(n+k-1)^{n_1},1^{n+n_1-2}],
\end{equation}
and we have $2N=n_1(n+k)+n$ and $2N+m=n+k$. The flavor central charge for flavor groups are computed by using the fact $[z]={2N\over n+k}$ and equation \ref{Ctype}, 
\begin{equation}
\begin{split}
& k_{Sp(n+n_1-2)}={n+2n_1\over 2}-{n_1(n+k)+n\over 2(n+k)}={n+n_1}-{n\over2(n+k)},\\
& k_{SO(n_1)} = n_1(n+k)+n-2-(n+k-1){n_1(n+k)+n\over n+k}=n_1-2+{n\over n+k}.
\end{split}
\end{equation}
which are the same as results from other description, see \ref{twistedflavor}. One can also check that the two configurations
have the same Coulomb branch spectrum. Therefore the corresponding VOA is the $W$ algebra
\begin{equation}
\boxed{\begin{split}VOA_C=&W^{k'}(sp_{n_1(n+k)+n-2},[(n+k-1)^{n_1}, 1^{n+n_1-2}]), \\&k'=-{n_1(n+k)+n\over2}+{n_1(n+k)+n\over 2(n+k)}
\end{split}
}
\end{equation}
From \ref{eq:centralC} the central charge of $VOA_C$ is
\begin{equation}
\label{eq:ccVOAC}
c(VOA_C)=-\half(n-2)(n-1)(n+k-1)-\half n_1(n+n_1-2)(3k+3n-2).
\end{equation}

\subsubsection{Twisted $\fsl_{2N}$ theories: $SO(2N+1) \times Sp(n_1)$ flavor symmetry}
Let's now consider twisted $\fsl_{2N}$ theory from which we can get $SO_{2N+1}$ flavor symmetry. The regular puncture is labeled by a Young Tableaux with size $2N+1$, and the constraint is that \textbf{even} parts has \textbf{even} multiplicities. Given a Young Tableaux $[r_1^{h_1},r_2^{h_2},\ldots]$, and the flavor symmetry is 
\begin{equation}
G_F=\prod_{h_i~even}Sp(r_i)\times \prod_{h_i~odd}SO(r_i).
\end{equation}
We are interested in punctures like $[m^q,1^s]$, with $m$ even, and $(q,s)$ odd. The flavor symmetry is $SO(s)\times Sp(q)$ with following central charge:
\begin{equation}
k_{SO(s)}=s+q-2-[z]/2,~~k_{Sp(q)}={s+mq-m[z]/2\over 2}.
\label{Bflavor}
\end{equation} 
Here $[z]$ is the scaling dimension of $z$ coordinate in spectral curve of twisted $sl_{2N}$ type.

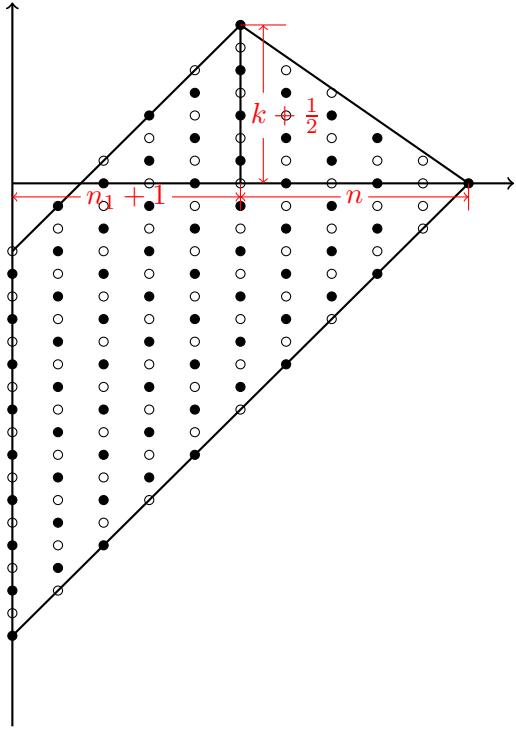
\begin{figure}
	\centering
	\begin{tikzpicture}[scale=0.6]
	
	\draw [thick,->] (1,-12) -- (1,4);
	\draw [thick,->] (1,0) -- (12,0);
	\draw [thick] (6,0) -- (6,3.5);
	\draw [thick] (6,3.5) -- (1, -1.5);
	\draw [thick] (6,3.5) -- (11,0);
	\draw [thick] (11,0) -- (1,-10);
	
	\draw [fill] (1,-10) circle [radius=0.1];
	\draw (1,-9.5) circle [radius=0.1];
	\draw [fill] (1,-9) circle [radius=0.1];
	\draw (1,-8.5) circle [radius=0.1];
	\draw [fill] (1,-8) circle [radius=0.1];
	\draw (1,-7.5) circle [radius=0.1];
	\draw [fill] (1,-7) circle [radius=0.1];
	\draw (1,-6.5) circle [radius=0.1];
	\draw [fill] (1,-6) circle [radius=0.1];
	\draw (1,-5.5) circle [radius=0.1];
	\draw [fill] (1,-5) circle [radius=0.1];
	\draw (1,-4.5) circle [radius=0.1];
	\draw [fill] (1,-4) circle [radius=0.1];
	\draw (1,-3.5) circle [radius=0.1];
	\draw [fill] (1,-3) circle [radius=0.1];
	\draw (1,-2.5) circle [radius=0.1];
	\draw [fill] (1,-2) circle [radius=0.1];
	\draw (1,-1.5) circle [radius=0.1];
	
	\draw (2,-9) circle [radius=0.1];
	\draw [fill] (2,-8.5) circle [radius=0.1];
	\draw (2,-8) circle [radius=0.1];
	\draw [fill] (2,-7.5) circle [radius=0.1];
	\draw (2,-7) circle [radius=0.1];
	\draw [fill] (2,-6.5) circle [radius=0.1];
	\draw (2,-6) circle [radius=0.1];
	\draw [fill] (2,-5.5) circle [radius=0.1];
	\draw (2,-5) circle [radius=0.1];
	\draw [fill] (2,-4.5) circle [radius=0.1];
	\draw (2,-4) circle [radius=0.1];
	\draw [fill] (2,-3.5) circle [radius=0.1];
	\draw (2,-3) circle [radius=0.1];
	\draw [fill] (2,-2.5) circle [radius=0.1];
	\draw (2,-2) circle [radius=0.1];
	\draw [fill] (2,-1.5) circle [radius=0.1];
	\draw (2,-1) circle [radius=0.1];
	\draw [fill] (2,-0.5) circle [radius=0.1];
	
	\draw [fill] (3,-8) circle [radius=0.1];
	\draw (3,-7.5) circle [radius=0.1];
	\draw [fill] (3,-7) circle [radius=0.1];
	\draw (3,-6.5) circle [radius=0.1];
	\draw [fill] (3,-6) circle [radius=0.1];
	\draw (3,-5.5) circle [radius=0.1];
	\draw [fill] (3,-5) circle [radius=0.1];
	\draw (3,-4.5) circle [radius=0.1];
	\draw [fill] (3,-4) circle [radius=0.1];
	\draw (3,-3.5) circle [radius=0.1];
	\draw [fill] (3,-3) circle [radius=0.1];
	\draw (3,-2.5) circle [radius=0.1];
	\draw [fill] (3,-2) circle [radius=0.1];
	\draw (3,-1.5) circle [radius=0.1];
	\draw [fill] (3,-1) circle [radius=0.1];
	\draw (3,-0.5) circle [radius=0.1];
	\draw [fill] (3,0) circle [radius=0.1];
	\draw (3,0.5) circle [radius=0.1];
	
	\draw (4,-7) circle [radius=0.1];
	\draw [fill] (4,-6.5) circle [radius=0.1];
	\draw (4,-6) circle [radius=0.1];
	\draw [fill] (4,-5.5) circle [radius=0.1];
	\draw (4,-5) circle [radius=0.1];
	\draw [fill] (4,-4.5) circle [radius=0.1];
	\draw (4,-4) circle [radius=0.1];
	\draw [fill] (4,-3.5) circle [radius=0.1];
	\draw (4,-3) circle [radius=0.1];
	\draw [fill] (4,-2.5) circle [radius=0.1];
	\draw (4,-2) circle [radius=0.1];
	\draw [fill] (4,-1.5) circle [radius=0.1];
	\draw (4,-1) circle [radius=0.1];
	\draw [fill] (4,-0.5) circle [radius=0.1];
	\draw (4,0) circle [radius=0.1];
	\draw [fill] (4,0.5) circle [radius=0.1];
	\draw (4,1) circle [radius=0.1];
	\draw [fill] (4,1.5) circle [radius=0.1];
	
	\draw [fill] (5,-6) circle [radius=0.1];
	\draw (5,-5.5) circle [radius=0.1];
	\draw [fill] (5,-5) circle [radius=0.1];
	\draw (5,-4.5) circle [radius=0.1];
	\draw [fill] (5,-4) circle [radius=0.1];
	\draw (5,-3.5) circle [radius=0.1];
	\draw [fill] (5,-3) circle [radius=0.1];
	\draw (5,-2.5) circle [radius=0.1];
	\draw [fill] (5,-2) circle [radius=0.1];
	\draw (5,-1.5) circle [radius=0.1];
	\draw [fill] (5,-1) circle [radius=0.1];
	\draw (5,-0.5) circle [radius=0.1];
	\draw [fill] (5,0) circle [radius=0.1];
	\draw (5,0.5) circle [radius=0.1];
	\draw [fill] (5,1) circle [radius=0.1];
	\draw (5,1.5) circle [radius=0.1];
	\draw [fill] (5,2) circle [radius=0.1];
	\draw (5,2.5) circle [radius=0.1];
	
	\draw (6,-5) circle [radius=0.1];
	\draw [fill] (6,-4.5) circle [radius=0.1];
	\draw (6,-4) circle [radius=0.1];
	\draw [fill] (6,-3.5) circle [radius=0.1];
	\draw (6,-3) circle [radius=0.1];
	\draw [fill] (6,-2.5) circle [radius=0.1];
	\draw (6,-2) circle [radius=0.1];
	\draw [fill] (6,-1.5) circle [radius=0.1];
	\draw (6,-1) circle [radius=0.1];
	\draw [fill] (6,-0.5) circle [radius=0.1];
	\draw (6,0) circle [radius=0.1];
	\draw [fill] (6,0.5) circle [radius=0.1];
	\draw (6,1) circle [radius=0.1];
	\draw [fill] (6,1.5) circle [radius=0.1];
	\draw (6,2) circle [radius=0.1];
	\draw [fill] (6,2.5) circle [radius=0.1];
	\draw (6,3) circle [radius=0.1];
	\draw [fill] (6,3.5) circle [radius=0.1];
	
	\draw [fill] (7,-4) circle [radius=0.1];
	\draw (7,-3.5) circle [radius=0.1];
	\draw [fill] (7,-3) circle [radius=0.1];
	\draw (7,-2.5) circle [radius=0.1];
	\draw [fill] (7,-2) circle [radius=0.1];
	\draw (7,-1.5) circle [radius=0.1];
	\draw [fill] (7,-1) circle [radius=0.1];
	\draw (7,-0.5) circle [radius=0.1];
	\draw [fill] (7,0) circle [radius=0.1];
	\draw (7,0.5) circle [radius=0.1];
	\draw [fill] (7,1) circle [radius=0.1];
	\draw (7,1.5) circle [radius=0.1];
	\draw [fill] (7,2) circle [radius=0.1];
	\draw (7,2.5) circle [radius=0.1];
	
	\draw (8,-3) circle [radius=0.1];
	\draw [fill] (8,-2.5) circle [radius=0.1];
	\draw (8,-2) circle [radius=0.1];
	\draw [fill] (8,-1.5) circle [radius=0.1];
	\draw (8,-1) circle [radius=0.1];
	\draw [fill] (8,-0.5) circle [radius=0.1];
	\draw (8,0) circle [radius=0.1];
	\draw [fill] (8,0.5) circle [radius=0.1];
	\draw (8,1) circle [radius=0.1];
	\draw [fill] (8,1.5) circle [radius=0.1];
	\draw (8,2) circle [radius=0.1];
	
	\draw [fill] (9,-2) circle [radius=0.1];
	\draw (9,-1.5) circle [radius=0.1];
	\draw [fill] (9,-1) circle [radius=0.1];
	\draw (9,-0.5) circle [radius=0.1];
	\draw [fill] (9,0) circle [radius=0.1];
	\draw (9,0.5) circle [radius=0.1];
	\draw [fill] (9,1) circle [radius=0.1];
	
	\draw (10,-1) circle [radius=0.1];
	\draw (10,-0.5) circle [radius=0.1];
	\draw (10,0) circle [radius=0.1];
	\draw (10,0.5) circle [radius=0.1];
	
	\draw [fill] (11,0) circle [radius=0.1];
	
	\node [red] at (3.5,-0.3) {$n_1+1$};
	\draw [red,->] (2.5,-0.3) -- (1,-0.3);
	\draw [red,->] (4.5,-0.3) -- (6,-0.3);
	\draw [red] (6,0) -- (6,-0.6);
	
	\node [red] at (8.5,-0.3) {$n$};
	\draw [red,->] (8.2,-0.3) -- (6,-0.3);
	\draw [red,->] (8.8,-0.3) -- (11,-0.3);
	\draw [red] (11,0) -- (11,-0.6);
	
	\node [red, right] at (6, 1.5) {$k+\frac{1}{2}$};
	\draw [red,->] (6.5, 2) -- (6.5, 3.5);
	\draw [red,->] (6.5, 1) -- (6.5, 0);
	\draw [red] (6, 3.5) -- (7, 3.5);
	
	
	\end{tikzpicture}
	\caption{Newton polygon for twisted $sl_{2N}$ theory. The integral points on $x=even$ axis are kept, while 
	the half integral points on odd $x=odd$ axis are kept, here $x$ is the horizonal coordinate.}
\label{twistedsl2n}
\end{figure}

The defining data for our theory is ,
\begin{equation}
\fg=(sl_{n+n_1+1})_{z_2},~~\Phi={T^t\over z^{2+{k+1/2\over n}}},~~f=[1^{n+n_1+2}].
\end{equation}
Here  $n$ is odd and $n_1$ is even (cf. figure \ref{twistedsl2n}). The flavor symmetry is $SO(n+n_1+2)\times Sp(n_1)$ with following flavor central charges,
\begin{equation}
k_{SO(n+n_1+2)}=n+n_1-{n\over 2n+2k+1},~~~k_{Sp(n_1)}={n_1+2\over 2}+{n\over 2(2n+2k+1)}.
\label{Btype}
\end{equation}
The VOA is found by identifying following equivalent configuration,
\begin{equation}
\fg=(\fsl_{n_1(2n+2k+1)+n+1})_{z_2},~~\Phi={T^t\over z^{2+{m+{1\over2}\over 2N-1}}},~~f=[(2n+2k)^{n_1}, 1^{n+n_1+2}].
\end{equation}
with $2N=n_1(2n+2k+1)+n+1$ and $2N-1+m+{1\over2}=n+k+{1\over2}$.
To compute the flavor central charges for $SO(n+n_1+2)\times Sp(n_1)$ flavor symmetry, use the fact that $[z]/2={n_1(2n+2k+1)+n\over 2n+2k+1}$,
\begin{equation}
\begin{split}
& k_{So(n+n_1+2)}=2n_1+n-{n_1(2n+2k+1)+n\over 2n+2k+1}=n+n_1-{n\over 2n+2k+1},   \\
& k_{Sp(n_1)}={1\over 2}[n_1(2n+2k+1)+n+2-(2n+2k){n_1(2n+2k+1)+n\over 2n+2k+1}]={1\over2}(n_1+2+{n\over 2n+2k+1}),
\end{split}
\end{equation}
which are exactly the same as that computed in other description \ref{Btype}. One can also check that the two configurations give the same Coulomb branch spectrum. Hence the corresponding $W$ algebra is 
\begin{equation}
\boxed{\begin{split}
VOA_D=& W^{k'}(so_{n_1(2n+2k+1)+n+2},[(2n+2k)^{n_1},1^{n+n_1+2}]), \\
&k'=-(n_1(2n+2k+1)+n)+{n_1(2n+2k+1)+n\over 2n+2k+1}.
\end{split}}
\end{equation}
Using equation \ref{eq:centralB}, the central charge is
\begin{equation}
\label{eq:ccVOAD}
c(VOA_D)=-(n+1)(n+2)(k+n)-\half n_1(n+n_1+2)(6k+6n+1).
\end{equation}

\subsubsection{Twisted $\fsl_{2N+1}$ theories: $Sp(2N)\times SO(n_1+1)$ flavor symmetry}
\begin{figure}
	\centering
	\begin{tikzpicture}
	
	\draw [thick,->] (0+14,-9-2) -- (0+14,6-2);
	\draw [thick,->] (0+14,0-2) -- (10+14,0-2);
	\draw [thick] (6+14,0-2) -- (6+14,5.5-2);
	\draw [thick] (6+14,5.5-2) -- (0+14, -0.5-2);
	\draw [thick] (6+14,5.5-2) -- (9+14,0-2);
	\draw [thick] (9+14,0-2) -- (0+14,-9-2);
	
	\draw (0+14,-9-2) circle [radius=0.1];
	\draw [fill] (0+14,-8.5-2) circle [radius=0.1];
	\draw (0+14,-8-2) circle [radius=0.1];
	\draw [fill] (0+14,-7.5-2) circle [radius=0.1];
	\draw (0+14,-7-2) circle [radius=0.1];
	\draw [fill] (0+14,-6.5-2) circle [radius=0.1];
	\draw (0+14,-6-2) circle [radius=0.1];
	\draw [fill] (0+14,-5.5-2) circle [radius=0.1];
	\draw (0+14,-5-2) circle [radius=0.1];
	\draw [fill] (0+14,-4.5-2) circle [radius=0.1];
	\draw (0+14,-4-2) circle [radius=0.1];
	\draw [fill] (0+14,-3.5-2) circle [radius=0.1];
	\draw (0+14,-3-2) circle [radius=0.1];
	\draw [fill] (0+14,-2.5-2) circle [radius=0.1];
	\draw (0+14,-2-2) circle [radius=0.1];
	\draw [fill] (0+14,-1.5-2) circle [radius=0.1];
	\draw (0+14,-1-2) circle [radius=0.1];
	\draw [fill] (0+14,-0.5-2) circle [radius=0.1];
	
	\draw [fill] (1+14,-8-2) circle [radius=0.1];
	\draw (1+14,-7.5-2) circle [radius=0.1];
	\draw [fill] (1+14,-7-2) circle [radius=0.1];
	\draw (1+14,-6.5-2) circle [radius=0.1];
	\draw [fill] (1+14,-6-2) circle [radius=0.1];
	\draw (1+14,-5.5-2) circle [radius=0.1];
	\draw [fill] (1+14,-5-2) circle [radius=0.1];
	\draw (1+14,-4.5-2) circle [radius=0.1];
	\draw [fill] (1+14,-4-2) circle [radius=0.1];
	\draw (1+14,-3.5-2) circle [radius=0.1];
	\draw [fill] (1+14,-3-2) circle [radius=0.1];
	\draw (1+14,-2.5-2) circle [radius=0.1];
	\draw [fill] (1+14,-2-2) circle [radius=0.1];
	\draw (1+14,-1.5-2) circle [radius=0.1];
	\draw [fill] (1+14,-1-2) circle [radius=0.1];
	\draw (1+14,-0.5-2) circle [radius=0.1];
	\draw [fill] (1+14,0-2) circle [radius=0.1];
	\draw (1+14,0.5-2) circle [radius=0.1];
	
	\draw (2+14,-7-2) circle [radius=0.1];
	\draw [fill] (2+14,-6.5-2) circle [radius=0.1];
	\draw (2+14,-6-2) circle [radius=0.1];
	\draw [fill] (2+14,-5.5-2) circle [radius=0.1];
	\draw (2+14,-5-2) circle [radius=0.1];
	\draw [fill] (2+14,-4.5-2) circle [radius=0.1];
	\draw (2+14,-4-2) circle [radius=0.1];
	\draw [fill] (2+14,-3.5-2) circle [radius=0.1];
	\draw (2+14,-3-2) circle [radius=0.1];
	\draw [fill] (2+14,-2.5-2) circle [radius=0.1];
	\draw (2+14,-2-2) circle [radius=0.1];
	\draw [fill] (2+14,-1.5-2) circle [radius=0.1];
	\draw (2+14,-1-2) circle [radius=0.1];
	\draw [fill] (2+14,-0.5-2) circle [radius=0.1];
	\draw (2+14,0-2) circle [radius=0.1];
	\draw [fill] (2+14,0.5-2) circle [radius=0.1];
	\draw (2+14,1-2) circle [radius=0.1];
	\draw [fill] (2+14,1.5-2) circle [radius=0.1];
	
	\draw [fill] (3+14,-6-2) circle [radius=0.1];
	\draw (3+14,-5.5-2) circle [radius=0.1];
	\draw [fill] (3+14,-5-2) circle [radius=0.1];
	\draw (3+14,-4.5-2) circle [radius=0.1];
	\draw [fill] (3+14,-4-2) circle [radius=0.1];
	\draw (3+14,-3.5-2) circle [radius=0.1];
	\draw [fill] (3+14,-3-2) circle [radius=0.1];
	\draw (3+14,-2.5-2) circle [radius=0.1];
	\draw [fill] (3+14,-2-2) circle [radius=0.1];
	\draw (3+14,-1.5-2) circle [radius=0.1];
	\draw [fill] (3+14,-1-2) circle [radius=0.1];
	\draw (3+14,-0.5-2) circle [radius=0.1];
	\draw [fill] (3+14,0-2) circle [radius=0.1];
	\draw (3+14,0.5-2) circle [radius=0.1];
	\draw [fill] (3+14,1-2) circle [radius=0.1];
	\draw (3+14,1.5-2) circle [radius=0.1];
	\draw [fill] (3+14,2-2) circle [radius=0.1];
	\draw (3+14,2.5-2) circle [radius=0.1];
	
	\draw (4+14,-5-2) circle [radius=0.1];
	\draw [fill] (4+14,-4.5-2) circle [radius=0.1];
	\draw (4+14,-4-2) circle [radius=0.1];
	\draw [fill] (4+14,-3.5-2) circle [radius=0.1];
	\draw (4+14,-3-2) circle [radius=0.1];
	\draw [fill] (4+14,-2.5-2) circle [radius=0.1];
	\draw (4+14,-2-2) circle [radius=0.1];
	\draw [fill] (4+14,-1.5-2) circle [radius=0.1];
	\draw (4+14,-1-2) circle [radius=0.1];
	\draw [fill] (4+14,-0.5-2) circle [radius=0.1];
	\draw (4+14,0-2) circle [radius=0.1];
	\draw [fill] (4+14,0.5-2) circle [radius=0.1];
	\draw (4+14,1-2) circle [radius=0.1];
	\draw [fill] (4+14,1.5-2) circle [radius=0.1];
	\draw (4+14,2-2) circle [radius=0.1];
	\draw [fill] (4+14,2.5-2) circle [radius=0.1];
	\draw (4+14,3-2) circle [radius=0.1];
	\draw [fill] (4+14,3.5-2) circle [radius=0.1];
	
	\draw [fill] (5+14,-4-2) circle [radius=0.1];
	\draw (5+14,-3.5-2) circle [radius=0.1];
	\draw [fill] (5+14,-3-2) circle [radius=0.1];
	\draw (5+14,-2.5-2) circle [radius=0.1];
	\draw [fill] (5+14,-2-2) circle [radius=0.1];
	\draw (5+14,-1.5-2) circle [radius=0.1];
	\draw [fill] (5+14,-1-2) circle [radius=0.1];
	\draw (5+14,-0.5-2) circle [radius=0.1];
	\draw [fill] (5+14,0-2) circle [radius=0.1];
	\draw (5+14,0.5-2) circle [radius=0.1];
	\draw [fill] (5+14,1-2) circle [radius=0.1];
	\draw (5+14,1.5-2) circle [radius=0.1];
	\draw [fill] (5+14,2-2) circle [radius=0.1];
	\draw (5+14,2.5-2) circle [radius=0.1];
	\draw [fill] (5+14,3-2) circle [radius=0.1];
	\draw (5+14,3.5-2) circle [radius=0.1];
	\draw [fill] (5+14,4-2) circle [radius=0.1];
	\draw (5+14,4.5-2) circle [radius=0.1];
	
	\draw (6+14,-3-2) circle [radius=0.1];
	\draw [fill] (6+14,-2.5-2) circle [radius=0.1];
	\draw (6+14,-2-2) circle [radius=0.1];
	\draw [fill] (6+14,-1.5-2) circle [radius=0.1];
	\draw (6+14,-1-2) circle [radius=0.1];
	\draw [fill] (6+14,-0.5-2) circle [radius=0.1];
	\draw (6+14,0-2) circle [radius=0.1];
	\draw [fill] (6+14,0.5-2) circle [radius=0.1];
	\draw (6+14,1-2) circle [radius=0.1];
	\draw [fill] (6+14,1.5-2) circle [radius=0.1];
	\draw (6+14,2-2) circle [radius=0.1];
	\draw [fill] (6+14,2.5-2) circle [radius=0.1];
	\draw (6+14,3-2) circle [radius=0.1];
	\draw [fill] (6+14,3.5-2) circle [radius=0.1];
	\draw (6+14,4-2) circle [radius=0.1];
	\draw [fill] (6+14,4.5-2) circle [radius=0.1];
	\draw (6+14,5-2) circle [radius=0.1];
	\draw [fill] (6+14,5.5-2) circle [radius=0.1];
	
	\draw [fill] (7+14,-2-2) circle [radius=0.1];
	\draw (7+14,-1.5-2) circle [radius=0.1];
	\draw [fill] (7+14,-1-2) circle [radius=0.1];
	\draw (7+14,-0.5-2) circle [radius=0.1];
	\draw [fill] (7+14,0-2) circle [radius=0.1];
	\draw (7+14,0.5-2) circle [radius=0.1];
	\draw [fill] (7+14,1-2) circle [radius=0.1];
	\draw (7+14,1.5-2) circle [radius=0.1];
	\draw [fill] (7+14,2-2) circle [radius=0.1];
	\draw (7+14,2.5-2) circle [radius=0.1];
	\draw [fill] (7+14,3-2) circle [radius=0.1];
	\draw (7+14,3.5-2) circle [radius=0.1];
	
	\draw (8+14,-1-2) circle [radius=0.1];
	\draw (8+14,-0.5-2) circle [radius=0.1];
	\draw (8+14,0-2) circle [radius=0.1];
	\draw (8+14,0.5-2) circle [radius=0.1];
	\draw (8+14,1-2) circle [radius=0.1];
	\draw (8+14,1.5-2) circle [radius=0.1];
	
	\draw [fill] (9+14,0-2) circle [radius=0.1];
	
	\node [red] at (3+14,-0.3-2) {$n_1$};
	\draw [red, ->] (2.5+14,-0.3-2) -- (0+14,-0.3-2);
	\draw [red, ->] (3.5+14,-0.3-2) -- (6+14,-0.3-2);
	\draw [red] (6+14,0-2) -- (6+14,-0.6-2);
	
	\node [red] at (7.5+14,-0.3-2) {$n$};
	\draw [red, ->] (7+14,-0.3-2) -- (6+14,-0.3-2);
	\draw [red, ->] (8+14,-0.3-2) -- (9+14,-0.3-2);
	\draw [red] (9+14,0-2) -- (9+14,-0.6-2);
	
	\node [red, right] at (6+14, 2.5-2) {$k+\frac{1}{2}$};
	\draw [red,->] (6.5+14, 3-2) -- (6.5+14, 5.5-2);
	\draw [red,->] (6.5+14, 2-2) -- (6.5+14, 0-2);
	\draw [red] (6+14, 5.5-2) -- (7+14, 5.5-2);
	
	
	\end{tikzpicture}
	
	\caption{Newton polygons for twisted $Sl_{2N+1}$ theory.The integral points on  $x=odd$ axis are kept, while 
		the half integral points on  $x=even$ axis are kept, and here $x$ is the horizonal coordinate.}
	\label{twistedsl2n1}
\end{figure}
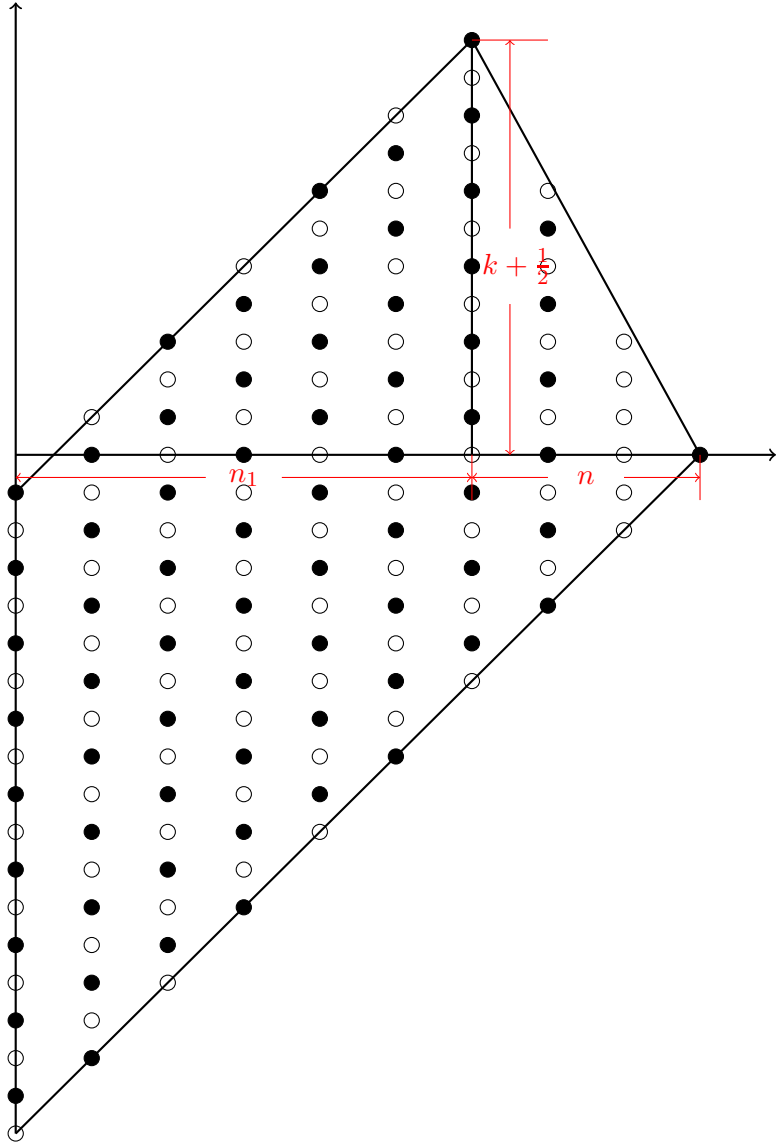

Now consider twisted $\fsl_{2N+1}$ theory from which we can also get $Sp(2N)$ flavor symmetry, but this time we
will also get another $B$ type flavor symmetry, which is different from twisted $D$ type theory. The twisted regular puncture is classified by a Young Tableaux with size $2N$. Given a Young Tableax 
$[r_1^{h_1},r_2^{h_2},\ldots]$, the flavor symmetry is,
\begin{equation}
G_F=\prod_{h_i~even}SO(r_i)\times \prod_{h_i~odd}Sp(r_i).
\end{equation}
Given the partition $[m^q,1^s]$ with $(q,m,s)$ even, the flavor groups are $SO(q)\times Sp(s)$ with following central charge:  
\begin{equation}
k_{Sp(s)}={(s+q+2)\over 2}-{[z]\over4},~~k_{SO(q)}=s+qm-{m[z]/2}.
\end{equation}
Here $[z]$ is the scaling dimension of the $z$ coordinate in spectral curve of twisted $sl_{2N+1}$ theory.

Consider a theory defined by following data, here we use the $z_2$ outerautomorphism of $sl_{n+n_1}$ theory:
\begin{equation}
\fg=(\fsl_{n+n_1})_{z_2},~~\Phi={T\over z^{2+{k+{1/2}\over n}}},~~f^t=[1^{n+n_1-1}].
\end{equation}
Here $n$ is odd and $n_1$ is even. The flavor symmetry is $Sp(n+n_1-1)\times SO(n_1+1)$ and flavor central charges are,
\begin{equation}
k_{Sp(n+n_1-1)}={n+n_1+1\over2}-{n\over 2 (2n+2k+1)},~~~k_{SO(n_1+1)}=n_1-1+{n\over 2n+2k+1)}.
\label{ctwist}
\end{equation}
For above theory, we find following equivalent realization,
\begin{equation}
\fg=(so_{{2N}})_{z_2},~~\Phi={T^t\over z^{2+{2m+1 \over 2N}}}+..,~f^t=[1^{2N-2}].
\end{equation}
Here $2N=(n_1+1)(2n+2k+1)+n$, and $2N+2m+1=2n+2k+1$. Notice that
we found a realization from a {\bf different} type of twisted theory! The flavor central charges for $Sp(n+n_1-1)\times SO(n_1+1)$ are (Here ${[z]\over2}={(n_1+1)(2n+2k+1)+n\over 2( 2n+2k+1)}$.)
\begin{equation}
\begin{split}
&k_{Sp(n+n_1-1)}={1\over 2}(n+2n_1+2)-{1\over2}{(n_1+1)(2n+2k+1)+n\over 2n+2k+1}={n+n_1+1\over2}-{n\over 2 (2n+2k+1)}, \\ 
&k_{So(n_1+1)}= (n_1+1)(2n+2k+1)+n-2-(2n+2k){(n_1+1)(2n+2k+1)+n\over 2n+2k+1}\\
&=n_1-1+{n\over 2n+2k+1},
\end{split}
\end{equation}
which agrees with the result shown in \ref{ctwist}. The corresponding VOA is following $W$ algebra:
\begin{equation}
\boxed{\begin{split}
VOA_E=&W^{k'}(\fsp_{(n_1+1)(2n+2k+1)+n-2},[(2n+2k)^{n_1+1},1^{n+n_1-1}]), \\
&k'=-{(n_1+1)(2n+2k+1)+n\over2}+{{(n_1+1)(2n+2k+1)+n}\over 2(2n+2k+1)}.
\end{split}}
\end{equation}
Using equation \ref{eq:centralC}, the central charge of $VOA_E$ is
\begin{equation}
\label{eq:ccVOAE}
c(VOA_E)=-\half(n-1)[2n^2+2(k+1)(n+1)-1]-\half n_1(n+n_1)(6k+6n+1).
\end{equation}

\subsubsection{Conformal embedding} \label{summary}
In previous discussions, we have found $W$ algebras corresponding to AD matters with two distinct type of non-abelian flavor symmetries. The 4d flavor symmetries give the 
AKM subalgebras of 2d VOAs with $k_{2d}=-k_{4d}$, so we obtain following embeddings of AKM algebra into $W$ algebra: 
\begin{equation}
\label{eq:embeddingOfAKM}
\begin{split}
&A:~\mathfrak{u}(1)\times V_{-n_1-\frac{n}{n+k}}(\fsu_{n_1})\times V_{-n_1-n+{n\over n+k}}(\fsu_{n_1+n})\subset VOA_A,\\
&B:~V_{-{n_1+2\over2}-{n\over 2(n+k)}}(\fsp_{n_1})\times V_{-n-n_1+{n\over n+k}}(\fso_{n+n_1+2}) \subset VOA_B,
\\
&C:~V_{-n_1+2-{n\over n+k}}(\fso_{n_1})\times V_{-{n+n_1\over2}+{n\over 2( n+k)}}(\fsp_{n+n_1-2})\subset VOA_C,\\
&D:~V_{-{n_1+2\over 2}-{n\over 2(2n+2k+1)}}(\fsp_{n_1})\times V_{-n-n_1+{n\over 2n+2k+1}}(\fso_{n+n_1+2}) \subset VOA_D,\\
&E:~V_{-n_1+1-{n\over 2n+2k+1}}(\fso_{n_1+1})\times V_{-{n+n_1+1\over2}+{n\over 2(2n+2k+1)}}(\fsp_{n+n_1-1})\subset VOA_E.
\end{split}
\end{equation} 
The righthandside are following  $W$ algebras:
\begin{equation}
\begin{split}
& VOA_A:~~W^{-h^{\vee}+{h^{\vee}\over n+k}}(\fsl_{n_1(n+k)+n},[(n+k-1)^{n_1},1^{n+n_1}]),~~~~~~~~~~~~~~~~~~~~~~~~~~~~h^{\vee}=n_1(n+k)+n,\\
&VOA_B:~~W^{-h^{\vee}+{h^{\vee}\over n+k}}(\fso_{n_1(n+k)+n+2},[(n+k-1)^{n_1},1^{n+n_1+2}]),~n~even,~n_1~even,~h^{\vee}=n_1(n+k)+n,\\
&VOA_C:~~W^{-h^{\vee}+{h^{\vee}\over n+k}}(\fsp_{n_1(n+k)+n-2},[(n+k-1)^{n_1},1^{n+n_1-2}]),~n~even,~n_1~even,~h^{\vee}={n_1(n+k)+n\over2},\\
&VOA_D:~~W^{-h^{\vee}+{h^{\vee}\over 2n+2k+1}}(\fso_{n_1(2n+2k+1)+n+2},[(2n+2k)^{n_1},1^{n+n_1+2}]),~n~odd,~n_1~even,~\\
&~~~~~~~~~~~~~~~~~~~~~~~~~~~~~~~~~~~~~~~~~~~~~~~~~~~~~~~~~~~~~~~~~~~~~~~~~~~~~~~~~~~~~~~~~~~~h^{\vee}=n_1(2n+2k+1)+n,\\
&VOA_E:~~W^{-h^{\vee}+{h^{\vee}\over 2n+2k+1}}(\fsp_{(n_1+1)(2n+2k+1)+n-2},[(2n+2k)^{n_1+1},1^{n+n_1-1}]),~n~odd,~n_1~even,\\
&~~~~~~~~~~~~~~~~~~~~~~~~~~~~~~~~~~~~~~~~~~~~~~~~~~~~~~~~~~~~~~~~~~~~~~~~~~~~~~~~~~~~~h^{\vee}={(n_1+1)(2n+2k+1)+n\over2}, \\
\end{split}
\end{equation}
with explicit 2d central charges given by equations  \ref{eq:ccVOAA}, \ref{eq:ccVOAB}, \ref{eq:ccVOAC}, \ref{eq:ccVOAD} and \ref{eq:ccVOAE},
\begin{equation}
\begin{split}
c(VOA_A)&=-\left(n^2-1\right) (k+n-1)-n_1(n+n_1) (3 k+3 n-2),\\
c(VOA_B)&=-\half(n+1)(n+2)(n+k-1)-\half n_1(n+n_1+2)(3k+3n-2),\\
c(VOA_C)&=-\half(n-2)(n-1)(n+k-1)-\half n_1(n+n_1-2)(3k+3n-2),\\
c(VOA_D)&=-(n+1)(n+2)(k+n)-\half n_1(n+n_1+2)(6k+6n+1),\\
c(VOA_E)&=-\half(n-1)[2n^2+2(k+1)(n+1)-1]-\half n_1(n+n_1)(6k+6n+1).
\end{split}
\end{equation}
Recalling the central charge of AKM algebra $\fg_k$ is
\begin{equation}
c(\fg_k)=\frac{k \dim\fg}{k+h^{\vee}},
\end{equation}
with $h^{\vee}$ the dual Coxter number of $\fg$,
one can show that central charges of AKM subalgebras on the left hand side of equation \ref{eq:embeddingOfAKM} are {\bf equal} to central charges of $W$ algebras on the right hand side. It would be interesting to check whether they 
are indeed conformal embeddings.

\subsection{Exceptional Lie algebra}
Now  consider AD theories constructed using exceptional 6d $(2,0)$ theory. In general, we 
 have a theory with flavor symmetry $E_n\times G$, where $G$ is some subgroup of $E_n$. 
Unfortunately, we do not know how to realize the above theory using the configuration presented in section \ref{sec:results}. 
Instead we analyze its Coulomb branch when a generic regular singularity is present, and check whether one can find 
the same Coulomb branch spectrum using constructions presented in section \ref{sec:results}. We do not attempt to do a general analysis, and only 
give some examples here:
\begin{itemize}
\item Start with $\fe_8$ theory and look at the regular puncture whose Nahm label is $[A_1]$ \cite{Chacaltana:2012zy}, then the flavor symmetry is $E_7$. The flavor central charge is, 
\begin{equation}
k_{E_7}=24-{30\over 30+k}, 	
\end{equation}
We would like to realize this theory by using a 6d $(2,0)$ $\fe_7$ theory with a full puncture of $E_7$ type. The flavor central charge in $\fe_7$ construction is, 
\begin{equation}
k_{E_7}=18-{18\over 18+k'}, 	
\end{equation}
We find a solution with $30+k=2=18+k'=2$. Both theory has the Coulomb branch spectrum  $[9,5,3]$ (These numbers can be derived using Newton polygon of $E_8$ type theory \cite{Xie:2017aqx} and the pole structure of $E_8$ nilpotent orbit $[A_1]$ \cite{Chcaltana:2018zag}.). However the $E_7$ description has 
an extra $U(1)$ flavor symmetry. The VOA is 
\begin{equation}
W^{-15}(\fe_8,[A_1])
\end{equation}
Here $[A_1]$ denotes minimal nilpotent orbit of $\fe_8$ Lie algebra. The interpretation is that the $\fe_8$ description misses the $U(1)$ flavor symmetry.

\item Now look at $\fe_7$ theory with regular puncture whose Nahm label is $[A_1]$. The flavor symmetry is $SO(12)$, and the flavor central charge is,
\begin{equation}
k_{SO(12)}=18-{18\over 18+k}.
\end{equation}
On the other hand, starting with the $\fso_{12}$ realization, the flavor central charge is,
\begin{equation}
k_{SO(12)}=14-{10\over 10+k'}.
\end{equation}
Again, we find solution $18+k=18+k'=2$. For this value, $E_7$ 
configuration has an extra $U(1)$ flavor symmetry. The $\fso_{12}$ 
descrition also has an extra $U(1)$ flavor symmetry. The Coulomb branch spectrum is {\color{red}$[5,3]$}, which is read from Newton polygon. We can not find its VOA because of this extra $U(1)$ flavor symmetry.

\item Finally we look at an $\fe_6$ theory with a regular puncture whose Nahm label is $[A_1]$. The flavor symmetry is $SU(6)$, and the flavor central charge is,
\begin{equation}
k_{SO(10)}=9-{12\over 12+k}.
\end{equation}
In an $\fsu_6$ realization, the flavor central charge is,
\begin{equation}
k_{SU(6)}=6-{6\over 6+k'}.
\end{equation}
The matching of flavor central charges $k_{SO(10)}=k_{SU(6)}$ requires $12+k=6+k'=2$. Now in the $SU(6)$ description, 
there is an extra $U(1)$ flavor symmetry, and it is just the flavor symmetry of the 
$\mathcal{N}=2$ $SU(3)$ superQCD (SQCD) with six fundamental hypermultiplets. The $\fe_6$ description has just $SU(6)$ manifest flavor symmetry. The VOA is 
\begin{equation}
W^{-6}(\fe_6,[A_1])
\end{equation}
We claim that this $W$ algebra is the VOA for $SU(3)$ SQCD with six fundamental flavors. One simple check is that the central charge of this $W$ algebra is $-34$ which is equal to $-12 c_{4d}$, where $c_{4d}={34/12}$ is the central charge of $SU(3)$ SQCD with six fundamental flavor. It is interesting to notice 
that there is an emerging $U(1)$ flavor symmetry for the $W$ algebra.

\end{itemize} 

Now we move on to more interesting examples with exceptional flavor symmetries. 
Some rank one theories with following data were found in \cite{Xie:2017obm},
\begin{equation}
\begin{split}
&G=B_3,~~k_{B_3}=2,~ u=[2],~~~~V_{-2}(B_3),\\
&G=G_2,~~k_{G_2}=2,~u=[2],~~~~V_{-2}(G_2),\\
&G=F_4,~~k_{F_4}=3,~~u=[3],~~~~V_{-3}(F_4).
\end{split}
\end{equation}	
The $(a,c)$ central charges of $B_3$ and $G_2$ theory are the same as the $\mathcal{N}=2$ $SU(2)$ SQCD with four flavors. The $F_4$ theory also has the same $(a,c)$ central charge as the $E_6$ Minahan-Nemeschansky theory \cite{Minahan:1996fg}. There are
interesting relations between corresponding VOAs of these 4d theories. In fact, following conformal embeddings were proven in \cite{adamovic2018classification},
\begin{equation}
V_{-2}(B_3) \subset V_{-2}(D_4),~~V_{-2}(G_2) \subset V_{-2}(B_3),~~V_{-3}(F_4)\subset V_{-3}(E_6).
\end{equation} 
It would be interesting to study further the relation between these VOAs and what they imply for 4d theories.

\subsection{Comments on collapsing levels}
We now make some remarks on collapsing levels of a $W$ algebra into its affine piece, namely finding the proper nilpotent orbit $f$ of the Lie algebra $\fg$ and level $k'$ such that the following equivalence between two VOAs holds,
\begin{equation}
W^{k'}(\fg,f)=V_k(\fg').
\end{equation}
We determine
$f$ and $k'$ by matching the Coulomb branch spectrum and other data of corresponding 4d theories. In practice, important insights can be gained if one first requires that flavor central charges should be equal for two descriptions. First consider some examples in detail. The $(2,0)$ configuration is the following,
\begin{equation}
g',~~\Phi={T\over z^{2+k/h^{\vee}}},~~f=f_{trivial},
\end{equation}
where $k$ is integer-valued for $g'=ADE$,  and half-integral valued for twisted theories. $k$ is restricted such that 
there is no flavor symmetry associated with the irregular singularity.
\begin{itemize}
\item Consider $\fg'=\fso_{2N}$, and the 4d flavor central charge $-k=(2N-2)-{2N-2\over k+2N-2}$.
 To find a $W$ algebra whose 4d partner has the same Coulomb branch spectrum, we consider $\fso_{qm+2N}$ theory with the regular singularity  $f=[q^m,1^{2N}]$. Naively, this puncture has the flavor symmetry $SO(2N)\times SO(m)$. The 4d flavor central charge $-k'$ of $SO(2N)$ flavor group in this description is 
\begin{equation}
-k'=m+2N-2-{qm+2N -2 \over 2N-2+k}.
\end{equation}
Matching $-k'$ with $-k=(2N-2)-{2N-2\over k+2N-2}$ leaves the requriement
\begin{equation}
q=2N-2+k,
\end{equation}
which is odd, so there might be the following equality of VOAs,
\begin{equation}
W^{-h^{\vee}+{h^{\vee}\over 2N-2+k}}(\fso_{(m(2N-2+k)+2N},[(2N-2+k)^m,1^{2N}])=V_{-(2N-2)+{2N-2\over 2N-2+k}},(\fso_{2N}).
\end{equation}
with $h^{\vee}=m(2N-2+k)+2N-2$. The flavor central charge for the $\SO(m)$ flavor symmetry is then,
\begin{equation}
m(2N-2+k)+2N-2-(2N-2+k){m(2N-2+k)+2N-2\over 2N-2+k}=0.
\end{equation}
which implies that the $\SO(m)$ flavor symmetry is decoupled in the IR. One can check that above two configurations give the same Coulomb branch spectrum.
\item Given $\fg'=\fsp_{2N-2}$, one has the following collapsing levels and nilpotent orbits,
\begin{equation}
W^{-h^{\vee}+{h^{\vee}\over 2N+2k+1}}(\fsp_{(m(2N+2k+1)+2N-2},[(2N+2k+1)^m,1^{2N-2}])=V_{-N+{N\over 2N+2k+1}},(\fsp_{2N-2})
\end{equation}
and $h^{\vee}={m(2N+2k+1)+2N\over2}$.
\item Taking $\fg'=\fso_{2N+1}$, one finds the following collapsing levels and nilpotent orbits,
\begin{equation}
W^{-h^{\vee}+{h^{\vee}\over 4N+2k-1}}(\fsp_{(m(4N+2k-1)+2N+1},[(4N+2k-1)^m,1^{2N+1}])=V_{-(2N-1)+{2N-1\over 4N+2k-1}}(\fso_{2N+1}),
\end{equation}
with $h^{\vee}=m(4N+2k-1)+2N-1$. 
\end{itemize}
In above cases, we can choose more general puncture on AKM side, and we then have collapsing of one $W$ algebra into another $W$ algebra.

We interpret our results as follows. Taking $\fg=\fsl_N$ (other cases are similar) and the irregular singularity,
\begin{equation}
\fg=\fsl_N,~~~~\Phi={T\over z^{2+{k\over N}}},~~f~\mathrm{generic},
\end{equation}
the Seiberg-Witten curve takes the following form
\begin{equation}
x^N+\sum_{i=2}\phi_i(z)x^{N-i}=0.
\end{equation}
If there is no Coulomb branch operators in $\phi_N$,  it is possible to 
find a $(2,0)$ realization with a lower rank Lie algebra. Assume we have a generic 
regular puncture, and denote $h_N(f)$ the height of the $N$th box in Young Tableaux $f$ with $h_N\leq N$. 
Denote $u$ as the Coulomb branch operator in $\phi_N$, then its scaling dimension is $[u]=N-{h_N N\over k+N}$.
To have a reduced theory (no Coulomb branch operators in $\phi_N(z)$), $[u]$ should have 
scaling dimension less or equal to one, so $\phi_N$ is zero and the above is factorized as,
\begin{equation}
x(x^{N-1}+\sum_{i=2}\phi_i(z)x^{N-1-i})=0.
\end{equation}
Then it is possible to find a description with lower rank $(2,0)$ theory. 
The constraint on $h_N$ is then,
\begin{equation}
[u]=N-{h_N N\over k+N}\leq 1 \rightarrow h_N\geq {(-1 + N) (k + N)\over N}.
\end{equation}
Since $h_N\leq N$, the above equation has solution if $k<0$. So if we have following situation 
\begin{equation}
W^{k'}(\fsl_N,f),~~k'=-N+{N\over k+N},~~k<0.
\end{equation}
Then if $h_N(f)\geq {(-1 + N) (k + N)\over N}$, there is a collapsing of $\fsl_{N}$ type $W$ algebra into a $\fsl_{N'}$ type $W$ algebra with $N'<N$.

\section{The Higgs branch}
\label{sec:Higgs}

The Coulomb branch spectrum of theories studied above can be found from Newton polygon, their dimensions are listed here for later uses ($ABCDE$ label theories studied in section \ref{sec:ADwithTwoFlavors}),
\begin{equation}
\begin{split}
&A:~n_C={(n+k-1)(2n_1+n-1)\over 2},~~\\
&B:~n_C={(n+k-1)(2n_1+2+n)\over4},\\
&C:~n_C={(n+k-1)(2n_1-2+n)\over4},\\
&D:~n_C={(n+k)(2n_1+n+1)\over2},\\
&E:~n_C={(n+k)(2n_1+n-1)\over2}.
\end{split}
\label{coulomb}
\end{equation}

We already know the flavor symmetry on the Higgs branch of 4d theories we studied in this paper, in this section we will use the associated variety of their corresponding
VOAs to learn the Higgs branch chiral ring of these theories.

\subsection{The Higgs branch as the associated variety of the VOA}
The $W$ algebras appear in section \ref{sec:ADwithTwoFlavors} take following form, $W^{k'}(\fg,f)$ with $f=[m^{q},1^s]$.
The associated variety of the above $W$ algebra is given by following formula \cite{MR3456698},
\begin{equation}
S_f\cap X_M.
\end{equation}
Here $S_f$ is the Slowdoy slice associated with the nilpotent orbit $f$, and 
$X_M$ is the associated variety of the affine vertex operator algebra $\fg$ with the level $k'$. 
If the level $k'$ is admissible, the associated variety of AKM is found in \cite{MR3456698}. We list the result below (we only show result for $k>2$ here, interested readers can work out the general case):
\begin{equation}
\begin{split}
& A:~~X_M=[(n+k)^{n_1},n], \\
& B:~~X_M=[(n+k)^{n_1},n+2], \\
& C:~~X_M=[(n+k)^{n_1},n-2], \\
& D:~~X_M=[(2n+2k+1)^{n_1},n+2], \\
& E:~~X_M=[(2n+2k+1)^{n_1+1},n-2].
\end{split}
\end{equation}
Here $X_M$ is the nilpotent orbit specified by the listed partition.
In the following, we will describe these Higgs branches in some detail. 

\subsection{The Higgs branch as a quiver variety}
Let us compactify our 4d theory on a circle and flow to  
deep IR to get a 3d $\mathcal{N}=4$ SCFT. The Higgs 
branch of the 3d theory is the same as the 4d theory, so it is 
described by the associated variety of the corresponding VOA. Meanwhile the 3d theory  has a Coulomb branch which is also 
a hyperkahler manifold.

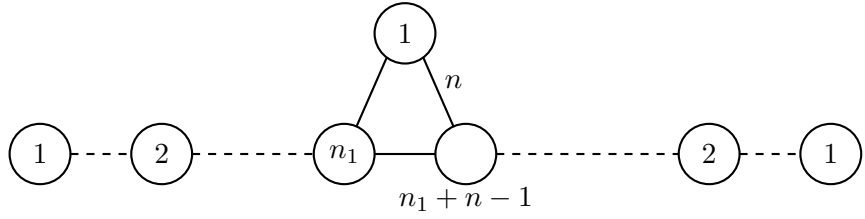
\begin{figure}
	\centering
	\begin{tikzpicture}[scale=0.8]
	
	\draw [thick] (-5,0) circle [radius=0.5];
	\node at (-5,0) {$1$};
	\draw [thick, dashed] (-4.5,0) -- (-3.5,0);
	\draw [thick] (-3,0) circle [radius=0.5];
	\node at (-3,0) {$2$};
	\draw [thick, dashed] (-2.5,0) -- (-0.5,0);
	\draw [thick] (0,0) circle [radius=0.5];
	\node at (0,0) {$n_1$};
	\draw [thick] (0.5,0) -- (1.5,0);
	\draw [thick] (2,0) circle [radius=0.5];
	\node at (2,-0.75) {$n_1+n-1$};
	\draw [thick, dashed] (2.5,0) -- (5.5,0);
	\draw [thick] (6,0) circle [radius=0.5];
	\node at (6,0) {$2$};
	\draw [thick, dashed] (6.5,0) -- (7.5,0);
	\draw [thick] (8,0) circle [radius=0.5];
	\node at (8,0) {$1$};
	
	\draw [thick] (1,2) circle [radius=0.5];
	\node at (1,2) {$1$};
	\draw [thick] (0.2,0.45) -- (0.7, 1.6);
	\draw [thick] (1.8,0.45) -- (1.3, 1.6);
	\node at (1.8,1.2) {$n$};	
	\end{tikzpicture}
	\caption{3d mirror for 4d theory whose associated VOA is \ref{VOAA} with $k=1$. From this quiver, one can actually read another Hitchin system description which uses type III irregular singularity as discussed in \cite{Xie:2012hs}, and we check the Coulomb branch spectrum which is the same as the one found using the construction in section 3.}
	\label{3dmirror}
\end{figure}

The 3d $\mathcal{N}=4$ theory has an interesting mirror symmetry: there is a mirror theory $B$ whose Higgs branch is the Coulomb branch of theory A, and vice versa. If the mirror theory has a Lagrangian description, its Higgs branch is described by the classical hyperkahler quotient. Let's look at our class $A$ theory whose VOA is:
\begin{equation}
W^{-n_1(n+k)-n+{n_1(n+k)+n\over n+k}}(\fsl_{n_1(n+k)+n},[(n+k-1)^{n_1},1^{n+n_1}])
\end{equation}
For the special case $k=1$, we conjecture that the mirror theory is given by the quiver in figure \ref{3dmirror}. The simple counting of dimensions of Coulomb and Higgs branch of mirror quiver is:
\begin{equation}
n_H=nn_1+{1\over2}n(n-1),~~n_C={{1\over2}n(n-1)}+n_1(n+n_1 ).
\end{equation}
The dimension $n_H$ of 3d mirror is equal to the Coulomb branch dimension of original 4d theory, see \ref{coulomb}.

\begin{figure}
	
	\tikzset{every picture/.style={line width=0.75pt}} 
	
	\begin{tikzpicture}[x=0.75pt,y=0.75pt,yscale=-1,xscale=1]
	
	\draw   (41,90) -- (91,90) -- (91,140) -- (41,140) -- cycle ;
	\draw   (111,116) .. controls (111,102.19) and (122.19,91) .. (136,91) .. controls (149.81,91) and (161,102.19) .. (161,116) .. controls (161,129.81) and (149.81,141) .. (136,141) .. controls (122.19,141) and (111,129.81) .. (111,116) -- cycle ;
	\draw   (181,115) .. controls (181,101.19) and (192.19,90) .. (206,90) .. controls (219.81,90) and (231,101.19) .. (231,115) .. controls (231,128.81) and (219.81,140) .. (206,140) .. controls (192.19,140) and (181,128.81) .. (181,115) -- cycle ;
	\draw    (92.5,115.92) -- (111,116) ;

	\draw    (161,116) -- (179.5,116.08) ;

	\draw    (231,115) -- (249.5,115.08) ;

	\draw  [dash pattern={on 4.5pt off 4.5pt}]  (249.5,115.08) -- (290.5,114.92) ;

	\draw   (310,114) .. controls (310,100.19) and (321.19,89) .. (335,89) .. controls (348.81,89) and (360,100.19) .. (360,114) .. controls (360,127.81) and (348.81,139) .. (335,139) .. controls (321.19,139) and (310,127.81) .. (310,114) -- cycle ;
	\draw    (290,115) -- (308.5,115.08) ;

	\draw   (380,114) .. controls (380,100.19) and (391.19,89) .. (405,89) .. controls (418.81,89) and (430,100.19) .. (430,114) .. controls (430,127.81) and (418.81,139) .. (405,139) .. controls (391.19,139) and (380,127.81) .. (380,114) -- cycle ;
	\draw    (360,115) -- (378.5,115.08) ;

	\draw    (429,115) -- (447.5,115.08) ;

	\draw  [dash pattern={on 4.5pt off 4.5pt}]  (447.5,115.08) -- (488.5,114.92) ;

	\draw    (488,115) -- (506.5,115.08) ;

	\draw   (507,115) .. controls (507,101.19) and (518.19,90) .. (532,90) .. controls (545.81,90) and (557,101.19) .. (557,115) .. controls (557,128.81) and (545.81,140) .. (532,140) .. controls (518.19,140) and (507,128.81) .. (507,115) -- cycle ;
	\draw   (576,90) -- (626,90) -- (626,140) -- (576,140) -- cycle ;
	\draw    (557,115) -- (577.5,114.92) ;

	\draw (66,115) node [scale=0.7]  {$n+n_{1}$};
	\draw (138,116) node [scale=0.7]  {$n+n_{1} -1$};
	\draw (208,115) node [scale=0.7]  {$n+n_{1} -2$};
	\draw (337,114) node [scale=0.7]  {$n_{1} +1$};
	\draw (407,114) node [scale=0.7]  {$n_{1}$};
	\draw (534,115) node [scale=0.7]  {$n_{1}$};
	\draw (601,115) node [scale=0.7]  {$n_{1}$};
	\draw (470,160) node   {$\underbrace{\ \ \ \ \ \ \ \ \ \ \ \ \ \ \ \ \ \\ \ \ \ \ \ \ \ \ \ \ \ \ \ \ \ \ \ }_{k}$};
	\end{tikzpicture}
	
	\caption{\label{fig:quiverS_fX_M}The quiver of $S_f\cap X_M$ with $f=[(n+k-1)^{n_1},1^{n+n_1}]$ and $M=[(n+k)^{n_1},n]$.}
\end{figure}
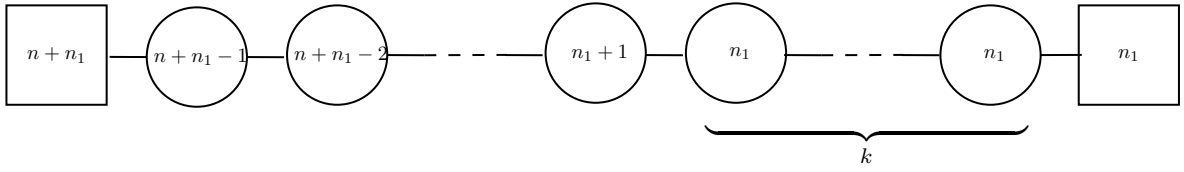

In the following, we use the methods proposed in \cite{Gaiotto:2008ak} to describe the Higgs branch as 
a quiver variety. 
Considering a nilpotent orbit $f=[(n+k-1)^{n_1},1^{n+n_1}]$ and $X_M$ with $M=[(n+k)^{n_1},n]$, we would like to find the  variety of $S_f\cap X_M$. Firstly we need the transpose of $M$, $M^t=[(n_1+1)^{n},(n_1)^{k}]$. Using  $f$ and $M^t$, one can form a $D5-NS5-D3$ systems \cite{Gaiotto:2008ak} as shown in figure \ref{fig:BraneConstruction}. After moving $D5$ branes according to rules in \cite{Gaiotto:2008ak},
one obtain an equivalent brane configuration shown in figure \ref{fig:BraneConstructionHWMove} which leads to the quiver gauge theory of $S_f\cap X_M$ as shown in figure \ref{fig:quiverS_fX_M}. The dimension of $S_f\cap X_M$ can be easily read from the Higgs branch of the quiver \ref{fig:quiverS_fX_M}, 
\begin{equation}
\dim(S_f\cap X_M)=n_H={{1\over2}n(n-1)}+n_1(n+n_1),
\end{equation}
which is actually independent of $k$.

\begin{figure}
	\tikzset{every picture/.style={line width=0.75pt}} 
	\begin{subfigure}[b]{1\linewidth}
		\centering
		\begin{tikzpicture}[x=0.75pt,y=0.75pt,yscale=-1,xscale=1]
		
		\draw  [dash pattern={on 4.5pt off 4.5pt}]  (241.5,98.92) -- (241.5,129.92) ;

		\draw    (51,59.92) -- (51,170.92) ;

		\draw    (90,59.92) -- (90,170.92) ;

		\draw    (130,59.92) -- (130,170.92) ;

		\draw    (170,59.92) -- (170,170.92) ;

		\draw    (51,115.42) -- (90,115.42) ;

		\draw    (91,120.42) -- (130,120.42) ;

		\draw    (91,111.42) -- (130,111.42) ;

		\draw    (131,125.42) -- (170,125.42) ;

		\draw    (131,116.42) -- (170,116.42) ;

		\draw    (131,106.42) -- (170,106.42) ;

		\draw  [dash pattern={on 4.5pt off 4.5pt}]  (175.5,115.92) -- (215.5,115.92) ;

		\draw    (221,59.92) -- (221,170.92) ;

		\draw    (261,59.92) -- (261,170.92) ;

		\draw    (222,90.42) -- (261,90.42) ;

		\draw    (222,137.42) -- (261,137.42) ;

		\draw  [dash pattern={on 4.5pt off 4.5pt}]  (281.5,82.92) -- (281.5,148.92) ;

		\draw    (301,59.92) -- (301,170.92) ;

		\draw    (262,75.42) -- (301,75.42) ;

		\draw    (262,155.42) -- (301,155.42) ;

		\draw  [dash pattern={on 4.5pt off 4.5pt}]  (369.5,82.92) -- (369.5,148.92) ;

		\draw    (350,75.42) -- (389,75.42) ;

		\draw    (350,155.42) -- (389,155.42) ;

		\draw    (350,59.92) -- (350,170.92) ;

		\draw  [dash pattern={on 4.5pt off 4.5pt}]  (305,115.42) -- (345,115.42) ;

		\draw (231,116) node [scale=0.7,rotate=-90]  {$n+n_{1} \ $};
		\draw (272.5,118.92) node [scale=0.7,rotate=-90]  {$2n+n_{1} +k-1\ $};
		\draw (360.5,118.92) node [scale=0.7,rotate=-90]  {$n_{1} \ ( n+k) +n$};
		\draw (53,41) node [scale=0.7]  {$D5$};
		\draw (90,41) node [scale=0.7]  {$D5$};
		\draw (130,41) node [scale=0.7]  {$D5$};
		\draw (170,41) node [scale=0.7]  {$D5$};
		\draw (222,41) node [scale=0.7]  {$D5$};
		\draw (262,41) node [scale=0.7]  {$D5$};
		\draw (302,41) node [scale=0.7]  {$D5$};
		\draw (351,41) node [scale=0.7]  {$D5$};
		\draw (70,123) node [scale=0.7]  {$D3$};
		\draw (110,129) node [scale=0.7]  {$D3$};
		\draw (150,134) node [scale=0.7]  {$D3$};
		\draw (240,145) node [scale=0.7]  {$D3$};
		\draw (281,164) node [scale=0.7]  {$D3$};
		\draw (371,164) node [scale=0.7]  {$D3$};
		\end{tikzpicture}
		\caption{}
		\label{subfig:S_f}
	\end{subfigure}\\%
	\begin{subfigure}[b]{1\linewidth}
		\centering
		\begin{tikzpicture}[x=0.75pt,y=0.75pt,yscale=-1,xscale=1]
		
		\draw  [dash pattern={on 0.84pt off 2.51pt}]  (143.59,103.17) -- (143.59,118.17) ;

		\draw    (110.26,102.42) -- (176.92,101.92) ;

		\draw    (110.26,119.42) -- (176.92,118.92) ;

		\draw   (175,110.25) .. controls (175,102.93) and (180.93,97) .. (188.25,97) .. controls (195.57,97) and (201.5,102.93) .. (201.5,110.25) .. controls (201.5,117.57) and (195.57,123.5) .. (188.25,123.5) .. controls (180.93,123.5) and (175,117.57) .. (175,110.25) -- cycle ; \draw   (178.88,100.88) -- (197.62,119.62) ; \draw   (197.62,100.88) -- (178.88,119.62) ;
		\draw  [dash pattern={on 4.5pt off 4.5pt}]  (201.5,110.25) -- (264.5,110.92) ;

		\draw   (86,111.25) .. controls (86,103.93) and (91.93,98) .. (99.25,98) .. controls (106.57,98) and (112.5,103.93) .. (112.5,111.25) .. controls (112.5,118.57) and (106.57,124.5) .. (99.25,124.5) .. controls (91.93,124.5) and (86,118.57) .. (86,111.25) -- cycle ; \draw   (89.88,101.88) -- (108.62,120.62) ; \draw   (108.62,101.88) -- (89.88,120.62) ;
		\draw  [dash pattern={on 0.84pt off 2.51pt}]  (54.59,104.17) -- (54.59,119.17) ;

		\draw    (21.26,103.42) -- (87.92,102.92) ;

		\draw    (21.26,120.42) -- (87.92,119.92) ;

		\draw   (264.5,110.92) .. controls (264.5,103.6) and (270.43,97.67) .. (277.75,97.67) .. controls (285.07,97.67) and (291,103.6) .. (291,110.92) .. controls (291,118.24) and (285.07,124.17) .. (277.75,124.17) .. controls (270.43,124.17) and (264.5,118.24) .. (264.5,110.92) -- cycle ; \draw   (268.38,101.55) -- (287.12,120.29) ; \draw   (287.12,101.55) -- (268.38,120.29) ;
		\draw  [dash pattern={on 0.84pt off 2.51pt}]  (410.59,102.17) -- (410.59,117.17) ;

		\draw    (377.26,101.42) -- (443.92,100.92) ;

		\draw    (377.26,118.42) -- (443.92,117.92) ;

		\draw   (442,109.25) .. controls (442,101.93) and (447.93,96) .. (455.25,96) .. controls (462.57,96) and (468.5,101.93) .. (468.5,109.25) .. controls (468.5,116.57) and (462.57,122.5) .. (455.25,122.5) .. controls (447.93,122.5) and (442,116.57) .. (442,109.25) -- cycle ; \draw   (445.88,99.88) -- (464.62,118.62) ; \draw   (464.62,99.88) -- (445.88,118.62) ;
		\draw  [dash pattern={on 4.5pt off 4.5pt}]  (468.5,109.25) -- (531.5,109.92) ;

		\draw   (353,110.25) .. controls (353,102.93) and (358.93,97) .. (366.25,97) .. controls (373.57,97) and (379.5,102.93) .. (379.5,110.25) .. controls (379.5,117.57) and (373.57,123.5) .. (366.25,123.5) .. controls (358.93,123.5) and (353,117.57) .. (353,110.25) -- cycle ; \draw   (356.88,100.88) -- (375.62,119.62) ; \draw   (375.62,100.88) -- (356.88,119.62) ;
		\draw  [dash pattern={on 0.84pt off 2.51pt}]  (321.59,103.17) -- (321.59,118.17) ;

		\draw    (288.26,102.42) -- (354.92,101.92) ;

		\draw    (288.26,119.42) -- (354.92,118.92) ;

		\draw   (531.5,109.92) .. controls (531.5,102.6) and (537.43,96.67) .. (544.75,96.67) .. controls (552.07,96.67) and (558,102.6) .. (558,109.92) .. controls (558,117.24) and (552.07,123.17) .. (544.75,123.17) .. controls (537.43,123.17) and (531.5,117.24) .. (531.5,109.92) -- cycle ; \draw   (535.38,100.55) -- (554.12,119.29) ; \draw   (554.12,100.55) -- (535.38,119.29) ;
		\draw  [dash pattern={on 0.84pt off 2.51pt}]  (587.59,101.17) -- (587.59,116.17) ;

		\draw    (554.26,100.42) -- (620.92,99.92) ;

		\draw    (554.26,117.42) -- (620.92,116.92) ;

		\draw   (619,108.25) .. controls (619,100.93) and (624.93,95) .. (632.25,95) .. controls (639.57,95) and (645.5,100.93) .. (645.5,108.25) .. controls (645.5,115.57) and (639.57,121.5) .. (632.25,121.5) .. controls (624.93,121.5) and (619,115.57) .. (619,108.25) -- cycle ; \draw   (622.88,98.88) -- (641.62,117.62) ; \draw   (641.62,98.88) -- (622.88,117.62) ;
		
		\draw (147.5,131.92) node [scale=0.5]  {$n_{1} \ ( n+k-1) +n-1$};
		\draw (144.23,85) node [scale=0.7]  {$D3$};
		\draw (56.5,132.92) node [scale=0.7]  {$n_{1} \ ( n+k) +n$};
		\draw (55.23,85) node [scale=0.7]  {$D3$};
		\draw (412.5,131.92) node [scale=0.7]  {$( k-1) n_{1}$};
		\draw (411.23,85) node [scale=0.7]  {$D3$};
		\draw (323.5,131.92) node [scale=0.7]  {$kn_{1}$};
		\draw (322.23,85) node [scale=0.7]  {$D3$};
		\draw (589.5,130.92) node [scale=0.7]  {$n_{1}$};
		\draw (588.23,85) node [scale=0.7]  {$D3$};
		\draw (100,81) node [scale=0.7]  {$NS5$};
		\draw (279,81) node [scale=0.7]  {$NS5$};
		\draw (190,81) node [scale=0.7]  {$NS5$};
		\draw (367,81) node [scale=0.7]  {$NS5$};
		\draw (456,81) node [scale=0.7]  {$NS5$};
		\draw (545,81) node [scale=0.7]  {$NS5$};
		\draw (633,81) node [scale=0.7]  {$NS5$};

		\end{tikzpicture}        %
		\caption{}
		\label{subfig:X_M}
	\end{subfigure}\\
	\begin{subfigure}[b]{1\linewidth}
		\centering
		\begin{tikzpicture}[x=0.75pt,y=0.75pt,yscale=-1,xscale=1]
		
		\draw  [dash pattern={on 0.84pt off 2.51pt}]  (410.59,105.17) -- (410.59,115.17) ;

		\draw    (377.26,103.42) -- (443.92,102.92) ;

		\draw    (377.26,115.42) -- (443.92,114.92) ;

		\draw   (442,109.25) .. controls (442,101.93) and (447.93,96) .. (455.25,96) .. controls (462.57,96) and (468.5,101.93) .. (468.5,109.25) .. controls (468.5,116.57) and (462.57,122.5) .. (455.25,122.5) .. controls (447.93,122.5) and (442,116.57) .. (442,109.25) -- cycle ; \draw   (445.88,99.88) -- (464.62,118.62) ; \draw   (464.62,99.88) -- (445.88,118.62) ;
		\draw  [dash pattern={on 4.5pt off 4.5pt}]  (468.5,109.25) -- (531.5,109.92) ;

		\draw   (353,110.25) .. controls (353,102.93) and (358.93,97) .. (366.25,97) .. controls (373.57,97) and (379.5,102.93) .. (379.5,110.25) .. controls (379.5,117.57) and (373.57,123.5) .. (366.25,123.5) .. controls (358.93,123.5) and (353,117.57) .. (353,110.25) -- cycle ; \draw   (356.88,100.88) -- (375.62,119.62) ; \draw   (375.62,100.88) -- (356.88,119.62) ;
		\draw  [dash pattern={on 0.84pt off 2.51pt}]  (321.59,103.17) -- (321.59,118.17) ;

		\draw    (288.26,102.42) -- (354.92,101.92) ;

		\draw    (288.26,119.42) -- (354.92,118.92) ;

		\draw   (531.5,109.92) .. controls (531.5,102.6) and (537.43,96.67) .. (544.75,96.67) .. controls (552.07,96.67) and (558,102.6) .. (558,109.92) .. controls (558,117.24) and (552.07,123.17) .. (544.75,123.17) .. controls (537.43,123.17) and (531.5,117.24) .. (531.5,109.92) -- cycle ; \draw   (535.38,100.55) -- (554.12,119.29) ; \draw   (554.12,100.55) -- (535.38,119.29) ;
		\draw  [dash pattern={on 0.84pt off 2.51pt}]  (587.59,106.17) -- (587.59,112.17) ;

		\draw    (556.26,105.42) -- (620.92,104.92) ;

		\draw    (557.5,112.92) -- (620.92,111.92) ;

		\draw   (619,108.25) .. controls (619,100.93) and (624.93,95) .. (632.25,95) .. controls (639.57,95) and (645.5,100.93) .. (645.5,108.25) .. controls (645.5,115.57) and (639.57,121.5) .. (632.25,121.5) .. controls (624.93,121.5) and (619,115.57) .. (619,108.25) -- cycle ; \draw   (622.88,98.88) -- (641.62,117.62) ; \draw   (641.62,98.88) -- (622.88,117.62) ;
		\draw    (289.5,50.92) -- (289.5,168.92) ;

		\draw  [dash pattern={on 0.84pt off 2.51pt}]  (255.59,106.17) -- (255.59,117.17) ;

		\draw    (222.26,105.42) -- (288.92,104.92) ;

		\draw    (222.26,118.42) -- (288.92,117.92) ;

		\draw    (222.5,51.92) -- (222.5,168.92) ;

		\draw  [dash pattern={on 4.5pt off 4.5pt}]  (159,109.75) -- (222,110.42) ;

		\draw    (159.5,51.25) -- (159.5,168.25) ;

		\draw    (92.35,110.25) -- (159,109.75) ;

		\draw    (92.85,51.75) -- (92.85,168.75) ;

		\draw (411.23,85) node [scale=0.7]  {$D3$};
		\draw (322.23,85) node [scale=0.7]  {$D3$};
		\draw (588.23,85) node [scale=0.7]  {$D3$};
		\draw (367,79) node [scale=0.7]  {$NS5$};
		\draw (456,78) node [scale=0.7]  {$NS5$};
		\draw (545,79) node [scale=0.7]  {$NS5$};
		\draw (633,77) node [scale=0.7]  {$NS5$};
		\draw (256.23,85) node [scale=0.7]  {$D3$};
		\draw (291.23,35) node [scale=0.7]  {$D5$};
		\draw (224.23,35) node [scale=0.7]  {$D5$};
		\draw (161.23,35) node [scale=0.7]  {$D5$};
		\draw (95.23,35) node [scale=0.7]  {$D5$};
		\draw (128.23,85) node [scale=0.7]  {$D3$};
		\draw (212,191) node   {$\underbrace{\ \ \ \ \ \ \ \ \ \ \ \ \ \ \ \  \ \ \ \ \  \ \ \ \ \ \ \ \ \ \ \ \ \ \ \ \ \ \ \ \ \ \ \ \ \ \ \ \ \ \ \ }_{S_{f}}$};
		\draw (488,43) node   {$\overbrace{\ \ \ \ \ \ \ \ \ \ \ \ \ \ \ \ \ \ \ \ \ \ \ \ \ \ \ \ \ \ \ \ \ \ \ \ \ \ \ \ \ \ \ \ \ \ \ \ \ \ \ \ \ \ \ \ \ \ \ \ \ \ \ \ \ \ \ \ }^{X_{M}}$};

		\end{tikzpicture}
		
		\caption{}
		\label{subfig:S_fX_M}
	\end{subfigure}
	
	\caption{(\subref{subfig:S_f}): The brane construction of $S_f$ with $f=[(n+k-1)^{n_1},1^{n+n_1}]$. The number of $D3$ branes between $i$-th and $i+1$-th $D5$ branes are $\sum_{j=n+2n_1-i+1}^{n+2n_1}f_j$. (\subref{subfig:X_M}): The brane construction of $X_M$ with $M^t=[(n_1+1)^{n},(n_1)^{k}]$. The number of $D3$ branes between $i$-th and $i+1$-th $NS5$ branes is $n_1(n+k)+n-\sum_{j=1}^{i}M^t_j$. (\subref{subfig:S_fX_M}): Schematics of brane construction of $S_f\cap X_M$ which just connects (\subref{subfig:S_f}) and (\subref{subfig:X_M}). One can connect these two brane configurations because the total number of boxes of $f$ and $M$ are the same, therefore the same amount of $D3$ branes.}
	\label{fig:BraneConstruction}
\end{figure}
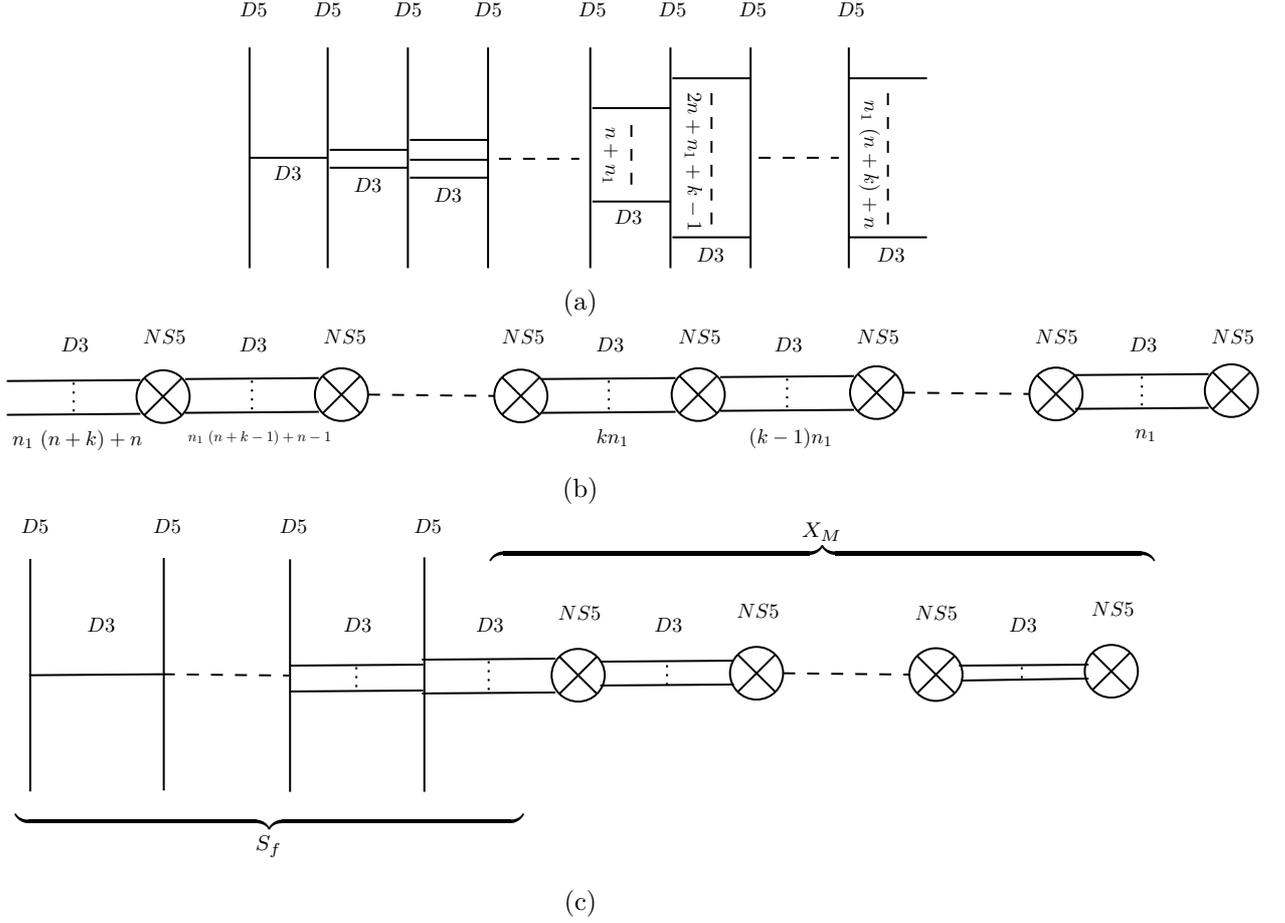

\begin{figure}
	\centering

	\tikzset{every picture/.style={line width=0.75pt}} 
	
	\begin{tikzpicture}[x=0.75pt,y=0.75pt,yscale=-1,xscale=1]
	
	\draw   (10,120.46) .. controls (10,109.71) and (18.71,101) .. (29.46,101) .. controls (40.21,101) and (48.92,109.71) .. (48.92,120.46) .. controls (48.92,131.21) and (40.21,139.92) .. (29.46,139.92) .. controls (18.71,139.92) and (10,131.21) .. (10,120.46) -- cycle ; \draw   (15.7,106.7) -- (43.22,134.22) ; \draw   (43.22,106.7) -- (15.7,134.22) ;
	\draw    (43.5,105.92) -- (174.66,105.92) ;

	\draw   (170,119.46) .. controls (170,108.71) and (178.71,100) .. (189.46,100) .. controls (200.21,100) and (208.92,108.71) .. (208.92,119.46) .. controls (208.92,130.21) and (200.21,138.92) .. (189.46,138.92) .. controls (178.71,138.92) and (170,130.21) .. (170,119.46) -- cycle ; \draw   (175.7,105.7) -- (203.22,133.22) ; \draw   (203.22,105.7) -- (175.7,133.22) ;
	\draw    (44.34,133.92) -- (175.5,133.92) ;

	\draw  [dash pattern={on 0.84pt off 2.51pt}]  (141.03+5,107.92) -- (141.03+5,131.92) ;

	\draw    (58.5,77.92) -- (58.5,161.92) ;

	\draw    (118.5,78.92) -- (118.5,160.92) ;

	\draw  [dash pattern={on 0.84pt off 2.51pt}]  (63.5,90.92) -- (114.5,90.92) ;

	\draw    (206.5,109.92) -- (254.5,109.92) ;

	\draw    (206.5,130.92) -- (254.5,130.92) ;

	\draw   (252,120.46) .. controls (252,109.71) and (260.71,101) .. (271.46,101) .. controls (282.21,101) and (290.92,109.71) .. (290.92,120.46) .. controls (290.92,131.21) and (282.21,139.92) .. (271.46,139.92) .. controls (260.71,139.92) and (252,131.21) .. (252,120.46) -- cycle ; \draw   (257.7,106.7) -- (285.22,134.22) ; \draw   (285.22,106.7) -- (257.7,134.22) ;
	\draw  [dash pattern={on 0.84pt off 2.51pt}]  (230.5,109.92) -- (230.5,130.92) ;

	\draw  [dash pattern={on 0.84pt off 2.51pt}]  (291.92,120.46) -- (327.5,120.92) ;

	\draw   (326.5,119.92) .. controls (326.5,109.17) and (335.21,100.46) .. (345.96,100.46) .. controls (356.71,100.46) and (365.42,109.17) .. (365.42,119.92) .. controls (365.42,130.67) and (356.71,139.38) .. (345.96,139.38) .. controls (335.21,139.38) and (326.5,130.67) .. (326.5,119.92) -- cycle ; \draw   (332.2,106.16) -- (359.72,133.68) ; \draw   (359.72,106.16) -- (332.2,133.68) ;
	\draw    (504.5,109.92) -- (609.5,109.92) ;

	\draw    (504.5,126.92) -- (609.5,126.92) ;

	\draw  [dash pattern={on 0.84pt off 2.51pt}]  (534,109.92) -- (534,126.92) ;

	\draw   (608.5,118.92) .. controls (608.5,108.17) and (617.21,99.46) .. (627.96,99.46) .. controls (638.71,99.46) and (647.42,108.17) .. (647.42,118.92) .. controls (647.42,129.67) and (638.71,138.38) .. (627.96,138.38) .. controls (617.21,138.38) and (608.5,129.67) .. (608.5,118.92) -- cycle ; \draw   (614.2,105.16) -- (641.72,132.68) ; \draw   (641.72,105.16) -- (614.2,132.68) ;
	\draw  [dash pattern={on 0.84pt off 2.51pt}]  (62.5,145.92) -- (113.5,145.92) ;

	\draw    (564.5,73.92) -- (564.5,155.92) ;

	\draw    (600.5,74.92) -- (600.5,155.92) ;

	\draw  [dash pattern={on 0.84pt off 2.51pt}]  (567.5,86.92) -- (598.1,86.92) ;

	\draw  [dash pattern={on 0.84pt off 2.51pt}]  (567.9,140.92) -- (598.5,140.92) ;

	\draw   (468.5,119.92) .. controls (468.5,109.17) and (477.21,100.46) .. (487.96,100.46) .. controls (498.71,100.46) and (507.42,109.17) .. (507.42,119.92) .. controls (507.42,130.67) and (498.71,139.38) .. (487.96,139.38) .. controls (477.21,139.38) and (468.5,130.67) .. (468.5,119.92) -- cycle ; \draw   (474.2,106.16) -- (501.72,133.68) ; \draw   (501.72,106.16) -- (474.2,133.68) ;
	\draw   (407.58-10,119.92) .. controls (407.58-10,109.17) and (416.29-10,100.46) .. (427.04-10,100.46) .. controls (437.79-10,100.46) and (446.5-10,109.17) .. (446.5-10,119.92) .. controls (446.5-10,130.67) and (437.79-10,139.38) .. (427.04-10,139.38) .. controls (416.29-10,139.38) and (407.58-10,130.67) .. (407.58-10,119.92) -- cycle ; \draw   (413.28-10,106.16) -- (440.8-10,133.68) ; \draw   (440.8-10,106.16) -- (413.28-10,133.68) ;
	\draw  [dash pattern={on 0.84pt off 2.51pt}]  (446.5-10,119.92) -- (469.5,119.92) ;

	\draw    (363.5,109.92) -- (411.5-10,109.92) ;

	\draw    (363.5,130.92) -- (411.5-10,130.92) ;

	\draw  [dash pattern={on 0.84pt off 2.51pt}]  (387.5-5,109.92) -- (387.5-5,130.92) ;

	\draw (148,150) node [scale=0.7]  {$ \begin{array}{l}
		n+n_{1} -1\\
		\ \ \ \ D3s
		\end{array}$};
	\draw (229,150) node [scale=0.7]  {$ \begin{array}{l}
		n+n_{1} -2\\
		\ \ \ \ D3s
		\end{array}$};
	\draw (533,145) node [scale=0.7]  {$n_{1} \ D3s$};
	\draw (89,62) node   {$\overbrace{\ \ \ \ \ \ \ \ \ \ \ \ \ }^{n+n_{1} \ D5s}$};
	\draw (582,60) node   {$\overbrace{\ \ \ \ \ \ \ \ }^{n_{1} \ D5s}$};
	\draw (32,85) node [scale=0.7]  {$NS5$};
	\draw (190,85) node [scale=0.7]  {$NS5$};
	\draw (274,85) node [scale=0.7]  {$NS5$};
	\draw (630,85) node [scale=0.7]  {$NS5$};
	\draw (381,145) node [scale=0.7]  {$n_{1} \ D3s$};
	\draw (416,88) node [scale=1]  {$\overbrace{\ \ \ \ \ \ \ \ \ \ \ \ \ \ \ \ \ \ \ \ \ \ \ \ \ \ \ \ \ \ \ \ \ \ \ \ }^{k\ NS5s}$};

	\end{tikzpicture}
	
	\caption{The brane construction after brane moves of $S_f\cap X_M$ with $f=[(n+k-1)^{n_1},1^{n+n_1}]$ and $M=[(n+k)^{n_1},n]$.}
	\label{fig:BraneConstructionHWMove}
\end{figure}
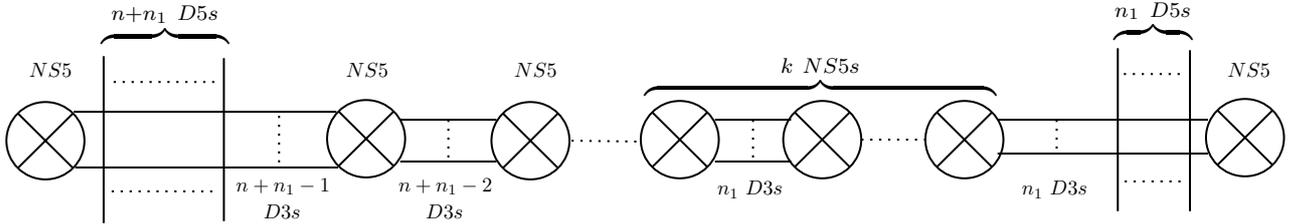

\newpage
\section{VOA for theories with exact marginal deformations}
\label{sec:VOAwithDef} 

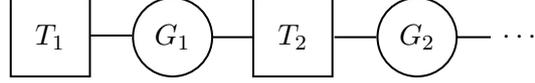
\begin{figure}
\centering

\tikzset{every picture/.style={line width=0.75pt}} 

\begin{tikzpicture}[x=0.75pt,y=0.75pt,yscale=-1,xscale=1]

\draw   (100,121) -- (139.5,121) -- (139.5,160.5) -- (100,160.5) -- cycle ;
\draw   (161,140.75) .. controls (161,129.84) and (169.84,121) .. (180.75,121) .. controls (191.66,121) and (200.5,129.84) .. (200.5,140.75) .. controls (200.5,151.66) and (191.66,160.5) .. (180.75,160.5) .. controls (169.84,160.5) and (161,151.66) .. (161,140.75) -- cycle ;
\draw    (139.5,139.92) -- (160.5,139.92) ;

\draw   (221,121) -- (260.5,121) -- (260.5,160.5) -- (221,160.5) -- cycle ;
\draw   (283,140.75) .. controls (283,129.84) and (291.84,121) .. (302.75,121) .. controls (313.66,121) and (322.5,129.84) .. (322.5,140.75) .. controls (322.5,151.66) and (313.66,160.5) .. (302.75,160.5) .. controls (291.84,160.5) and (283,151.66) .. (283,140.75) -- cycle ;
\draw    (200.5,140.75) -- (221.5,140.75) ;

\draw    (261.5,140.75) -- (282.5,140.75) ;

\draw    (322.5,140.75) -- (339.5,140.75) ;

\draw (119.75,140.75) node   {$T_{1}$};
\draw (180.75,140.75) node   {$G_{1}$};
\draw (240.75,140.75) node   {$T_{2}$};
\draw (302.75,140.75) node   {$G_{2}$};
\draw (355,140.75) node   {$\cdots $};

\end{tikzpicture}
\caption{A typical quiver for the weakly coupled gauge theory description of a 4d N=2 SCFT. $T_i$'s are matter systems with non-abelian flavor symmetries, and $G_i$'s are gauge groups.}
\label{fig:4dweakly}
\end{figure}

If a four dimensional $\mathcal{N}=2$ SCFT has exact marginal deformations, then it is possible to write down a weakly coupled gauge theory description which typically looks like figure \ref{fig:4dweakly}.
If we know the VOA $V_i$ for each matter (which should have an affine vertex subalgebra $ V_{k_i}(G_i)$),  
the VOA for the parent theory is given by the following coset,
\begin{equation}
V_0={V_1\oplus V_2 \oplus \ldots \over (G_1)_{-2h^{\vee }_1} \oplus (G_2)_{-2h^{\vee}_2} \oplus \ldots}.
\end{equation}
We use the fact that the conformal gauging condition implies that the sum of levels from 
the matter gauged by gauge group $G_i$ is $-2h_i^{\vee}$.
If there are more than one weakly coupled gauge theory descriptions, then we have found equivalence between non-trivial coset constructions.

\begin{figure}

	\tikzset{every picture/.style={line width=0.75pt}} 
	\centering
	\begin{tikzpicture}[x=0.75pt,y=0.75pt,xscale=0.7,yscale=0.7]
	
	\draw   (33,131.74) .. controls (33,95.99) and (88.18,67) .. (156.25,67) .. controls (224.32,67) and (279.5,95.99) .. (279.5,131.74) .. controls (279.5,167.5) and (224.32,196.48) .. (156.25,196.48) .. controls (88.18,196.48) and (33,167.5) .. (33,131.74) -- cycle ;
	\draw  [line width=0.75]  (93.41,101.97) -- (102.27,111.44)(102.17,102.09) -- (93.51,111.33) ;
	\draw   (135.41,88.97) -- (144.27,98.44)(144.17,89.09) -- (135.51,98.33) ;
	\draw   (175.41,91.97) -- (184.27,101.44)(184.17,92.09) -- (175.51,101.33) ;
	\draw   (214.41,100.97) -- (223.27,110.44)(223.17,101.09) -- (214.51,110.33) ;
	\draw  [color={rgb, 255:red, 1; green, 10; blue, 250 }  ,draw opacity=1 ][line width=1.5]  (59,137) -- (67.5,137) -- (67.5,146.48) -- (59,146.48) -- cycle ;
	\draw  [color={rgb, 255:red, 252; green, 2; blue, 32 }  ,draw opacity=1 ][line width=1.5]  (237,141.24) .. controls (237,138.06) and (239.58,135.48) .. (242.76,135.48) .. controls (245.94,135.48) and (248.52,138.06) .. (248.52,141.24) .. controls (248.52,144.42) and (245.94,147) .. (242.76,147) .. controls (239.58,147) and (237,144.42) .. (237,141.24) -- cycle ;
	\draw    (489,67.54) .. controls (540.5,67.54) and (511.5,119.35) .. (555.5,121.47) ;

	\draw    (440.5,122.53) .. controls (468.5,115.12) and (452.5,73.89) .. (489,67.54) ;

	\draw    (613.5,66.48) .. controls (656.5,66.48) and (638.5,124.64) .. (673.5,122.53) ;

	\draw    (555.5,121.47) .. controls (586.5,117.24) and (578.5,67.54) .. (613.5,66.48) ;

	\draw    (367.5,78.12) .. controls (411.5,52.74) and (419.5,124.64) .. (440.5,122.53) ;

	\draw    (673.5,122.53) .. controls (699.5,119.35) and (676.5,69.66) .. (728.5,68.6) ;

	\draw    (728.5,68.6) .. controls (790.5,63.31) and (802.5,193.37) .. (727.5,191.25) ;

	\draw    (373.5,189.14) .. controls (310.5,178.57) and (321.5,103.49) .. (367.5,78.12) ;

	\draw    (373.5,189.14) .. controls (424.5,191.25) and (431.5,127.81) .. (460.5,175.39) ;

	\draw    (460.5,175.39) .. controls (470.5,193.37) and (486.5,193.37) .. (496.5,193.37) ;

	\draw    (554.95,156.86) .. controls (581.5,155.3) and (582.5,195.48) .. (614.5,195.48) ;

	\draw    (614.5,195.48) .. controls (647.5,195.48) and (654.95,162.15) .. (672.95,157.92) ;

	\draw    (671.47,157.76) .. controls (701.5,156.36) and (682.5,192.31) .. (727.5,191.25) ;

	\draw  [line width=0.75]  (378.41,88.15) -- (387.27,98.16)(387.17,88.27) -- (378.51,98.04) ;
	\draw  [line width=0.75]  (490.41,88.15) -- (499.27,98.16)(499.17,88.27) -- (490.51,98.04) ;
	\draw  [line width=0.75]  (615.41,84.98) -- (624.27,94.99)(624.17,85.1) -- (615.51,94.87) ;
	\draw  [line width=0.75]  (731.41,83.92) -- (740.27,93.93)(740.17,84.04) -- (731.51,93.81) ;
	\draw  [color={rgb, 255:red, 1; green, 10; blue, 250 }  ,draw opacity=1 ][line width=1.5]  (349,137.87) -- (357.5,137.87) -- (357.5,147.9) -- (349,147.9) -- cycle ;
	\draw  [color={rgb, 255:red, 248; green, 4; blue, 4 }  ,draw opacity=1 ][line width=0.75]  (409.41,137.84) -- (418.27,147.86)(418.17,137.97) -- (409.51,147.73) ;
	\draw  [color={rgb, 255:red, 26; green, 2; blue, 252 }  ,draw opacity=1 ][line width=0.75]  (466.41,137.84) -- (475.27,147.86)(475.17,137.97) -- (466.51,147.73) ;
	\draw  [color={rgb, 255:red, 250; green, 2; blue, 2 }  ,draw opacity=1 ][line width=0.75]  (516.41,135.73) -- (525.27,145.74)(525.17,135.85) -- (516.51,145.62) ;
	\draw  [color={rgb, 255:red, 35; green, 2; blue, 250 }  ,draw opacity=1 ][line width=0.75]  (577.41,136.79) -- (586.27,146.8)(586.17,136.91) -- (577.51,146.68) ;
	\draw  [color={rgb, 255:red, 252; green, 3; blue, 3 }  ,draw opacity=1 ][line width=0.75]  (642.41,135.73) -- (651.27,145.74)(651.17,135.85) -- (642.51,145.62) ;
	\draw  [color={rgb, 255:red, 8; green, 1; blue, 252 }  ,draw opacity=1 ][line width=0.75]  (704.41,134.67) -- (713.27,144.69)(713.17,134.8) -- (704.51,144.56) ;
	\draw  [draw opacity=0] (439.93,157.74) .. controls (434.03,156.5) and (429.5,149.13) .. (429.5,140.22) .. controls (429.5,130.45) and (434.96,122.52) .. (441.69,122.52) .. controls (441.82,122.52) and (441.95,122.52) .. (442.08,122.53) -- (441.69,140.22) -- cycle ; \draw   (439.93,157.74) .. controls (434.03,156.5) and (429.5,149.13) .. (429.5,140.22) .. controls (429.5,130.45) and (434.96,122.52) .. (441.69,122.52) .. controls (441.82,122.52) and (441.95,122.52) .. (442.08,122.53) ;
	\draw  [draw opacity=0][dash pattern={on 4.5pt off 4.5pt}] (443.28,122.69) .. controls (449.19,123.86) and (453.8,131.18) .. (453.89,140.09) .. controls (453.98,149.86) and (448.6,157.85) .. (441.87,157.92) .. controls (441.74,157.92) and (441.61,157.92) .. (441.49,157.92) -- (441.69,140.22) -- cycle ; \draw  [dash pattern={on 4.5pt off 4.5pt}] (443.28,122.69) .. controls (449.19,123.86) and (453.8,131.18) .. (453.89,140.09) .. controls (453.98,149.86) and (448.6,157.85) .. (441.87,157.92) .. controls (441.74,157.92) and (441.61,157.92) .. (441.49,157.92) ;
	\draw  [draw opacity=0] (553.37,156.68) .. controls (547.48,155.44) and (542.94,148.07) .. (542.94,139.16) .. controls (542.94,129.39) and (548.4,121.46) .. (555.14,121.46) .. controls (555.27,121.46) and (555.39,121.46) .. (555.52,121.47) -- (555.14,139.16) -- cycle ; \draw   (553.37,156.68) .. controls (547.48,155.44) and (542.94,148.07) .. (542.94,139.16) .. controls (542.94,129.39) and (548.4,121.46) .. (555.14,121.46) .. controls (555.27,121.46) and (555.39,121.46) .. (555.52,121.47) ;
	\draw  [draw opacity=0][dash pattern={on 4.5pt off 4.5pt}] (556.73,121.63) .. controls (562.64,122.8) and (567.24,130.12) .. (567.33,139.03) .. controls (567.43,148.8) and (562.05,156.79) .. (555.31,156.86) .. controls (555.19,156.87) and (555.06,156.86) .. (554.93,156.86) -- (555.14,139.16) -- cycle ; \draw  [dash pattern={on 4.5pt off 4.5pt}] (556.73,121.63) .. controls (562.64,122.8) and (567.24,130.12) .. (567.33,139.03) .. controls (567.43,148.8) and (562.05,156.79) .. (555.31,156.86) .. controls (555.19,156.87) and (555.06,156.86) .. (554.93,156.86) ;
	\draw  [draw opacity=0] (671.37,157.74) .. controls (665.48,156.5) and (660.94,149.13) .. (660.94,140.22) .. controls (660.94,130.44) and (666.4,122.52) .. (673.14,122.52) .. controls (673.27,122.52) and (673.39,122.52) .. (673.52,122.53) -- (673.14,140.22) -- cycle ; \draw   (671.37,157.74) .. controls (665.48,156.5) and (660.94,149.13) .. (660.94,140.22) .. controls (660.94,130.44) and (666.4,122.52) .. (673.14,122.52) .. controls (673.27,122.52) and (673.39,122.52) .. (673.52,122.53) ;
	\draw  [draw opacity=0][dash pattern={on 4.5pt off 4.5pt}] (674.73,122.68) .. controls (680.64,123.86) and (685.24,131.18) .. (685.33,140.09) .. controls (685.43,149.86) and (680.05,157.85) .. (673.31,157.92) .. controls (673.19,157.92) and (673.06,157.92) .. (672.93,157.92) -- (673.14,140.22) -- cycle ; \draw  [dash pattern={on 4.5pt off 4.5pt}] (674.73,122.68) .. controls (680.64,123.86) and (685.24,131.18) .. (685.33,140.09) .. controls (685.43,149.86) and (680.05,157.85) .. (673.31,157.92) .. controls (673.19,157.92) and (673.06,157.92) .. (672.93,157.92) ;
	\draw    (496.5,193.37) .. controls (522.5,194.43) and (524.5,160.59) .. (554.95,156.86) ;

	\draw  [color={rgb, 255:red, 252; green, 2; blue, 32 }  ,draw opacity=1 ][line width=1.5]  (753,139.19) .. controls (753,135.82) and (755.58,133.1) .. (758.76,133.1) .. controls (761.94,133.1) and (764.52,135.82) .. (764.52,139.19) .. controls (764.52,142.55) and (761.94,145.28) .. (758.76,145.28) .. controls (755.58,145.28) and (753,142.55) .. (753,139.19) -- cycle ;
	\draw   (288,131.62) -- (309.3,131.62) -- (309.3,126) -- (323.5,137.24) -- (309.3,148.48) -- (309.3,142.86) -- (288,142.86) -- cycle ;
	
	\end{tikzpicture}
	\caption{Puncture Riemann surface and its degeneration: Each three punctured sphere represents a AD matter, and each tube represents a gauge group.}
	\label{pant}
	
\end{figure}
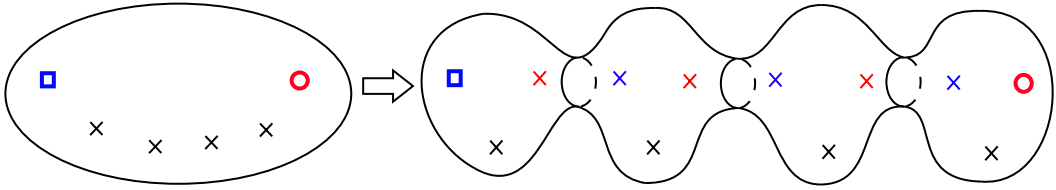

\subsection{Weakly coupled gauge theory descriptions for AD theories}
\label{sec:weaklycoupledAD}
Consider AD theory engineered using following data,
\begin{equation}
\fg,~~\Phi={T\over z^{2+{q k\over q n}}},~~~f,
\label{akan}
\end{equation} 
where $(k,n)$ is coprime, and $T$ is taken to be the most general matrix allowed by the grading. The corresponding AD theory often has exact marginal deformations, and 
its weakly coupled gauge theory description is found in \cite{Xie:2017vaf,Xie:2017aqx}. Here we
summarize the basic ideas:
\begin{itemize}
\item We first represent the above theory by an auxiliary punctured sphere $\Sigma$ with $n_a$ black marked points ($n_a$ is equal to 
the number of exact marginal deformation plus one), one blue 
marked point representing the irregular singularity with flavor symmetries, and one red point representing the regular singularity.
\item The weakly coupled description is found by finding a pair-of-pants decomposition of $\Sigma$ with the rules,
\begin{enumerate}[label=\alph*)]
\item In degenerating 
a tube, one create a pair of blue marked point and red marked point.
\item Each three punctured sphere in the pants decomposition 
has one black, one red and one blue puncture.  
\end{enumerate}
\end{itemize}
Now the crucial thing is to determine the puncture type created in the degeneration limit and the matter which is 
identified with the three punctured sphere. It turns out that the matter appearing in the above degeneration 
is exactly the matter studied in section \ref{sec:ADwithTwoFlavors}. Since we have already figured out the VOA for the matter part\footnote{In fact, we only have the full $VOA$ information for following cases: $n$ arbitrary of $A$ type theory; $n$ even of $D_N$ and $D_N$ twisted type theory; and $n$ odd for twisted $sl_{2N}$ and $sl_{2N+1}$ theory.}, 
we can now describe the VOA for the full theory as a coset. 

Now we discuss in more detail about VOA of general $A$ type theory. A general AD theory of $A$ type is represented by following configuration,
\begin{equation}
\fg=\fsl_{nq+n_1},~~\Phi={T\over z^{2+{kq\over n q }}},~~~f=[h_1^{r_1},\ldots, h_t^{r_t}].
\end{equation}
Here  $(k,n)=1$. $T=\diag(I_{nq\times nq},0_{n_1})$. This theory has 
flavor symmetries $U(n_1)\times U(1)^{q-1}\times G_F$\footnote{In general, we can choose a partition of size $n_1$ such that one can have more general flavor symmetries other than $U(n_1)$.}, where $G_F$ is the 
flavor symmetry associated to the regular puncture,
\begin{equation}
G_F=\prod U(r_i)/U(1).
\end{equation}
Weakly coupled gauge theory descriptions are given in \cite{Xie:2017aqx}. The idea is to represent our theory by an auxilliary punctured sphere $\Sigma$ with $q$ black marked points, one red marked point reprenting the data of 
regular singularity, and one blue marked point represent $U(n_1)$ flavor symmetry for the irregular singularity\footnote{For $n=1,~k\neq 1$, a blue puncture is equivalent to a red puncture. For $n=1, k=1$, all three type of punctures are the same\cite{Xie:2017aqx}.}. In particular, the theory considered in last section is represented by a three puncture sphere, whose VOA is identified as a $W$ algebra.

The weakly coupled gauge theory description is given by finding a pair-of-pants decomposition of $\Sigma$ such that each tube is connected by a 
blue and red punctrue. Moreover, each three punctured sphere 
has to have one blue, one red and one black puncture as shown in figure \ref{pant}. Now since we have 
already identified the VOA for each three punctured sphere, 
the VOA for the original theory is constructed from cosets of 
the VOA of the matter system.

\begin{figure}
\centering

\tikzset{every picture/.style={line width=0.75pt}} 

\begin{tikzpicture}[x=0.75pt,y=0.75pt,yscale=-1,xscale=1,scale=0.9]

\draw   (1,121) -- (40.5,121) -- (40.5,160.5) -- (1,160.5) -- cycle ;
\draw   (62,140.75) .. controls (62,129.84) and (70.84,121) .. (81.75,121) .. controls (92.66,121) and (101.5,129.84) .. (101.5,140.75) .. controls (101.5,151.66) and (92.66,160.5) .. (81.75,160.5) .. controls (70.84,160.5) and (62,151.66) .. (62,140.75) -- cycle ;
\draw    (40.5,139.92) -- (61.5,139.92) ;

\draw   (164,121) -- (203.5,121) -- (203.5,160.5) -- (164,160.5) -- cycle ;
\draw   (226,140.75) .. controls (226,129.84) and (234.84,121) .. (245.75,121) .. controls (256.66,121) and (265.5,129.84) .. (265.5,140.75) .. controls (265.5,151.66) and (256.66,160.5) .. (245.75,160.5) .. controls (234.84,160.5) and (226,151.66) .. (226,140.75) -- cycle ;
\draw    (101.5,140.75) -- (122.5,140.75) ;

\draw    (204.5,140.75) -- (225.5,140.75) ;

\draw    (265.5,140.75) -- (282.5,140.92) ;

\draw  [dash pattern={on 0.84pt off 2.51pt}]  (122.5,140.75) -- (143.5,140.75) ;

\draw    (143.5,140.75) -- (164.5,140.75) ;

\draw   (282,121) -- (321.5,121) -- (321.5,160.5) -- (282,160.5) -- cycle ;
\draw   (344,140.75) .. controls (344,129.84) and (352.84,121) .. (363.75,121) .. controls (374.66,121) and (383.5,129.84) .. (383.5,140.75) .. controls (383.5,151.66) and (374.66,160.5) .. (363.75,160.5) .. controls (352.84,160.5) and (344,151.66) .. (344,140.75) -- cycle ;
\draw    (323,140.75) -- (344,140.75) ;

\draw    (383.5,140.75) -- (400.5,140.92) ;

\draw   (400,121) -- (439.5,121) -- (439.5,160.5) -- (400,160.5) -- cycle ;
\draw   (462,140.75) .. controls (462,129.84) and (470.84,121) .. (481.75,121) .. controls (492.66,121) and (501.5,129.84) .. (501.5,140.75) .. controls (501.5,151.66) and (492.66,160.5) .. (481.75,160.5) .. controls (470.84,160.5) and (462,151.66) .. (462,140.75) -- cycle ;
\draw    (441,140.75) -- (462,140.75) ;

\draw    (501.5,140.75) -- (518.5,140.92) ;

\draw  [dash pattern={on 0.84pt off 2.51pt}]  (519.5,140.75) -- (540.5,140.75) ;

\draw    (540.5,140.75) -- (561.5,140.75) ;

\draw   (562,140.75) .. controls (562,129.84) and (570.84,121) .. (581.75,121) .. controls (592.66,121) and (601.5,129.84) .. (601.5,140.75) .. controls (601.5,151.66) and (592.66,160.5) .. (581.75,160.5) .. controls (570.84,160.5) and (562,151.66) .. (562,140.75) -- cycle ;
\draw    (601.5,140.75) -- (618.5,140.92) ;

\draw   (618,121) -- (657.5,121) -- (657.5,160.5) -- (618,160.5) -- cycle ;

\draw (20.75,140.75) node   {$T_{1}$};
\draw (81.75,140.75) node [scale=0.7]  {$SU( 1)$};
\draw (183.75,140.75) node   {$T_{a}$};
\draw (245.75,140.75) node [scale=0.7]  {$SU( an)$};
\draw (301.75,140.75) node   {$T_{a+1}$};
\draw (363.75,140.75) node [scale=0.7]  {$SU( bk)$};
\draw (419.75,140.75) node   {$L_{b}$};
\draw (481.75,140.75) node [scale=0.5]  {$SU( bk-k)$};
\draw (581.75,140.75) node [scale=0.7]  {$SU( k)$};
\draw (637.75,140.75) node   {$L_{1}$};

\end{tikzpicture}
\caption{The weakly coupled description of $(A_{nq-1}, A_{kq-1})$ theory with $a=[{q k\over n+k}]$ and $b=[{qn\over n+k}]$.}
\label{fig:AAtheory}
\end{figure}
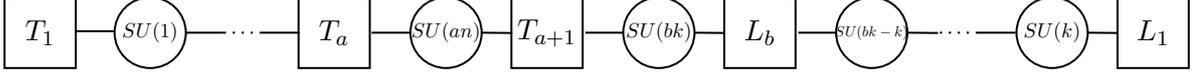

\textbf{Example}: Taking $n_1=0$ and $f=[nq]$ in \ref{akan}, this theory is also called the $(A_{nq-1}, A_{kq-1})$ theory. To find the weakly coupled gauge theory description, 
we represent our theory by a sphere $\Sigma$ with  $q$ black points, one trivial blue 
point and one trivial red point.
The weakly coupled description is described by taking a pants decomposition of $\Sigma$ such that each pant has a black puncture, a red punctrue, and a blue puncture. 
We assume $k\geq n$ without losing any generality. The weakly coupled gauge theory description is  shown in figure \ref{fig:AAtheory}, 
where
\begin{equation}
a=[{q k\over n+k}],~~b=[{qn\over n+k}].
\end{equation}
The square bracket means the integral part of the number inside. 
The flavor symmetry of the matter content is,
\begin{equation}
T_i:~~U((i-1)n)\times SU(in),~~L_i:~~U(ik)\times U((i-1)k),~~~T_{a+1}:~~U(an)\times U(bk)
\end{equation}
$T_i$ is the theory studied in section \ref{sec:ADwithTwoFlavors}, which is engineered by following configuration,
\begin{equation}
\fg=\fsl_{in},~~\Phi={T\over z^{2+{k\over n}}},~~f=[1^{in}],
\end{equation}
with $T=\diag(I_{n\times n},0_{(i-1)n})$.
The matter system $L_i$ is engineered by following configuration,
\begin{equation}
\fg=\fsl_{n+ik},~~\Phi={T\over z^{2+{k\over n}}},~~f=[n+k,1^{(i-1)k}],
\end{equation}
with $T=\diag(I_{n\times n},0_{ik})$.
For $L_i$'s however, a $U(1)$ flavor symmetry inside $U(ik)$ is decoupled. The reason is that the central charge of AKM part (with this extra $U(1)$s)
is bigger than the full central charge, so some flavor symmetries have to be decoupled. On the other hand, one can find a different realization of $L_i$ 
where the extra $U(1)$ flavor symmetry is indeed gone (see section \ref{collapsing}).

There is another descrition of the same theory: 
\begin{equation}
\fg=\fsl_{qk},~~\Phi={T\over z^{2+{qn\over qk}}},~~~f=[qk],
\end{equation}
which is just the level-rank dual of the original 4d theory. The weakly coupled gauge theory description is found in similar way, but the VOA
of each matter is described differently (which is just the level rank dual discussed in section. \ref{level}). 
So the level rank duality of above theory is the consequence of the level-rank duality of each AD matter. 

\subsection{Theory with Lagrangian description}
\begin{figure}
\centering

\tikzset{every picture/.style={line width=0.75pt}} 

\begin{tikzpicture}[x=0.75pt,y=0.75pt,yscale=-1,xscale=1]

\draw   (62,145.48) .. controls (62,131.96) and (72.96,121) .. (86.48,121) .. controls (100.01,121) and (110.97,131.96) .. (110.97,145.48) .. controls (110.97,159.01) and (100.01,169.97) .. (86.48,169.97) .. controls (72.96,169.97) and (62,159.01) .. (62,145.48) -- cycle ;
\draw   (135.14,145.48) .. controls (135.14,131.96) and (146.1,121) .. (159.63,121) .. controls (173.15,121) and (184.11,131.96) .. (184.11,145.48) .. controls (184.11,159.01) and (173.15,169.97) .. (159.63,169.97) .. controls (146.1,169.97) and (135.14,159.01) .. (135.14,145.48) -- cycle ;
\draw    (110.97,145.48) -- (134.52,145.48) ;

\draw    (184.11,145.48) -- (208.9,145.7) ;

\draw    (308.08,146.72) -- (329.15,146.94) ;

\draw   (328.53,122.24) -- (377.5,122.24) -- (377.5,171.21) -- (328.53,171.21) -- cycle ;
\draw   (259.11,145.48) .. controls (259.11,131.96) and (270.07,121) .. (283.59,121) .. controls (297.12,121) and (308.08,131.96) .. (308.08,145.48) .. controls (308.08,159.01) and (297.12,169.97) .. (283.59,169.97) .. controls (270.07,169.97) and (259.11,159.01) .. (259.11,145.48) -- cycle ;
\draw  [dash pattern={on 0.84pt off 2.51pt}]  (208.9,145.7) -- (234.94,145.7) ;

\draw    (234.94,145.7) -- (258.49,145.7) ;

\draw (86.48,145.48) node [scale=0.7]  {$SU( k)$};
\draw (159.63,145.48) node [scale=0.7]  {$SU( 2k)$};
\draw (353.02,146.72) node   {$kN$};
\draw (283.59,145.48) node [scale=0.5]  {$SU( Nk-k)$};

\end{tikzpicture}
\caption{The Lagrangian description of the theory \ref{eq:ADwithLagrangian}.}
\label{fig:LagrangianOfAD}
\end{figure}
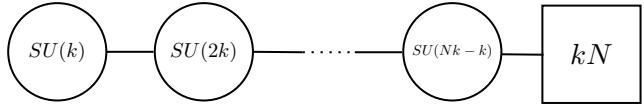

Some theories have Lagrangian descriptions, therefore one can find 
the VOA by cosets of symplectic bosons. Here are some examples,
\begin{equation}
\label{eq:ADwithLagrangian}
\fg=\fsl_{kN},~~\Phi={T\over z^{2+{-N+1\over N}}},~~f=[1^{kN}].
\end{equation}
This theory has a Lagrangian description given by the quiver in figure \ref{fig:LagrangianOfAD}.
Now each matter is just bifundamental hypers and its VOA is a set of symplectic bosons, and 
the VOA of the full theory is just the coset of symplectic bosons.

\section{More on the conformal embedding}
\label{sec:ConformalEmbedding}

VOAs for AD matters considered in this paper have the interesting property that their AKM subalgebras have the same central charges as full VOAs, so they define possible conformal embeddings. In this section, we will show that such possible conformal embeddings are much more general in our theory space. Consider following configuration,
\begin{equation}
\Phi={T\over z^{2+{k\over b}}},~~~ f=\mathrm{trivial},
\end{equation}
where $T$ is given by a principle grading, and the classification of $b$ is summarized in section \ref{sec:results}. The flavor symmetry group is $U(1)^{f_0}\times G_F$, where  $f_0$  is the number of mass parameters encoded in irregular singularity and $G_F$ is the flavor symmetry from the regular singularity. It was noticed in \cite{Xie:2016evu} that the 4d central charge has the following form:
\begin{equation}
c_{4d}={1\over 12}\left({k_G\dim(G)\over -k_G+h^{\vee}}-f_0\right).
\end{equation} 
Here $k_G$ is the flavor central charge of flavor group $G_F$. So the corresponding central charge of 2d VOA is 
\begin{equation}
c_{2d}=-12 c_{4d}={k_{2d} \dim(G)\over k_{2d}+h^{\vee}}+f_0. 
\end{equation}
Here we use the correspondence $k_{2d}=-k_G$.
We know that the 2d VOA has a AKM subalgebra  $V_k(G)\times U(1)^{f_0}$, and 
the central charge of AKM sector is exactly the central charge of full VOA. Therefore, 
potentially we have a conformal embedding of AKM subalgebra $V_k(G)\times U(1)^{f_0}$ into 
the VOA. If we can indeed prove the above conformal embedding, we could define 
the full VOA as a reducible module of AKM subalgebra, which would provide 
us a definition of the full VOA. 

The conformal embedding of the AKM algebra into the full VOA is generally not true if we 
change the regular singularity to a generic one. The conformal embedding is 
possible only for very special choice of level and nilpotent orbit $f$, see examples in section \ref{sec:ADwithTwoFlavors}.

Now consider a theory which has exact marginal deformations, and one can find a weakly coupled gauge theory description. Assuming that each matter  has a conformal embedding of AKM, we now prove that the full theory  also  has a possible conformal embedding (at least the total  central charges of AKM pieces and that of  the full VOA are the same.). Assume that  
the weakly coupled gauge theory description has the form,
\begin{equation}
T_L-\boxed{G}-T_R,
\end{equation}
and also assume that $T_L$ part has flavor symmetry $G_L\times G$, and $T_R$ part has flavor symmetry $G_R\times G$, here $G$ is a simple factor of flavor symmetry group of two matter systems for simplicity. The full theory then has the flavor symmetry $G_L\times G_R$.
 The central charge of 4d theory is,
\begin{equation}
c_{4d}=c_{T_L}+{2\dim(G)\over 12}+c_{T_R}.
\label{4d}
\end{equation}
If each individual piece has a conformal embedding, we have 
\begin{equation}
c_{T_L}=c(G_L)+{1\over 12}{k_{LG}\dim(G)\over -k_{LG}+h^{\vee}},~~~c_{T_R}=c(G_R)+{1\over 12}{k_{RG}\dim(G)\over -k_{RG}+h^{\vee}},
\label{piece}
\end{equation}
where  $c(G_L)$ and $c(G_R)$ is the AKM central charge from the flavor symmetry $G_L$ and $G_R$ respectively, and $k_{LG}$ and $k_{RG}$ are flavor central charges. Conformal gauging requires,
\begin{equation}
k_{LG}+k_{RG}=2h^{\vee}.
\label{conformal}
\end{equation}
Substiting \ref{piece} and \ref{conformal} into \ref{4d}, we find that 
\begin{equation}
c_{4d}=c(G_L)+c(G_R), 
\end{equation}
So the full theory also has a possible conformal embedding. 
We have concluded that the gauged system has the conformal embedding if each matter
piece has the conformal embedding. On the other hand, if we assume that one piece 
of matter and the gauged system has the conformal embedding, the other piece of matter
would also have the conformal embedding.

The general AD matter has three non-abelian flavor symmetries \cite{Xie:2016uqq,Xie:2017vaf}, and the addition of a 
third non-abelian flavor symmetries would not change the flavor central charge of the other two non-abelian flavor symmetries. 
We have shown in section \ref{sec:ADwithTwoFlavors} that conformal embedding of two non-abelian 
AKMs into a $W$ algebra $W^{k'}(g,f)$ is possible, one might wonder whether it is possible to have a conformal embedding of general AD theory into a W algebra. 
However the analysis of section \ref{collapsing} shows that 
the conformal embedding of three AKM into a $W$ algebra is not possible (the third non-abelian flavor symmetry would have flavor 
central charge zero and is decoupled).

\section{Conclusion}
\label{sec:conclusion}
We have identified VOAs of a class of AD matters with two distinct non-abelian flavor 
symmetries as $W$ algebras. Using  weakly coupled gauge theory descriptions  
formed by gauging above types of AD matters, we found the VOA for more general AD theories engineered from 6d $(2,0)$ SCFTs, i.e. VOA for 
general $(A_{N-1}, A_{k-1})$ theory is found.

One usually learns many interesting properties of 4d theory by using properties of 2d VOA, since 4d theory is strongly coupled and little is known
about their spectrum while many aspects of 2d VOA are  much more well understood. Therefore it is pleasant that 4d theory can actually predict many interesting
features about 2d VOAs. In this paper, we show that the simple fact that a single 4d SCFT can be engineered by different 6d configurations can often teach us very interesting lessons about VOAs. For example, we find new level-rank 
duality, coset descriptions,  possible conformal embeddings, and etc. Although VOA is mainly about the Schur sector which includes the Higgs branch information,
the usage of Coulomb branch data is often very useful in telling whether two configurations are the same or not, which in turn teaches us interesting lessons of VOA.

One of interesting lesson we learned in this paper is that the flavor symmetries of SCFTs defined by 
a 6d $(2,0)$ construction is a subtle issue. There are situations where the naive flavor symmetry is 
actually decoupled in the IR, which corresponds to collapsing levels of 2d VOA. There are also 
situations where there is extra flavor symmetries which are not manifest in certain 6d descriptions. 
Therefore it is interesting to understand the emergency of symmetry from the VOA point of view.

The general AD matter has three non-abelian flavor symmetries, and the remaining task is to identify VOAs for them. Once we find  VOAs for these AD matters, we can find the VOA 
for all SCFTs constructed from 6d $(2,0)$ construction.  VOAs defined using  junctions of $\mathcal{N}=4$ boundary conditions are studied in \cite{Gaiotto:2017euk,Creutzig:2017uxh}. It appears that 
they have similar structures involving two or three Lie algebras, and it would be interesting to figure out whether these VOAs have anything to do with VOAs 
studied in this paper. 

We mainly identify the 4d/2d pair in this paper, and a detailed study of characters and its physical implication will be given in a follow-up paper.

\section*{Acknowledgements}
The authors would like to thank Tomoyuki Arakawa for helpful discussions. DX and WY are supported by Yau mathematical science center at Tsinghua University. WY is also supported by the Young overseas high-level talents introduction plan. D.X would like to thank Caltech for hospitality at the final stage of completion of this paper.


\appendix

\section{Hitchin system descriptions for $(G,G')$ and $D_p(G)$ theory}
\label{sec:app:ADs}
There are various class of four dimensional $\mathcal{N}=2$ AD SCFTs found in the literature, and they have different labels which might 
cause some confusions. Here we  provide a  mapping between these labels and our theories. 
There are three class of theories:
\begin{enumerate}
\item Theories with label $(G,G')$ \cite{Cecotti:2010fi}. This class of theories are engineered by following 3-fold singularity:
\begin{equation}
f_{G}(x,y)+f_{G'}(z,w)=0.
\end{equation}	
Here $G=ADE$ and $f_G(x,y)$ are following polynomials:
\begin{equation}
f_{A_{N}}=x^2+y^{N+1},~f_{D_{N}}=x^{N-1}+x y^2,~f_{E_6}=x^3+y^4,~f_{E_7}=x^3+xy^3,~f_{E_8}=x^3+y^5.
\end{equation}
There is a symmetry exchanging $G$ and $G'$ in the definition of the 3d singularity so that the $(G,G')$ theory 
is the same as the $(G',G)$ theory. This class of theories include the original AD theory found in \cite{Argyres:1995jj} (It is the $(A_1,A_2)$ theory.), 
and the later $ADE$ generalizations \cite{Eguchi:1996vu} (They are $(A_1, G)$ type theories) with $G=ADE$. This class of theories typically 
do not have any non-abelian flavor symmetries, although they could have abelian flavor symmetries.

\item Theories with label $D_p(G)$ \cite{Cecotti:2012jx}, where $p$ is a positive integer and $G=ADE$. For $G=A_{N}$, 
they are called type IV theory in \cite{Xie:2012hs}. This class of theories has a flavor symmetry group $G$ and possibly some more abelian flavor symmetry depending on value of $p$.

\item Theories with label $(J^{(b)}[k],f)$ in \cite{Wang:2015mra}, with $k>-b$.
They were studied in \cite{Xie:2012hs,Wang:2015mra} and are defined using 6d $(2,0)$ SCFT with following data,
\begin{equation}
J=ADE,~~\Phi={T\over z^{2+{k\over b}}},~~f.
\end{equation}
Here $f$ is a nilpotent orbit of $J=ADE$\footnote{Here we use Nahm labels so that a regular nilpotent orbit  gives no flavor symmetry, while the trivial nilpotent orbit gives $G$ flavor symmetry with $G$ the Lie group of $\fg$.}, and $T$ is a regular semi-simple matrix whose form depending on value $b$. $b$ takes a finite set of numbers as in table 
\ref{table:sing}, and in particular $b$ can always take the value $h^{\vee}$ which is the dual Coxeter number. 
For $J=A_{N-1}, b=N$, it is called type I theory in \cite{Xie:2012hs}, and for $b=N-1$, it is called type II theory in \cite{Xie:2012hs}. 
\end{enumerate}	
We have the following mapping between the third class of theories and the first two class of theories:
\begin{equation}
(J^{h^{\vee}}[k],f_{reg}) = (J,A_{k-1}),~~~~~(J^{h^{\vee}}[k],f_{trivial}) = D_{k+h^{\vee}}(J).
\end{equation}
Here $h^{\vee}$ is the dual Coxeter number.

\section{Coulomb branch spectrum from the Newton polygon}
\label{sec:app:Newton}

 Let us now briefly review 
how to find the Coulomb branch spectrum from the Newton polygon:
\begin{itemize}
\item The SW curve at SCFT point is 
\begin{equation}
x^{n+n_1}+x^{n_1}z^k=0.
\end{equation}
The scaling dimension of $x$ and $z$ coordinates can be found as follows. 
Each term in the above equation has the same scaling dimension, so we have $n[x]=k[z]$. The SW differential $\lambda=xdz$ has scaling dimension one, therefore$[x]+[z]=1$. We then have,
\begin{equation}
[x]={k\over n+k},~~[z]={n\over n+k}.
\end{equation}
\item The full SW curve takes the following form,
\begin{equation}
x^{n+n_1}+x^{n_1}z^k+\sum_{i,j}u_{ij}x^{n+n_1-i}z^j=0.
\end{equation}
We include all monomials in the Newton polygon (including boundary points)
and the coefficients $u_{ij}$ are parameters of 4d $\mathcal{N}=2$ theory including vevs of Coulomb branch operators, coupling constants and mass parameters.  
The scaling dimension of $u_{ij}$ is computed by the requirement that each term has same 
scaling dimension, therefore,
\begin{equation}
[u_{ij}]=i[x]-j[z]={i k\over n+k}-{j n\over n+k}.
\end{equation}
The Coulomb branch spectrum of a theory is defined as subsets of $u_{ij}$ whose scaling dimension is bigger than one (We only consider the lattice points inside Newton polygon, and the boundary points whose scaling dimensions are bigger than one  are actually mass parameters), and we have the following Coulomb branch spectrum,
\begin{equation}
\begin{split}
&\{l-{j n\over n+k}\},~~~~l=2,\ldots, n,~ j=1, \ldots,[{(l-1)(n+k)\over n}],   \\
& \{l-{j n\over n+k}\},~~~l=n+1,\ldots, n+n_1,~j=1,\ldots, n+k-1.
\end{split}
\label{CoulombA}
\end{equation}

Now if there is a generic puncture with Young Tableaux of size $n+n_1$, we label the boxes of Young Tableaux from one to $n+n_1$ row by row, and record the height of the $l$th box as $h(l)$\footnote{For a full puncture with Young Tableaux $[1,\ldots,1]$, we have $h(l)=[1,\ldots,1]$.}, then the Coulomb branch spectrum has the following description,
\begin{equation}
\begin{split}
&\{l-{j n\over n+k}\},~~~~l=2,\ldots, n,~ j=h(l), \ldots,[{(l-1)(n+k)\over n}],   \\
& \{l-{j n\over n+k}\},~~~l=n+1,\ldots, n+n_1,~j=h(l),\ldots, n+k-1.
\end{split}
\label{CoulombAA}
\end{equation}

\end{itemize}

\section{The central charge of $W^k(\fg,f)$}
\label{sec:app:central}

To any $\fsl_2$-triple $\{f, x, e\}$ in $\fg$, where $[x,f]=-f$, $[x,e]=e$, one associates a W-algebra $W^k(\fg,f)$ through the quantum Hamiltonian reduction from the vacuum $\hat{\fg}$-module of level $k$. The central charge of $W^k(\fg,f)$ is
\begin{equation}
\label{eq:Wcentralcharge}
c(W^k(\fg,f))=\mathrm{dim}\fg_0-\frac{1}{2}\mathrm{dim}\fg_{\half}-\frac{12}{k+h^{\vee}}|\rho-(k+h^{\vee})x_0|^2,
\end{equation}
where
\begin{equation}
x_0=\frac{x}{2},
\end{equation}
and
\begin{equation}
\fg_j=\{g\in\fg|[x_0,g]=jg\}.
\end{equation}

For $\fg=\fsl_n=A_{n-1}$, the Cartan subalgebra $\fh$ is the set of traceless diagonal $n$ by $n$ matrices. Define linear functionals $e_i\in \fh^{\ast}$ by $e_i(H)=i^{\mathrm{th}}$ diagonal entry of $H$ where $1\leq i\leq n$. Then the root system of $\fg$ is
\begin{equation}
\{e_i-e_j\,|\, 1\leq i, j\leq n, i\neq j\}.
\end{equation}
The set of positive roots is $\{e_i-e_j\,|\, i< j\}$. The $(e_i-e_j)$-root space is spanned by the elementary matrix $E_{i,j}$ with its $ij$-entry $1$ and zeros otherwise.

Nilpotent orbits in $A_{n-1}$ are labelled by partitions of $n$ (or Young tableaux of $n$ boxes). Following the notation and recipe in \cite{Collingwood:1993rr}, for the partition $Y=[d_1,\cdots,d_i,\cdots,d_l]$, choose a block of consecutive indices $\{N_i+1,\cdots,N_i+d_i\}$ in such way that disjoint blocks are attached to different $d_i$'s. Then define the set of simple roots for each $d_i$,
\begin{equation}
\mathcal{C}^+(d_i)=\{e_{N_i+1}-e_{N_i+2},\cdots, e_{N_i+d_i-1}-d_{N_i+d_i}\},
\end{equation}
with $\mathcal{C}^+$ empty whenever $d_i=1$. One choice of the standard triple $\{H, X, Y\}$ for $Y$ is,
\begin{equation}
H=\sum_{1\leq i\leq l} H_{\mathcal{C}(d_i)}=\sum_{1\leq i\leq l}\sum_{1\leq j\leq d_i} (d_i-2j+1)E_{N_i+j, N_i+j},
\end{equation}
and
\begin{equation}
\begin{split}
X&=\sum_{\alpha\in\cup_i \mathcal{C}^+(d_i)} X_\alpha,\\
Y&=\sum_{\alpha\in\cup_i \mathcal{C}^+(d_i)} X_{-\alpha},
\end{split}
\end{equation}
where $X_\alpha$ is the $\alpha$-root vector. Finally $x$ is the diagonal matrix derived from $H$ via Weyl group of $A_{n-1}$ satisfying the $\Delta$-dominant condition,
\begin{equation}
x_1\geq x_2\geq \cdots\geq x_{n}.
\end{equation}
For example, given the tomahawk tableaux $Y=[2,2,1,1]$ for $\fg=sl_6=A_5$, $H$ is
\begin{equation}
H=\mathrm{diag}(1,-1,1,-1,0,0),
\end{equation}
and $x$ is 
\begin{equation}
x=\mathrm{diag}(1,1,0,0,-1,-1).
\end{equation}

Notice the diagonal entry of $x$ is also its coordinates in orthogonal basis of $A_{n-1}$, and the coordinates of $\rho$ in orthogonal basis is,
\begin{equation}
\rho = \half \left( n-1, n-3, \cdots, -n+3, -n+1\right).
\end{equation}
One can then easily compute $(\rho, \rho)$, $(\rho, x_0)$ and $(x_0,x_0)$ because they are just ordinary scalar product in orthogonal basis. For tomahawk tableaux $Y=[q^m,1^{n-qm}]$, explicit results are,
\begin{equation}
\begin{split}
(\rho, \rho)&=\frac{1}{12}(n^3-n),\\
(x_0,x_0)&=\frac{1}{12}m(q^3-q),\\
(\rho, x_0)&=\left\{
\begin{array}{cc}
\frac{1}{24}mq(3nq-m(2+q^2)),&q\,\,\mathrm{even},\\
\frac{1}{24}m(3n-mq)(q^2-1),&q\,\,\mathrm{odd}.
\end{array}
\right.
\end{split}
\end{equation}

$\mathrm{dim} \fg_0$ and $\mathrm{dim}\fg_{\half}$ can be solved easily using the explicit expression of $x_0$. For tomahawk tablaeux $Y=[q^m,1^{n+1-qm}]$, the explicit expressions are,
\begin{equation}
\mathrm{dim}\fg_0 =
\left\{
\begin{array}{cc}
(n-mq)^2+m^2q-1,&q\,\,\mathrm{even},\\
(n-mq+m)^2+m^2(q-1)-1,&q\,\,\mathrm{odd},
\end{array}
\right.
\end{equation}
and
\begin{equation}
\mathrm{dim}\fg_{\half}=
\left\{
\begin{array}{cc}
2m(n-mq),&q\,\,\mathrm{even},\\
0,&q\,\,\mathrm{odd}.
\end{array}
\right.
\end{equation}

Plugging results in the central charge formula \ref{eq:Wcentralcharge}, the central charge for $Y=[q^m,1^{n-mq}]$ is,
\begin{equation}
\label{eq:centralA}
c(W^k(\fsl_n,[q^m,1^{n-mq}])
=mq(k-n+(m+3n)q-(k+m+n)q^2)
-\frac{k+k(m-n)n+mn^2}{k+n}.
\end{equation}

For other Lie algebra $\fg$, partitions will be specified to particular cases used in the main context. The general recipe for $sl_2$-triple can be found in \cite{Collingwood:1993rr}. In the case of $\fg=\mathfrak{so}_{2n}=D_n$ and the partition $Y=[q^m,1^{2n-qm}]$ with $q$ and $m$ even, one can work out $x_0$,
\begin{equation}
x_0=\half\left(
\begin{array}{cc}
D & 0 \\
0 & -D 
\end{array}
\right),
\end{equation}
with
\begin{equation}
D=\mathrm{diag}(\underbrace{q-1,\ldots,q-1}_m,\underbrace{q-3,\ldots,q-3}_m,\ldots,\underbrace{1,\ldots,1}_m,0,\ldots,0),
\end{equation}
therefore
\begin{equation}
\begin{split}
(\rho,\rho)&=\frac{1}{6}(n-1)n(2n-1),\\
(x_0,x_0)&=\frac{1}{24}mq(q^2-1),\\
(\rho, x_0)&=-\frac{1}{48}mq(m(q^2+2)+q(-6n+3)),
\end{split}
\end{equation}
and also 
\begin{equation}
\begin{split}
\mathrm{dim}\fg_0&=\half q m^2+\left(n-\half mq\right)^2+\left(n-\half mq\right)\left(n-\half mq-1\right),\\
\mathrm{dim}\fg_{\half}&=2m\left(n-\half mq\right).
\end{split}
\end{equation}
Combining all results, the central charge is,
\begin{equation}
\label{eq:centralD}
\begin{split}
c(W^k(\mathfrak{so}_{2n},[q^m,1^{2n-mq}])
=&-\half mq((k+m+2n-2)q^2-q(6n+m-3)+2n-k+1)\\
&-\frac{kn(m-2n+1)+2mn(n-1)}{k+2n-2}.
\end{split}
\end{equation}

In the case of $\fg=\mathfrak{so}_{2n+1}=B_n$ and the partition $Y=[q^m,1^{2n+1-qm}]$ with $q$ and $m$ even, one can work out $x_0$,
\begin{equation}
x_0=\half\left(
\begin{array}{ccc}
0 & 0 & 0\\
0 &D & 0 \\
0 & 0 & -D 
\end{array}
\right),
\end{equation}
with
\begin{equation}
D=\mathrm{diag}(\underbrace{q-1,\ldots,q-1}_m,\underbrace{q-3,\ldots,q-3}_m,\ldots,\underbrace{1,\ldots,1}_m,0,\ldots,0),
\end{equation}
therefore
\begin{equation}
\begin{split}
(\rho,\rho)&=\frac{1}{12}n(2n-1)(2n+1),\\
(x_0,x_0)&=\frac{1}{24}mq(q^2-1),\\
(\rho, x_0)&=-\frac{1}{48}mq(m(q^2+2)-6nq),
\end{split}
\end{equation}
and also 
\begin{equation}
\begin{split}
\mathrm{dim}\fg_0&=\half q m^2+\left(n-\half mq\right)^2+\left(n-\half mq\right)\left(n-\half mq-1\right)+2\left(n-\half mq\right),\\
\mathrm{dim}\fg_{\half}&=2m\left(n-\half mq\right)+m.
\end{split}
\end{equation}
Combining all results, the central charge is,
\begin{equation}
\label{eq:centralB}
\begin{split}
c(W^k(\mathfrak{so}_{2n+1},[q^m,1^{2n+1-mq}])
=&-\half mq((k+m+2n-1)q^2-q(6n+m)+2n-k+2)\\
&-\frac{(2n+1)(m(k+2n-1)-2kn)}{2(k+2n-1)}.
\end{split}
\end{equation}

In the case of $\fg=\mathfrak{sp}_{2n}=C_n$ and the partition $Y=[q^m,1^{2n-qm}]$ with $q$ even, one can work out $x_0$,
\begin{equation}
x_0=\left(
\begin{array}{cc}
D & 0 \\
0 & -D 
\end{array}
\right),
\end{equation}
with
\begin{equation}
D=\mathrm{diag}(\underbrace{q-1,\ldots,q-1}_m,\underbrace{q-3,\ldots,q-3}_m,\ldots,\underbrace{1,\ldots,1}_m,0,\ldots,0),
\end{equation}
therefore
\begin{equation}
\begin{split}
(\rho,\rho)&=\frac{1}{12}n(n+1)(2n+1),\\
(x_0,x_0)&=\frac{1}{12}mq(q^2-1),\\
(\rho, x_0)&=-\frac{1}{48}mq(m(q^2+2)-3nq(2n+1)),
\end{split}
\end{equation}
and also 
\begin{equation}
\begin{split}
\mathrm{dim}\fg_0&=\half q m^2+\left(n-\half mq\right)^2+\left(n-\half mq\right)\left(n-\half mq+1\right),\\
\mathrm{dim}\fg_{\half}&=2m\left(n-\half mq\right).
\end{split}
\end{equation}
Combining all results, the central charge is,
\begin{equation}
\label{eq:centralC}
\begin{split}
c(W^k(\mathfrak{sp}_{2n},[q^m,1^{2n-mq}])
=&-\half mq((2k+m+2n+2)q^2-q(6n+m+3)+2n-2k-1)\\
&-\frac{n(m(n+1)-k(2n-m+1))}{k+n+1}.
\end{split}
\end{equation}


\bibliographystyle{utphys}

\bibliography{ADhigher}

\providecommand{\href}[2]{#2}\begingroup\raggedright\begin{thebibliography}{10}

\bibitem{Beem:2013sza}
C.~Beem, M.~Lemos, P.~Liendo, W.~Peelaers, L.~Rastelli, and B.~C. van Rees,
  ``{Infinite Chiral Symmetry in Four Dimensions},''
  \href{http://dx.doi.org/10.1007/s00220-014-2272-x}{{\em Commun. Math. Phys.}
  {\bf 336} (2015) no.~3, 1359--1433},
\href{http://arxiv.org/abs/1312.5344}{{\tt arXiv:1312.5344 [hep-th]}}.

\bibitem{Beem:2014rza}
C.~Beem, W.~Peelaers, L.~Rastelli, and B.~C. van Rees, ``{Chiral algebras of
  class S},'' \href{http://dx.doi.org/10.1007/JHEP05(2015)020}{{\em JHEP} {\bf
  1505} (2015)  020},
\href{http://arxiv.org/abs/1408.6522}{{\tt arXiv:1408.6522 [hep-th]}}.

\bibitem{Lemos:2014lua}
M.~Lemos and W.~Peelaers, ``{Chiral Algebras for Trinion Theories},''
  \href{http://dx.doi.org/10.1007/JHEP02(2015)113}{{\em JHEP} {\bf 02} (2015)
  113},
\href{http://arxiv.org/abs/1411.3252}{{\tt arXiv:1411.3252 [hep-th]}}.

\bibitem{Lemos:2015orc}
M.~Lemos and P.~Liendo, ``{$\mathcal{N}=2$ central charge bounds from $2d$
  chiral algebras},'' \href{http://dx.doi.org/10.1007/JHEP04(2016)004}{{\em
  JHEP} {\bf 04} (2016)  004},
\href{http://arxiv.org/abs/1511.07449}{{\tt arXiv:1511.07449 [hep-th]}}.

\bibitem{Cecotti:2015lab}
S.~Cecotti, J.~Song, C.~Vafa, and W.~Yan, ``{Superconformal Index, BPS
  Monodromy and Chiral Algebras},''
  \href{http://dx.doi.org/10.1007/JHEP11(2017)013}{{\em JHEP} {\bf 11} (2017)
  013},
\href{http://arxiv.org/abs/1511.01516}{{\tt arXiv:1511.01516 [hep-th]}}.

\bibitem{Arakawa:2016hkg}
T.~Arakawa and K.~Kawasetsu, ``{Quasi-lisse vertex algebras and modular linear
  differential equations},''
\href{http://arxiv.org/abs/1610.05865}{{\tt arXiv:1610.05865 [math.QA]}}.

\bibitem{Bonetti:2016nma}
F.~Bonetti and L.~Rastelli, ``{Supersymmetric localization in AdS$_{5}$ and the
  protected chiral algebra},''
  \href{http://dx.doi.org/10.1007/JHEP08(2018)098}{{\em JHEP} {\bf 08} (2018)
  098},
\href{http://arxiv.org/abs/1612.06514}{{\tt arXiv:1612.06514 [hep-th]}}.

\bibitem{Song:2016yfd}
J.~Song, ``{Macdonald Index and Chiral Algebra},''
  \href{http://dx.doi.org/10.1007/JHEP08(2017)044}{{\em JHEP} {\bf 08} (2017)
  044},
\href{http://arxiv.org/abs/1612.08956}{{\tt arXiv:1612.08956 [hep-th]}}.

\bibitem{Fredrickson:2017yka}
L.~Fredrickson, D.~Pei, W.~Yan, and K.~Ye, ``{Argyres-Douglas Theories, Chiral
  Algebras and Wild Hitchin Characters},''
\href{http://arxiv.org/abs/1701.08782}{{\tt arXiv:1701.08782 [hep-th]}}.

\bibitem{Cordova:2017mhb}
C.~Cordova, D.~Gaiotto, and S.-H. Shao, ``{Surface Defects and Chiral
  Algebras},'' \href{http://dx.doi.org/10.1007/JHEP05(2017)140}{{\em JHEP} {\bf
  05} (2017)  140},
\href{http://arxiv.org/abs/1704.01955}{{\tt arXiv:1704.01955 [hep-th]}}.

\bibitem{Song:2017oew}
J.~Song, D.~Xie, and W.~Yan, ``{Vertex operator algebras of Argyres-Douglas
  theories from M5-branes},''
  \href{http://dx.doi.org/10.1007/JHEP12(2017)123}{{\em JHEP} {\bf 12} (2017)
  123},
\href{http://arxiv.org/abs/1706.01607}{{\tt arXiv:1706.01607 [hep-th]}}.

\bibitem{Buican:2017fiq}
M.~Buican, Z.~Laczko, and T.~Nishinaka, ``{$ \mathcal{N} $ = 2 S-duality
  revisited},'' \href{http://dx.doi.org/10.1007/JHEP09(2017)087}{{\em JHEP}
  {\bf 09} (2017)  087},
\href{http://arxiv.org/abs/1706.03797}{{\tt arXiv:1706.03797 [hep-th]}}.

\bibitem{Beem:2017ooy}
C.~Beem and L.~Rastelli, ``{Vertex operator algebras, Higgs branches, and
  modular differential equations},''
  \href{http://dx.doi.org/10.1007/JHEP08(2018)114}{{\em JHEP} {\bf 08} (2018)
  114},
\href{http://arxiv.org/abs/1707.07679}{{\tt arXiv:1707.07679 [hep-th]}}.

\bibitem{Pan:2017zie}
Y.~Pan and W.~Peelaers, ``{Chiral Algebras, Localization and Surface
  Defects},'' \href{http://dx.doi.org/10.1007/JHEP02(2018)138}{{\em JHEP} {\bf
  02} (2018)  138},
\href{http://arxiv.org/abs/1710.04306}{{\tt arXiv:1710.04306 [hep-th]}}.

\bibitem{Fluder:2017oxm}
M.~Fluder and J.~Song, ``{Four-dimensional Lens Space Index from
  Two-dimensional Chiral Algebra},''
  \href{http://dx.doi.org/10.1007/JHEP07(2018)073}{{\em JHEP} {\bf 07} (2018)
  073},
\href{http://arxiv.org/abs/1710.06029}{{\tt arXiv:1710.06029 [hep-th]}}.

\bibitem{Choi:2017nur}
J.~Choi and T.~Nishinaka, ``{On the chiral algebra of Argyres-Douglas theories
  and S-duality},'' \href{http://dx.doi.org/10.1007/JHEP04(2018)004}{{\em JHEP}
  {\bf 04} (2018)  004},
\href{http://arxiv.org/abs/1711.07941}{{\tt arXiv:1711.07941 [hep-th]}}.

\bibitem{Arakawa:2017fdq}
T.~Arakawa, ``{Representation theory of W-algebras and Higgs branch
  conjecture},'' in {\em {International Congress of Mathematicians (ICM 2018)
  Rio de Janeiro, Brazil, August 1-9, 2018}}.
\newblock 2017.
\newblock
\href{http://arxiv.org/abs/1712.07331}{{\tt arXiv:1712.07331 [math.RT]}}.
\newblock

\bibitem{Niarchos:2018mvl}
V.~Niarchos, ``{Geometry of Higgs-branch superconformal primary bundles},''
  \href{http://dx.doi.org/10.1103/PhysRevD.98.065012}{{\em Phys. Rev.} {\bf
  D98} (2018) no.~6, 065012},
\href{http://arxiv.org/abs/1807.04296}{{\tt arXiv:1807.04296 [hep-th]}}.

\bibitem{Feigin:2018bkf}
B.~Feigin and S.~Gukov, ``{VOA[$M_4$]},''
\href{http://arxiv.org/abs/1806.02470}{{\tt arXiv:1806.02470 [hep-th]}}.

\bibitem{Creutzig:2018lbc}
T.~Creutzig, ``{Logarithmic W-algebras and Argyres-Douglas theories at higher
  rank},'' \href{http://dx.doi.org/10.1007/JHEP11(2018)188}{{\em JHEP} {\bf 11}
  (2018)  188},
\href{http://arxiv.org/abs/1809.01725}{{\tt arXiv:1809.01725 [hep-th]}}.

\bibitem{Bonetti:2018fqz}
F.~Bonetti, C.~Meneghelli, and L.~Rastelli, ``{VOAs labelled by complex
  reflection groups and $4d$ SCFTs},''
\href{http://arxiv.org/abs/1810.03612}{{\tt arXiv:1810.03612 [hep-th]}}.

\bibitem{arakawa2018quasi}
T.~Arakawa and K.~Kawasetsu, ``Quasi-lisse vertex algebras and modular linear
  differential equations,'' in {\em Lie Groups, Geometry, and Representation
  Theory}, pp.~41--57.
\newblock Springer, 2018.

\bibitem{Gaiotto:2009we}
D.~Gaiotto, ``{N=2 dualities},''
  \href{http://dx.doi.org/10.1007/JHEP08(2012)034}{{\em JHEP} {\bf 1208} (2012)
   034},
\href{http://arxiv.org/abs/0904.2715}{{\tt arXiv:0904.2715 [hep-th]}}.

\bibitem{Gaiotto:2009hg}
D.~Gaiotto, G.~W. Moore, and A.~Neitzke, ``{Wall-crossing, Hitchin Systems, and
  the WKB Approximation},''
\href{http://arxiv.org/abs/0907.3987}{{\tt arXiv:0907.3987 [hep-th]}}.

\bibitem{Xie:2012hs}
D.~Xie, ``{General Argyres-Douglas Theory},''
  \href{http://dx.doi.org/10.1007/JHEP01(2013)100}{{\em JHEP} {\bf 1301} (2013)
   100},
\href{http://arxiv.org/abs/1204.2270}{{\tt arXiv:1204.2270 [hep-th]}}.

\bibitem{Wang:2015mra}
Y.~Wang and D.~Xie, ``{Classification of Argyres-Douglas theories from M5
  branes},''
\href{http://arxiv.org/abs/1509.00847}{{\tt arXiv:1509.00847 [hep-th]}}.

\bibitem{Wang:2018gvb}
Y.~Wang and D.~Xie, ``{Codimension-two defects and Argyres-Douglas theories
  from outer-automorphism twist in 6d $(2,0)$ theories},''
\href{http://arxiv.org/abs/1805.08839}{{\tt arXiv:1805.08839 [hep-th]}}.

\bibitem{arakawa2018chiral}
T.~Arakawa, ``Chiral algebras of class s and moore-tachikawa symplectic
  varieties,'' {\em arXiv preprint arXiv:1811.01577} (2018)  .

\bibitem{Buican:2015hsa}
M.~Buican and T.~Nishinaka, ``{Argyres--Douglas theories, S$^1$ reductions, and
  topological symmetries},''
  \href{http://dx.doi.org/10.1088/1751-8113/49/4/045401}{{\em J. Phys.} {\bf
  A49} (2016) no.~4, 045401},
\href{http://arxiv.org/abs/1505.06205}{{\tt arXiv:1505.06205 [hep-th]}}.

\bibitem{Buican:2015tda}
M.~Buican and T.~Nishinaka, ``{Argyres-Douglas Theories, the Macdonald Index,
  and an RG Inequality},''
  \href{http://dx.doi.org/10.1007/JHEP02(2016)159}{{\em JHEP} {\bf 02} (2016)
  159},
\href{http://arxiv.org/abs/1509.05402}{{\tt arXiv:1509.05402 [hep-th]}}.

\bibitem{Song:2015wta}
J.~Song, ``{Superconformal indices of generalized Argyres-Douglas theories from
  2d TQFT},'' \href{http://dx.doi.org/10.1007/JHEP02(2016)045}{{\em JHEP} {\bf
  02} (2016)  045},
\href{http://arxiv.org/abs/1509.06730}{{\tt arXiv:1509.06730 [hep-th]}}.

\bibitem{Buican:2016arp}
M.~Buican and T.~Nishinaka, ``{Conformal Manifolds in Four Dimensions and
  Chiral Algebras},''
  \href{http://dx.doi.org/10.1088/1751-8113/49/46/465401}{{\em J. Phys.} {\bf
  A49} (2016) no.~46, 465401},
\href{http://arxiv.org/abs/1603.00887}{{\tt arXiv:1603.00887 [hep-th]}}.

\bibitem{Cordova:2015nma}
C.~Cordova and S.-H. Shao, ``{Schur Indices, BPS Particles, and Argyres-Douglas
  Theories},'' \href{http://dx.doi.org/10.1007/JHEP01(2016)040}{{\em JHEP} {\bf
  01} (2016)  040},
\href{http://arxiv.org/abs/1506.00265}{{\tt arXiv:1506.00265 [hep-th]}}.

\bibitem{Xie:2016evu}
D.~Xie, W.~Yan, and S.-T. Yau, ``{Chiral algebra of Argyres-Douglas theory from
  M5 brane},''
\href{http://arxiv.org/abs/1604.02155}{{\tt arXiv:1604.02155 [hep-th]}}.

\bibitem{Cordova:2016uwk}
C.~Cordova, D.~Gaiotto, and S.-H. Shao, ``{Infrared Computations of Defect
  Schur Indices},'' \href{http://dx.doi.org/10.1007/JHEP11(2016)106}{{\em JHEP}
  {\bf 11} (2016)  106},
\href{http://arxiv.org/abs/1606.08429}{{\tt arXiv:1606.08429 [hep-th]}}.

\bibitem{Creutzig:2017qyf}
T.~Creutzig, ``{W-algebras for Argyres-Douglas theories},''
\href{http://arxiv.org/abs/1701.05926}{{\tt arXiv:1701.05926 [hep-th]}}.

\bibitem{Cordova:2017ohl}
C.~Cordova, D.~Gaiotto, and S.-H. Shao, ``{Surface Defect Indices and 2d-4d BPS
  States},''
\href{http://arxiv.org/abs/1703.02525}{{\tt arXiv:1703.02525 [hep-th]}}.

\bibitem{Buican:2017uka}
M.~Buican and T.~Nishinaka, ``{On Irregular Singularity Wave Functions and
  Superconformal Indices},''
  \href{http://dx.doi.org/10.1007/JHEP09(2017)066}{{\em JHEP} {\bf 09} (2017)
  066},
\href{http://arxiv.org/abs/1705.07173}{{\tt arXiv:1705.07173 [hep-th]}}.

\bibitem{Frenkel:1992ju}
E.~Frenkel, V.~Kac, and M.~Wakimoto, ``{Characters and fusion rules for W
  algebras via quantized Drinfeld-Sokolov reductions},''
\href{http://dx.doi.org/10.1007/BF02096589}{{\em Commun. Math. Phys.} {\bf 147}
  (1992)  295--328}.

\bibitem{Xie:2016uqq}
D.~Xie and S.-T. Yau, ``{New N = 2 dualities},''
\href{http://arxiv.org/abs/1602.03529}{{\tt arXiv:1602.03529 [hep-th]}}.

\bibitem{Xie:2017vaf}
D.~Xie and S.-T. Yau, ``{Argyres-Douglas matter and N=2 dualities},''
\href{http://arxiv.org/abs/1701.01123}{{\tt arXiv:1701.01123 [hep-th]}}.

\bibitem{Xie:2017aqx}
D.~Xie and K.~Ye, ``{Argyres-Douglas matter and S-duality: Part II},''
  \href{http://dx.doi.org/10.1007/JHEP03(2018)186}{{\em JHEP} {\bf 03} (2018)
  186},
\href{http://arxiv.org/abs/1711.06684}{{\tt arXiv:1711.06684 [hep-th]}}.

\bibitem{Dolan:2002zh}
F.~Dolan and H.~Osborn, ``{On short and semi-short representations for
  four-dimensional superconformal symmetry},''
  \href{http://dx.doi.org/10.1016/S0003-4916(03)00074-5}{{\em Annals Phys.}
  {\bf 307} (2003)  41--89},
\href{http://arxiv.org/abs/hep-th/0209056}{{\tt arXiv:hep-th/0209056
  [hep-th]}}.

\bibitem{Seiberg:1994rs}
N.~Seiberg and E.~Witten, ``{Electric - magnetic duality, monopole
  condensation, and confinement in N=2 supersymmetric Yang-Mills theory},''
  \href{http://dx.doi.org/10.1016/0550-3213(94)90124-4,
  10.1016/0550-3213(94)00449-8}{{\em Nucl. Phys.} {\bf B426} (1994)  19--52},
  \href{http://arxiv.org/abs/hep-th/9407087}{{\tt arXiv:hep-th/9407087
  [hep-th]}}.
[Erratum: Nucl. Phys.B430,485(1994)].

\bibitem{Seiberg:1994aj}
N.~Seiberg and E.~Witten, ``{Monopoles, duality and chiral symmetry breaking in
  N=2 supersymmetric QCD},''
  \href{http://dx.doi.org/10.1016/0550-3213(94)90214-3}{{\em Nucl. Phys.} {\bf
  B431} (1994)  484--550},
\href{http://arxiv.org/abs/hep-th/9408099}{{\tt arXiv:hep-th/9408099
  [hep-th]}}.

\bibitem{Gadde:2011uv}
A.~Gadde, L.~Rastelli, S.~S. Razamat, and W.~Yan, ``{Gauge Theories and
  Macdonald Polynomials},''
  \href{http://dx.doi.org/10.1007/s00220-012-1607-8}{{\em Commun.Math.Phys.}
  {\bf 319} (2013)  147--193},
\href{http://arxiv.org/abs/1110.3740}{{\tt arXiv:1110.3740 [hep-th]}}.

\bibitem{Lemos:2015awa}
M.~Lemos and P.~Liendo, ``{Bootstrapping ${\mathcal N}=2$ chiral
  correlators},''
\href{http://arxiv.org/abs/1510.03866}{{\tt arXiv:1510.03866 [hep-th]}}.

\bibitem{Lemos:2016xke}
M.~Lemos, P.~Liendo, C.~Meneghelli, and V.~Mitev, ``{Bootstrapping
  $\mathcal{N}=3$ superconformal theories},''
  \href{http://dx.doi.org/10.1007/JHEP04(2017)032}{{\em JHEP} {\bf 04} (2017)
  032},
\href{http://arxiv.org/abs/1612.01536}{{\tt arXiv:1612.01536 [hep-th]}}.

\bibitem{Buican:2017rya}
M.~Buican and Z.~Laczko, ``{Nonunitary Lagrangians and unitary non-Lagrangian
  conformal field theories},''
  \href{http://dx.doi.org/10.1103/PhysRevLett.120.081601}{{\em Phys. Rev.
  Lett.} {\bf 120} (2018) no.~8, 081601},
\href{http://arxiv.org/abs/1711.09949}{{\tt arXiv:1711.09949 [hep-th]}}.

\bibitem{Buican:2019huq}
M.~Buican and Z.~Laczko, ``{Rationalizing CFTs and Anyonic Imprints on Higgs
  Branches},''
\href{http://arxiv.org/abs/1901.07591}{{\tt arXiv:1901.07591 [hep-th]}}.

\bibitem{Beem:2018duj}
C.~Beem, ``{Flavor symmetries and unitarity bounds in ${\mathcal N}=2$
  SCFTs},''
\href{http://arxiv.org/abs/1812.06099}{{\tt arXiv:1812.06099 [hep-th]}}.

\bibitem{Kozcaz:2018usv}
C.~Koz\c{c}az, S.~Shakirov, and W.~Yan, ``{Argyres-Douglas Theories, Modularity
  of Minimal Models and Refined Chern-Simons},''
\href{http://arxiv.org/abs/1801.08316}{{\tt arXiv:1801.08316 [hep-th]}}.

\bibitem{Agarwal:2018zqi}
P.~Agarwal, S.~Lee, and J.~Song, ``{Vanishing OPE Coefficients in 4d $N=2$
  SCFTs},''
\href{http://arxiv.org/abs/1812.04743}{{\tt arXiv:1812.04743 [hep-th]}}.

\bibitem{Nishinaka:2016hbw}
T.~Nishinaka and Y.~Tachikawa, ``{On 4d rank-one $ \mathcal{N}=3 $
  superconformal field theories},''
  \href{http://dx.doi.org/10.1007/JHEP09(2016)116}{{\em JHEP} {\bf 09} (2016)
  116},
\href{http://arxiv.org/abs/1602.01503}{{\tt arXiv:1602.01503 [hep-th]}}.

\bibitem{Nishinaka:2018zwq}
T.~Nishinaka, S.~Sasa, and R.-D. Zhu, ``{On the Correspondence between Surface
  Operators in Argyres-Douglas Theories and Modules of Chiral Algebra},''
\href{http://arxiv.org/abs/1811.11772}{{\tt arXiv:1811.11772 [hep-th]}}.

\bibitem{Kiyoshige:2018wol}
K.~Kiyoshige and T.~Nishinaka, ``{OPE Selection Rules for Schur Multiplets in
  4D $\mathcal{N}=2$ Superconformal Field Theories},''
\href{http://arxiv.org/abs/1812.06394}{{\tt arXiv:1812.06394 [hep-th]}}.

\bibitem{Dedushenko:2017tdw}
M.~Dedushenko, S.~Gukov, and P.~Putrov, ``{Vertex algebras and 4-manifold
  invariants},''
\href{http://arxiv.org/abs/1705.01645}{{\tt arXiv:1705.01645 [hep-th]}}.

\bibitem{Costello:2018fnz}
K.~Costello and D.~Gaiotto, ``{Vertex Operator Algebras and 3d $\mathcal N=4$
  gauge theories},''
\href{http://arxiv.org/abs/1804.06460}{{\tt arXiv:1804.06460 [hep-th]}}.

\bibitem{Costello:2018swh}
K.~Costello, T.~Creutzig, and D.~Gaiotto, ``{Higgs and Coulomb branches from
  vertex operator algebras},''
\href{http://arxiv.org/abs/1811.03958}{{\tt arXiv:1811.03958 [hep-th]}}.

\bibitem{kac1998vertex}
V.~G. Kac, {\em Vertex algebras for beginners}.
\newblock No.~10. American Mathematical Soc., 1998.

\bibitem{frenkel1992vertex}
I.~B. Frenkel and Y.~Zhu, ``Vertex operator algebras associated to
  representations of affine and virasoro algebras,'' {\em Duke Mathematical
  Journal} {\bf 66} (1992) no.~1, 123--168.

\bibitem{francesco2012conformal}
P.~Francesco, P.~Mathieu, and D.~S{\'e}n{\'e}chal, {\em Conformal field
  theory}.
\newblock Springer Science \& Business Media, 2012.

\bibitem{bouwknegt1993w}
P.~Bouwknegt and K.~Schoutens, ``W symmetry in conformal field theory,'' {\em
  Physics Reports} {\bf 223} (1993) no.~4, 183--276.

\bibitem{arakawa2017introduction}
T.~Arakawa, ``Introduction to w-algebras and their representation theory,'' in
  {\em Perspectives in Lie Theory}, pp.~179--250.
\newblock Springer, 2017.

\bibitem{reeder2012gradings}
M.~Reeder, P.~Levy, J.-K. Yu, and B.~H. Gross, ``Gradings of positive rank on
  simple lie algebras,'' {\em Transformation Groups} {\bf 17} (2012) no.~4,
  1123--1190.

\bibitem{Xie:2015rpa}
D.~Xie and S.-T. Yau, ``{4d N=2 SCFT and singularity theory Part I:
  Classification},''
\href{http://arxiv.org/abs/1510.01324}{{\tt arXiv:1510.01324 [hep-th]}}.

\bibitem{Chacaltana:2012zy}
O.~Chacaltana, J.~Distler, and Y.~Tachikawa, ``{Nilpotent orbits and
  codimension-two defects of 6d N=(2,0) theories},''
  \href{http://dx.doi.org/10.1142/S0217751X1340006X}{{\em Int. J. Mod. Phys.}
  {\bf A28} (2013)  1340006},
\href{http://arxiv.org/abs/1203.2930}{{\tt arXiv:1203.2930 [hep-th]}}.

\bibitem{kac2003quantum}
V.~Kac, S.-S. Roan, and M.~Wakimoto, ``Quantum reduction for affine
  superalgebras,'' {\em Communications in mathematical physics} {\bf 241}
  (2003) no.~2, 307--342.

\bibitem{hitchin1987stable}
N.~Hitchin {\em et al.}, ``Stable bundles and integrable systems,'' {\em Duke
  mathematical journal} {\bf 54} (1987) no.~1, 91--114.

\bibitem{Xie:2014yya}
D.~Xie, ``{N=1 Curve},''
\href{http://arxiv.org/abs/1409.8306}{{\tt arXiv:1409.8306 [hep-th]}}.

\bibitem{feigin2010logarithmic}
B.~Feigin and I.~Y. Tipunin, ``Logarithmic cfts connected with simple lie
  algebras,'' {\em arXiv preprint arXiv:1002.5047} (2010)  .

\bibitem{altschuler1990level}
D.~Altschuler, M.~Bauer, and H.~Saleur, ``Level-rank duality in nonunitary
  coset theories,'' {\em Journal of Physics A: Mathematical and General} {\bf
  23} (1990) no.~16, L789.

\bibitem{adamovic2017conformal}
D.~Adamovi{\'c}, V.~G. Kac, P.~M. Frajria, P.~Papi, and O.~Perse, ``{Conformal
  embeddings of affine vertex algebras in minimal W-algebras II:
  decompositions},'' {\em Japanese Journal of Mathematics} {\bf 12} (2017)
  no.~2, 261--315.

\bibitem{Chcaltana:2018zag}
O.~Chacaltana, J.~Distler, A.~Trimm, and Y.~Zhu, ``{Tinkertoys for the $E_8$
  Theory},''
\href{http://arxiv.org/abs/1802.09626}{{\tt arXiv:1802.09626 [hep-th]}}.

\bibitem{Xie:2017obm}
D.~Xie, ``{$\mathcal{N}=2$ SCFT with minimal flavor central charge},''
\href{http://arxiv.org/abs/1712.03244}{{\tt arXiv:1712.03244 [hep-th]}}.

\bibitem{Minahan:1996fg}
J.~A. Minahan and D.~Nemeschansky, ``{An N=2 superconformal fixed point with
  E(6) global symmetry},''
  \href{http://dx.doi.org/10.1016/S0550-3213(96)00552-4}{{\em Nucl. Phys.} {\bf
  B482} (1996)  142--152},
\href{http://arxiv.org/abs/hep-th/9608047}{{\tt arXiv:hep-th/9608047
  [hep-th]}}.

\bibitem{adamovic2018classification}
D.~Adamovi{\'c}, V.~G. Kac, P.~M. Frajria, P.~Papi, and O.~Perse, ``{On the
  classification of non-equal rank affine conformal embeddings and
  applications},'' {\em Selecta Mathematica} {\bf 24} (2018) no.~3, 2455--2498.

\bibitem{MR3456698}
T.~Arakawa, ``Associated varieties of modules over {K}ac-{M}oody algebras and
  {$C_2$}-cofiniteness of {$W$}-algebras,'' {\em Int. Math. Res. Not. IMRN}
  (2015) no.~22, 11605--11666.

\bibitem{Gaiotto:2008ak}
D.~Gaiotto and E.~Witten, ``{S-Duality of Boundary Conditions In N=4 Super
  Yang-Mills Theory},''
  \href{http://dx.doi.org/10.4310/ATMP.2009.v13.n3.a5}{{\em Adv. Theor. Math.
  Phys.} {\bf 13} (2009) no.~3, 721--896},
\href{http://arxiv.org/abs/0807.3720}{{\tt arXiv:0807.3720 [hep-th]}}.

\bibitem{Gaiotto:2017euk}
D.~Gaiotto and M.~Rap\c{c}\'{a}k, ``{Vertex Algebras at the Corner},''
  \href{http://dx.doi.org/10.1007/JHEP01(2019)160}{{\em JHEP} {\bf 01} (2019)
  160},
\href{http://arxiv.org/abs/1703.00982}{{\tt arXiv:1703.00982 [hep-th]}}.

\bibitem{Creutzig:2017uxh}
T.~Creutzig and D.~Gaiotto, ``{Vertex Algebras for S-duality},''
\href{http://arxiv.org/abs/1708.00875}{{\tt arXiv:1708.00875 [hep-th]}}.

\bibitem{Cecotti:2010fi}
S.~Cecotti, A.~Neitzke, and C.~Vafa, ``{R-Twisting and 4d/2d
  Correspondences},''
\href{http://arxiv.org/abs/1006.3435}{{\tt arXiv:1006.3435 [hep-th]}}.

\bibitem{Argyres:1995jj}
P.~C. Argyres and M.~R. Douglas, ``{New phenomena in SU(3) supersymmetric gauge
  theory},'' \href{http://dx.doi.org/10.1016/0550-3213(95)00281-V}{{\em Nucl.
  Phys.} {\bf B448} (1995)  93--126},
\href{http://arxiv.org/abs/hep-th/9505062}{{\tt arXiv:hep-th/9505062
  [hep-th]}}.

\bibitem{Eguchi:1996vu}
T.~Eguchi, K.~Hori, K.~Ito, and S.-K. Yang, ``{Study of N=2 superconformal
  field theories in four-dimensions},''
  \href{http://dx.doi.org/10.1016/0550-3213(96)00188-5}{{\em Nucl. Phys.} {\bf
  B471} (1996)  430--444},
\href{http://arxiv.org/abs/hep-th/9603002}{{\tt arXiv:hep-th/9603002
  [hep-th]}}.

\bibitem{Cecotti:2012jx}
S.~Cecotti and M.~Del~Zotto, ``{Infinitely many N=2 SCFT with ADE flavor
  symmetry},'' \href{http://dx.doi.org/10.1007/JHEP01(2013)191}{{\em JHEP} {\bf
  01} (2013)  191},
\href{http://arxiv.org/abs/1210.2886}{{\tt arXiv:1210.2886 [hep-th]}}.

\bibitem{Collingwood:1993rr}
D.~Collingwood and W.~McGovern, ``{Nilpotent orbits in semisimple lie
  algebra},''. VanNostrand Reinhold Math.Series, New York, 1993.

\end{thebibliography}\endgroup

\end{document}